\documentclass[twocolumn,trackchanges]{aastex631}
\usepackage{silence} 
\usepackage{natbib}
\usepackage{amsmath}
\usepackage{float}
\usepackage{times}
\usepackage{enumitem}
\usepackage{etoolbox}
\usepackage{graphicx}
\usepackage{amsmath}
\usepackage{bm}
\usepackage{graphics}
\usepackage{url}
\usepackage{epsfig}
\usepackage{wrapfig}
\usepackage{ulem}
\usepackage{verbatim}
\usepackage{float}
\usepackage{lastpage}
\usepackage{CJK}
\usepackage[whole]{bxcjkjatype} 
\usepackage{CJKutf8}
\usepackage[caption=false]{subfig}

\def\lsim{\hbox{\rlap{\raise 0.425ex\hbox{$<$}}\lower 0.65ex\hbox{$\sim$}}}
\def\gsim{\hbox{\rlap{\raise 0.425ex\hbox{$>$}}\lower 0.65ex\hbox{$\sim$}}}

\def\arcsec{\hbox{$^{\prime\prime}$}}
\def\arcdeg{\mbox{$^\circ$}}

\newcommand\N[1]{{\color{blue} \bf #1}} 

\shorttitle{Spectropolarimetry of SN\,2023ixf}
\shortauthors{Vasylyev et al.}

\begin{document}

\title{Spectropolarimetric Evolution of SN\,2023ixf: an Asymmetric Explosion in a Confined Aspherical Circumstellar Medium} 
\correspondingauthor{Sergiy~S.Vasylyev}
\author[0000-0002-4951-8762]{Sergiy~S.Vasylyev}
\email{sergiy\_vasylyev@berkeley.edu}
\affiliation{Department of Astronomy, University of California, Berkeley, CA 94720-3411, USA}

\author[0000-00000-0000-0000]{Luc Dessart}
\affiliation{Institut d'Astrophysique de Paris, CNRS-Sorbonne Universit\'e, 98 bis boulevard Arago, F-75014 Paris, France}

\correspondingauthor{Yi Yang}

\author[0000-0002-6535-8500]{Yi Yang
\begin{CJK}{UTF8}{gbsn}
(杨轶)
\end{CJK}}
\email{yi\_yang@mail.tsinghua.edu.cn}
\affiliation{Department of Astronomy, University of California, Berkeley, CA 94720-3411, USA}

\author[0000-0003-3460-0103]{Alexei~V.~Filippenko}
\affiliation{Department of Astronomy, University of California, Berkeley, CA 94720-3411, USA}
\affiliation{Hagler Institute for Advanced Study, Texas A\&M University, 3572 TAMU, College Station, TX 77843, USA}

\author[0000-0002-1092-6806]{Kishore C. Patra}
\affiliation{Department of Astronomy, University of California, Berkeley, CA 94720-3411, USA}
\affiliation{Department of Astronomy \& Astrophysics, University of California, Santa Cruz, CA 95064, USA}

\author[0000-0001-5955-2502]{Thomas~G.~Brink}
\affiliation{Department of Astronomy, University of California, Berkeley, CA 94720-3411, USA}

\author[0000-0001-7092-9374]{Lifan Wang}
\affiliation{George P.\ and Cynthia Woods Mitchell Institute for Fundamental Physics $\&$ Astronomy, Texas A$\&$M University, 4242 TAMU, College Station, TX 77843, USA}

\author[0000-0002-7706-5668]{Ryan Chornock}
\affiliation{Department of Astronomy, University of California, Berkeley, CA 94720-3411, USA}

\author[0000-0003-4768-7586]{Raffaella Margutti}
\affiliation{Department of Astronomy, University of California, Berkeley, CA 94720-3411, USA}
\affiliation{Department of Physics, University of California, Berkeley, CA 94720, USA}

\author[0000-0002-3739-0423]{Elinor~L.~Gates}
\affiliation{University of California Observatories/Lick Observatory, Mount Hamilton, CA 95140, USA}

\author[0000-0002-6523-9536]{Adam~J.~Burgasser}
\affiliation{Department of Astronomy \& Astrophysics, University of California, San Diego, La Jolla, CA 92093, USA}

\author[0000-0001-8023-4912]{Huei Sears}
\affiliation{Department of Physics and Astronomy, Rutgers, the State University of New Jersey, 136 Frelinghuysen Road, Piscataway, NJ 08854-8019, USA}
\author[0000-0002-1480-9041]{Preethi~R.~Karpoor}
\affiliation{Department of Astronomy \& Astrophysics, University of California, San Diego, La Jolla, CA 92093, USA}

\author[0000-0002-2249-0595]{Natalie~LeBaron}
\affiliation{Department of Astronomy, University of California, Berkeley, CA 94720-3411, USA}

\author[0000-0002-1420-1837]{Emma~Softich}
\affiliation{Department of Astronomy \& Astrophysics, University of California, San Diego, La Jolla, CA 92093, USA}

\author[0000-0002-9807-5435]{Christopher~A.~Theissen}
\affiliation{Department of Astronomy \& Astrophysics, University of California, San Diego, La Jolla, CA 92093, USA}



\author[0009-0002-4843-2913]{Eli~Wiston}
\affiliation{Department of Astronomy, University of California, Berkeley, CA 94720-3411, USA}

\author[0000-0002-2636-6508]{WeiKang Zheng},
\affiliation{Department of Astronomy, University of California, 
Berkeley, CA 94720-3411, USA}



\begin{abstract}
We present complete spectropolarimetric coverage of the Type II supernova (SN) 2023ixf ranging from 1 to 120 days after explosion. Polarimetry was obtained with the Kast double spectrograph on the Shane 3~m telescope at Lick Observatory. As the ejecta interact with circumstellar material (CSM) during the first week, the intrinsic polarization of SN\,2023ixf is initially high at $\lesssim$\,1\,\%, dropping steeply within days down to $\sim$\,0.4\,\% when the ejecta sweep up the optically-thick CSM. The continuum polarization stays low at $\sim$\,0.2\,\% thereafter, until it rises again to $\sim$\,0.6\,\% as the ejecta transition to the nebular phase.  We model this evolution using a combination of archival and newly-computed 2D polarized radiative-transfer models. In this context, we interpret the early-time polarization as arising from an aspherical CSM with a pole-to-equator density contrast $\gtrsim$\,3. We propose that the surge in polarization at late times originates from an asymmetric distribution of $^{56}$Ni deep in the ejecta. The distinct sources of asymmetries at early and late times are consistent with the temporal evolution of the observed polarization and the polarization angle in SN\,2023ixf. 

\end{abstract}



\section{Introduction}\label{sec:intro}
Type II supernovae (SNe II), a subclass of core-collapse SNe (CCSNe) with hydrogen lines in their spectra, offer insights into the terminal stages of massive stellar evolution. Over the last few decades, photometric and spectroscopic surveys from X-rays to radio wavelengths have revealed an emerging diversity among these events. Such a diversity in observed properties cannot be entirely accounted for by properties intrinsic to the progenitor. Instead, interaction between the SN ejecta and  circumstellar matter (CSM) can have observable effects on the light curve as well as on the total-flux spectrum \citep{valenti_diversity_2016,gutierrez_type_2017,morozova_unifying_2017,yaron_confined_2017,hillier_photometric_2019}, particularly in the ultraviolet  \citep[UV;][]{chevalier_emission_1994,ben-ami_ultraviolet_2015,dessart_explosion_2017,vasylyev_early-time_2022,dessart_modeling_2022,bostroem_early_2023}.

Based on the morphology of their light curves, SNe II are further categorized into two subclasses: Type II-P, characterized by a distinctive plateau phase lasting $\sim$\,100 days, and Type II-L, displaying a linearly declining light curve (in magnitudes), with a shorter plateau.
SNe II exhibiting relatively narrow emission lines (e.g., H\,\textsc{i}, He\,\textsc{ii}, N\,\textsc{iii}/\textsc{iv}, C\,\textsc{iii}/\textsc{iv}) superimposed on an otherwise featureless continuum are spectroscopically classified as Type IIn (see \citealt{filippenko_optical_1997} and \citealt{gal-yam_observational_2017} for  reviews of SN classification). The narrow cores of these lines, with full width at half-maximum intensity (FWHM) $\approx 100$ km s$^{-1}$, are produced by recombination photons in the CSM, while the broader (FWHM $\approx 1000$ km s$^{-1}$) Lorentzian wings are attributed to electron scattering of photons within the dense CSM \citep{chugai_broad_2001,khazov_flash_2016,yaron_confined_2017,huang__electron_2018,jacobson-galan_final_2024}. Typically, SNe with such ``flash" features in the first day to weeks after explosion evolve into a more normal SN II-P. However, there is clearly a diversity in the duration and strength of the SN~IIn features. For example, SN\,2013fs showed a short 5-day flash or CSM phase \citep{yaron_confined_2017}, while SN\,1998S had a 2-week phase \citep{fassia_optical_2001,shivvers_early_2015}, and SN\,2010jl exhibited SN~IIn features into the nebular phase \citep{zhang_type_2012}. The duration of these features is dependent on the radius and extent of the CSM \citep{morozova_unifying_2017,morozova_measuring_2018}. For reviews of interacting SNe II (IIn/II-CSM), see \citet{dessart_interacting_2024} and \citet{jacobson-galan_final_2024}. 

\citet{bruch_large_2021,bruch_prevalence_2023} find that a significant fraction of SNe~II exhibit transient flash features in their early-time spectra, indicating a common presence of CSM around red supergiant (RSG) progenitors. \citet{hillier_photometric_2019} show that modest amounts of CSM (up to 0.2\,M$_{\odot}$) are required to explain the early boost in luminosity 
as well as the color evolution for a comprehensive sample of SNe~II. Furthermore, there likely exists an observational bias against detecting SNe~II with confined CSM at radii less than $10^{14}$\,cm, as such observations would require capturing the event within a single day post-explosion, given the typical shock velocities of $\sim 10^4$\,km\,s$^{-1}$ \citep{khazov_flash_2016,yaron_confined_2017}. 

Although the origin and formation channels of the CSM surrounding SN~II progenitors remain debated, several mechanisms have been proposed to describe the mass-loss histories of the RSG stars that give rise to these events. These models include steady-state wind-driven mass loss \citep{mauron_mass-loss_2011}, enhanced mass loss through pulsational instabilities \citep{yoon_evolution_2010}, and eruptive mass-loss episodes \citep{smith_mass_2014}, each of which leaves a distinct imprint on the structure of the resulting CSM. These mass-loss mechanisms are inherently not spherically symmetric, resulting in the formation of an aspherical CSM. Departures from spherical symmetry can also arise from binary interactions \citep{smith_observed_2011}, rapid stellar rotation \citep{maeder_evolution_2010,georgy_circumstellar_2013}, or large-scale convective instabilities \citep{goldberg_shock_2022} in the stellar envelope such as effervescent zone \citep{soker_pre-explosion_2023} and boil-off \citep{fuller_boil-off_2024} mechanisms. 
The presence of aspherical CSM can significantly influence the observed properties of the subsequent explosion, including the light-curve evolution, spectral features, and polarization signatures \citep[e.g.,][]{leonard_non-spherical_2006, mauerhan_multi-epoch_2014}.

The early detection ($<1$ day) and subsequent multiwavelength observations of SN\,2023ixf by both professional and amateur astronomers provided a rare opportunity to probe the progenitor's immediate environment. High-resolution spectroscopy conducted in the first week after explosion revealed the presence and rapid evolution of the flash features on day-long timescales. These observations imply that the progenitor experienced enhanced mass loss leading up to the explosion, resulting in dense and structured CSM with an extent $\lesssim 10^{14}$\,cm \citep{smith_high-resolution_2023,jacobson-galan_sn_2023,bostroem_early_2023,dickinson_immediate_2024}. 
Additionally, photometric analyses show that the presence of such CSM is required to explain the fast rise and luminous peak of the light curve  \citep{hiramatsu_discovery_2023,singh_unravelling_2024,hsu_one_2024}. The structure of the CSM appeared to be complex, possibly comprised of multiple density profiles due to variable mass loss (see Fig. 3 of \citealt{zimmerman_complex_2024}). There is also evidence from ultraviolet (UV) spectroscopy of continuing interaction with an extended wind ($r > 5 \times 10^{14}$\,cm) during the photospheric phase \citep{bostroem_circumstellar_2024}.

X-ray observations provided insights into high-energy processes occurring shortly after the explosion, revealing interactions between the ejecta and the surrounding material \citep{grefenstette_nustar_2023,chandra_chandras_2023}. Additionally, radio observations and initial nondetections of SN\,2023ixf helped constrain the extent of the CSM \citep{berger_millimeter_2023}. 
A followup paper by \citet{nayana_dinosaur_2024} found a discrepancy between the electromagnetically (EM) inferred neutral hydrogen column density (NH$_{\text{int}}$) and the measured NH$_{\text{int}}$ from X-ray observations at early times (see their Figure 6). This discrepancy, along with the relative strength of the Fe K$\alpha$ line in the X-ray spectra, suggests an underestimation of the CSM density along the line of sight, possibly owing to the presence of clumps or global asymmetries.
Furthermore, the modeling of \citet{li_shock_2024} suggests that the sublimation of an optically thick dust shell occurred on a timescale only possible if the CSM departs from spherical symmetry.
Optical observations up to a year after explosion show evidence for continued interaction with a clumpy or asymmetric CSM \citep{hsu_one_2024,kumar_signatures_2025, zheng_sn_2025} 

Spectropolarimetric observations by \citet[][hereafter Paper I]{vasylyev_early_2023}  confirmed the presence of optically thick, radially confined, aspherical CSM around SN\,2023ixf. \citet{singh_unravelling_2024} and \citet{shrestha_spectropolarimetry_2024} corroborated these findings, reporting intrinsic polarization levels up to 1\% in the first few days after explosion. 
Spectropolarimetry provides a direct probe of SN geometry without requiring spatial resolution of the source \citep[e.g.,][]{wang_spectropolarimetry_2008, patat_introduction_2017}. The polarimetric measurements obtained in this work not only support the nonspherical CSM hypothesis derived from other observations, but can also be used to quantify the degree of asphericity. The high polarization levels observed in SN\,2023ixf are consistent with recent studies of SNe~II that have revealed diverse polarization properties. Some Type II-P/L SNe show low polarization during most of the photospheric phase before increasing to 1--2\% during the transition to the nebular phase \citep{leonard_non-spherical_2006,chornock_large_2010}, while others exhibit either consistently high photospheric polarization or remain weakly polarized throughout their evolution \citep{dessart_modeling_2022,mauerhan_asphericity_2017,nagao_aspherical_2019,nagao_evidence_2021, vasylyev_spectropolarimetry_2024, nagao_evidence_2024}. Type IIn SNe have been shown to reach high polarization values at various phases of their evolution (SN 1997eg, \citealt{hoffman_dual-axis_2008}; SN 1998S, \citealt{leonard_evidence_2000}; SN 2006tf, \citealt{smith_sn_2008}; SN 2010jl, \citealt{patat_asymmetries_2011}; SN 2009ip, \citealt{mauerhan_multi-epoch_2014}, \citealt{reilly_spectropolarimetry_2017}; SN 2012ab, \citealt{bilinski_sn2012ab_2018}; SN 2013fs, \citealt{bullivant_sn_2018}; SN 2014ab, \citealt{bilinski_sn_2020}; SN 2017hcc, \citealt{kumar_observational_2019,mauerhan_record-breaking_2024}). A systematic study of 14 SNe IIn with long-lasting CSM interaction (such that the nebular phase is not reached) by \citet{bilinski_multi-epoch_2024} found that the majority are highly intrinsically polarized, with more than half reaching or exceeding 1\% polarization within the first weeks.

The application of polarized radiative-transfer modeling for the interpretation of spectropolarimetric observations of CCSNe has been limited in the literature. This work is one of the few studies that employs physically consistent analytical models to interpret polarization signatures. Existing approaches for modeling polarization in SNe include Markov-Chain Monte Carlo (MCMC) radiative-transfer techniques. For example, the MCMC radiative-transfer code described by \citet{hoflich_asphericity_1991} was used to model the polarization in SNe~II \citep{hoeflich_analysis_1996}, while the SEDONA code, developed by \citet{kasen_time-dependent_2006}, has been used to model various complex configurations (e.g., disk, torus) in SNe~Ia \citep{kasen_analysis_2003}. These codes are capable of handling complex geometries in three dimensions. Analytical models have also been developed to complement MCMC approaches, as the latter are susceptible to Monte Carlo noise and require longer computation time. The analytical approach in this work uses the time-dependent one-dimensional (1D) radiative-transfer \texttt{CMFGEN} code \citep{hillier_cmfgen_2001,hillier_time-dependent_2012} in combination with a long characteristic code \texttt{LONG\_POL} \citep{hillier_calculation_1994,hillier_calculation_1996,dessart_multiepoch_2021} 
for polarized radiative transfer in 2D (see Section \ref{sec:2dpol} for discussion). Another common approach has been to use the analytic formalism developed by \citet{brown_polarisation_1977}, which assumes the ejecta behave as a single-scattering optically thin nebula around a point source. However, this approximation is inadequate for pre-nebular-phase polarimetry as proper treatment of multiple scattering is required \citep{dessart_synthetic_2011}. The \citet{brown_polarisation_1977} formalism is only applicable when the optical depth is well below unity, which is not the case for SN atmospheres during the photospheric phase. The method presented here accounts for such optical-depth effects.


This work serves as a follow-up to Paper I, which describes six epochs of spectropolarimetry obtained by our group at Lick Observatory in the first 15 days after first light. Here we present the full Lick spectropolarimetric sequence, extending to 120 days after explosion. 
The paper is organized as follows. In Section~\ref{sec:obs}, we present our complete spectropolarimetric observations of SN\,2023ixf.  
Section~\ref{sec:2dpol} describes the 2D polarized radiative-transfer modeling approach with \texttt{CMFGEN} and LONG\_POL. In Section~\ref{sec:results}, we present the results of our spectropolarimetric modeling of SN\,2023ixf. An in-depth discussion of the spectropolarimetric properties of SN\,2023ixf and comparison to other well-studied SNe is given in Section~\ref{sec:discussion}. We summarize our results in Section~\ref{sec:conclusion}. 

\section{Observations and Data Reduction}~\label{sec:obs}
SN\,2023ixf was observed using the Kast double spectrograph on the Shane 3\,m reflector at Lick Observatory \citep{miller_ccd_1988, miller_stone_1994}. 
A spectropolarimetric sequence was obtained throughout the first four months,
consisting of 15 epochs 
that spanned from +1.4 to +120 days after first light. The estimated time of first light is MJD~60082.75 \citep{2023arXiv230606097H},  as in Paper 1. The dataset presented in this work is one of the few spectropolarimetric datasets with such extensive coverage and cadence.

Observations and data reduction were carried out following the description provided by \citet{patra_spectropolarimetry_2021}. Telluric lines were removed through comparison with the flux spectrum of the standard star BD\,+262606 \citep{oke_secondary_1983}. 
The 300\,lines\,mm$^{-1}$ grating, a GG455 order-blocking filter, and the 3$\arcsec$-wide slit were adopted as in Paper I. All observations were conducted at airmass $\lesssim$ 1.8, minimizing atmospheric-dispersion effects \citep{filippenko_importance_1982} and enabling north-south slit alignment. High-polarization standard stars HD\,155528 and/or HD\,154445 were observed during each epoch to verify the accuracy and stability of the polarimeter calibration. Throughout these observations, the polarization remained within 0.1\% and the polarization angle within 3\degr\ of the expected values. 
A log of spectropolarimetric observations is presented in Table~\ref{tbl:specpol_log}. 
All spectra have been corrected to the rest frame.


The instrumental polarization is characterized using the methods described in Section 2 of Paper I. Each epoch of spectropolarimetry 
was found to have an instrumental polarization comparable to the typical systematic uncertainties expected from the Kast polarimeter and reduction procedure.
These systematic errors are well above the statistical uncertainties calculated for the Stokes parameters. 
A detailed discussion of the systematic effects expected from the Kast spectropolarimeter can be found in the Appendix of \citet{leonard_is_2001}. 

\begin{figure*}[t]
    \centering
\includegraphics[width=0.75\textwidth]{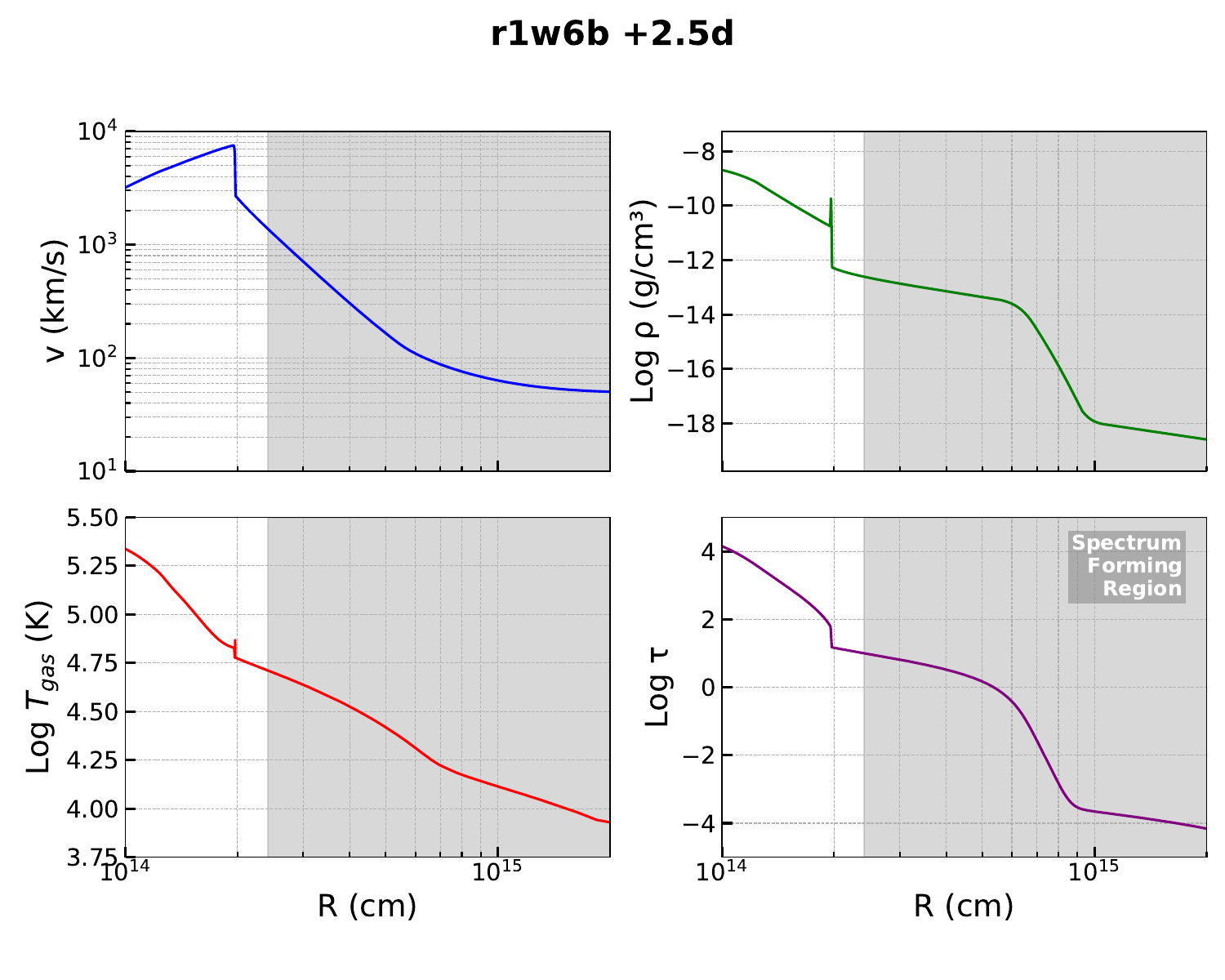}
    \caption{SN structure computed by HERACLES for model r1w6b at 2.5 days after explosion. The bulk of the spectrum forms in the outer layers, where the optical depth $\tau < 10$ (gray shaded region). 
}~\label{fig:r1w6b}
\end{figure*}

\section{2D Polarized Radiative Transfer}~\label{sec:2dpol}

The 2D polarized radiative-transfer models presented here are computed using the long characteristic code \texttt{Long\_Pol},
which solves the radiative-transfer equation for polarized light, computing the polarization as a function of wavelength for axisymmetric geometries at different viewing angles.
The results of the non-local-thermodynamic-equilibrium (non-LTE) time-dependent radiative-transfer code \texttt{CMFGEN} 
are used as input, providing the opacities, emissivities, and electron-density structure for the ejecta. The underlying radiation hydrodynamics was computed using the \texttt{HERACLES} \citep{gonzalez_heracles_2007} and V1D (\citealt{livne_implicit_1993}) codes for the early-flash/CSM phase (See Figure \ref{fig:r1w6b}) and later ejecta phases, respectively.  A limitation of \textsc{Long\_Pol} is its requirement for monotonic flows, specifically homologous expansion. To work within this constraint, we modify the velocity structure from the \textsc{HERACLES} snapshot used by CMFGEN. We impose homologous expansion by setting the velocity according to $v = R/t$, where $t$ is defined as $R_\mathrm{phot}/v_\mathrm{phot}$. 

The current modeling approach computes the full optical polarized spectrum (rather than individual lines) and enables flexible construction of the 2D ejecta geometry. Mirror symmetry about the equatorial plane is adopted to reduce computation time. The code calculates the emergent polarization spectrum for nine viewing angles, equally spaced in inclination angle from pole-on to equator-on (edge-on).
The 2D axisymmetric ejecta models are constructed in two phases as follows.
\begin{enumerate}
\item For the early ``flash'' or CSM phase ($t < 5$ days), a latitudinal density scaling of the 1D r1w6b model $\rho_\text{1D}(r)$ from \citet{jacobson-galan_sn_2023} is used. The r1w6b model assumes a CSM radius or extent of $R_\mathrm{CSM} = 8 \times 10^{14}$ cm and an instantaneous mass-loss rate of $\dot{M} = 10^{-2}$ M$_\odot$ yr$^{-1}$. The density at a given radius $r$ and polar angle $\theta$ is given by
\begin{equation}
\label{eq1}
\rho(r,\theta) = \rho_\text{1D}(r) \times (1 + A\cos^2\theta)\, ,
\end{equation}
where $A$ is the scaling parameter controlling the level of asymmetry. The level of asphericity will be quantified in terms of the pole-to-equator density contrast, parameterized as $(1+A)$; see Equation \ref{eq1}. The model is set up such that $A > 0$ ($A < 0$) corresponds to an prolate (oblate) geometry. This 2D  configuration is symmetric about the azimuthal direction. A more complete description of this method is presented by \citet{dessart_spectropolarimetric_2024}.

\item For the later ``ejecta'' phase ($t > 28$ days), the 1D SN~2012aw models from \citet{dessart_polarization_2021} are mapped onto the 2D grid. Although in our analysis for the ejecta phase we compare to a single bipolar model for varying inclination angles, its features are generally representative of the polarimetric behavior seen in many SNe~II that show similar photospheric-phase evolution to that of SN\,2012aw \citep{dessart_polarization_2021,dessart_evolution_2024}. Discovered in M95 ($\sim 9.9$\,Mpc), SN\,2012aw showed no evidence of flash features or early CSM interaction, instead maintaining broad P~Cygni Balmer lines typical of SNe~IIP/L throughout its 130\,day plateau phase \citep{bose_supernova_2013}. In particular, this approach yields a satisfactory match to SN\,2012aw, including the continuum polarization level rise at the end of the plateau phase. Distinct 1D models are assigned to different latitude ranges in order to simulate large-scale asymmetries in the ejecta structure. Specifically, we use a bipolar explosion geometry \citep{dessart_multiepoch_2021} constructed from two different 1D \texttt{CMFGEN} models: model 1D-X1 ($E_\mathrm{kin}=1.2\times10^{51}\,\mathrm{erg}$, $M_\mathrm{ej}=12.1\,M_\odot$, $M(^{56}\mathrm{Ni})=0.056\,M_\odot$) covering polar angles $22.5^\circ$ to $90^\circ$, and model 1D-X2b ($E_\mathrm{kin}=1.3\times10^{51}$\,erg, $M_\mathrm{ej}=12.4\,M_\odot$, $M(^{56}\mathrm{Ni})=0.047\,M_\odot$), which includes a $^{56}$Ni enhancement at 4000\,km\,s$^{-1}$ to capture stronger mixing and is assigned to polar angles $0^\circ$ to $\beta_{1/2}=22.5^\circ$. The half-opening angle of this inner cone, denoted as $\beta_{1/2}$, determines the boundary between these regions and thus the level of asymmetry in the ejecta structure.  Model 1D-X2b (along the poles) has more $^{56}$Ni at higher velocities, perturbing the shape of the electron-scattering photosphere via radioactive decay heating. Although for the inner ejecta we choose the model with enhanced polar $^{56}$Ni, the model with enhanced polar kinetic energy yields a similar polarized flux. Therefore, it is difficult to distinguish between these two possible sources in the late photospheric phase. \end{enumerate}

Both modeling approaches compute the polarized flux $F_Q$, which is rotated in the coordinate frame such that $F_U = 0$ due to the assumed left-right symmetry. The polarization $p$ (in percent) is then calculated as $100 \times |F_Q/F_I|$, where $F_I$ is the total flux. For  spherical geometry, $F_Q = F_U = 0$ and therefore $p = 0$. We adopt the sign convention where $F_Q > 0$ corresponds to polarization parallel to the projected symmetry axis on the sky, while $F_Q < 0$ indicates perpendicular polarization. The polarization position angle (PA) of the model can be arbitrarily rotated in the plane of the sky about the line of sight by rotating the symmetry axis (i.e., the polar direction) to match the observed PA. 

For model comparisons, we normalize time to the end of the plateau phase to account for differences in plateau duration and evolution rate among different SNe (see Section \ref{sec:comparison} for details). This approach is justified given that SNe II-P/L display comparable polarization evolution during this phase when observed at similar viewing angles and with comparable asphericity.

\begin{figure*}
    \centering
    \includegraphics[width=0.85\textwidth]{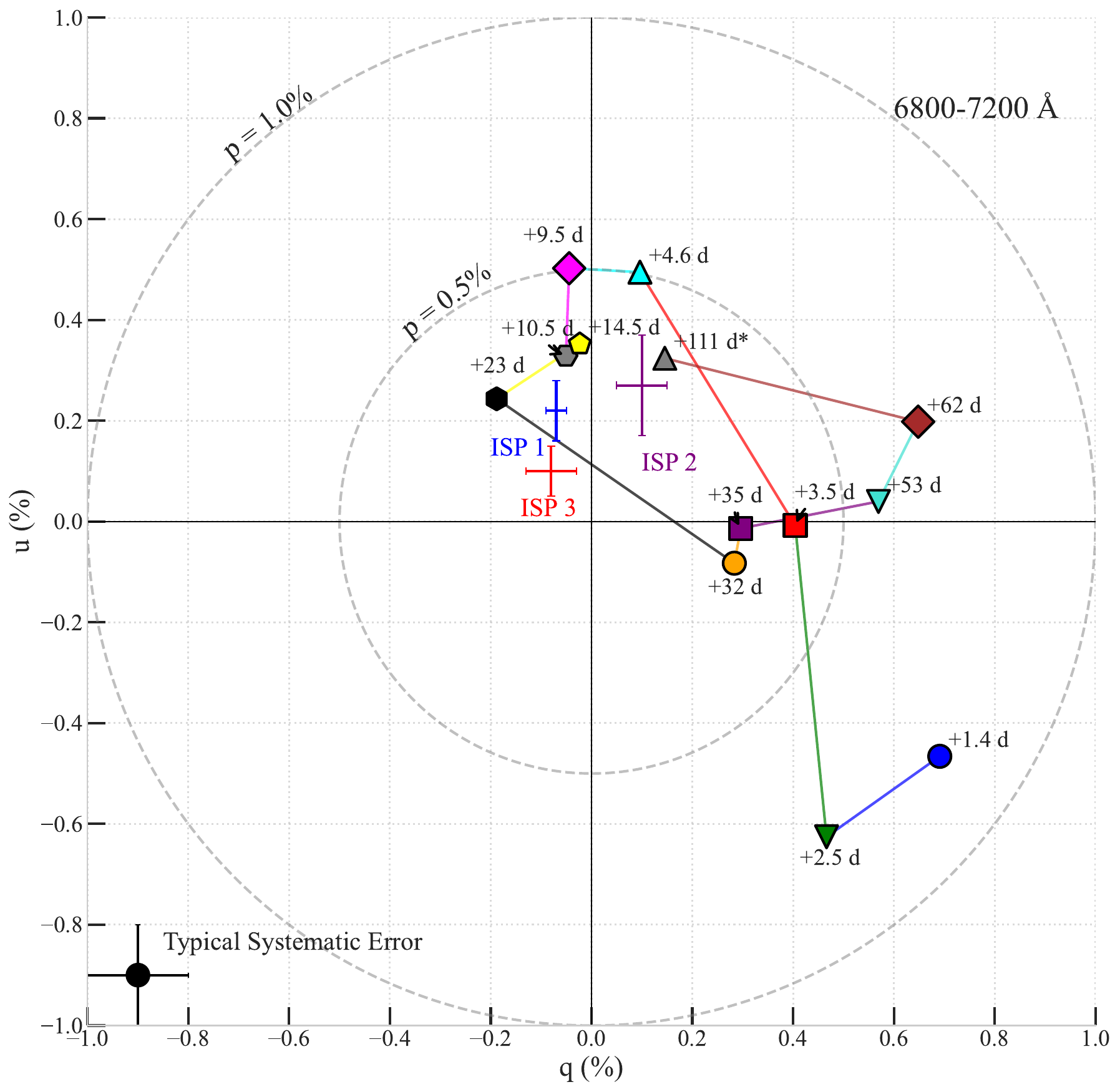}
    \caption{Temporal evolution of the continuum polarization of SN\,2023ixf from days +1.4 to +120 after first light displayed in the Stokes $q$--$u$ plane. The continuum is calculated over the wavelength range 6800--7200\,\AA\ to avoid contamination by strong lines. The phases of the measurements are relative to an estimated time of first light of MJD~60082.75 \citep{2023arXiv230606097H,2023arXiv230607964S}. The scale of the lower-left cross represents the typical systematic error as estimated from adopting various parameters during the extraction of the individual spectrum obtained during the polarization measurement. The outer and the inner dashed circles mark the 1.0\% and the 0.5\% contours of the polarization level, respectively. The blue cross (ISP~1) indicates the ISP estimated by \citet{singh_unravelling_2024}. The purple cross (ISP~2) corresponds to the ISP estimated in this paper, $q_{\text{ISP}} = 0.1 \pm 0.05\%$ and $u_{\text{ISP}} = 0.27 \pm 0.1\%$. The red cross (ISP~3) corresponds to the ISP estimated by \citet{shrestha_spectropolarimetry_2024}, $q_{\text{ISP}} = -0.08 \pm 0.05\%$ and $u_{\text{ISP}} = 0.10 \pm 0.05\%$. The error bars indicate $1\sigma$ uncertainties. 
    }
    \label{fig:story}
\end{figure*}

\begin{figure}
    \centering
\includegraphics[width=0.5\textwidth]{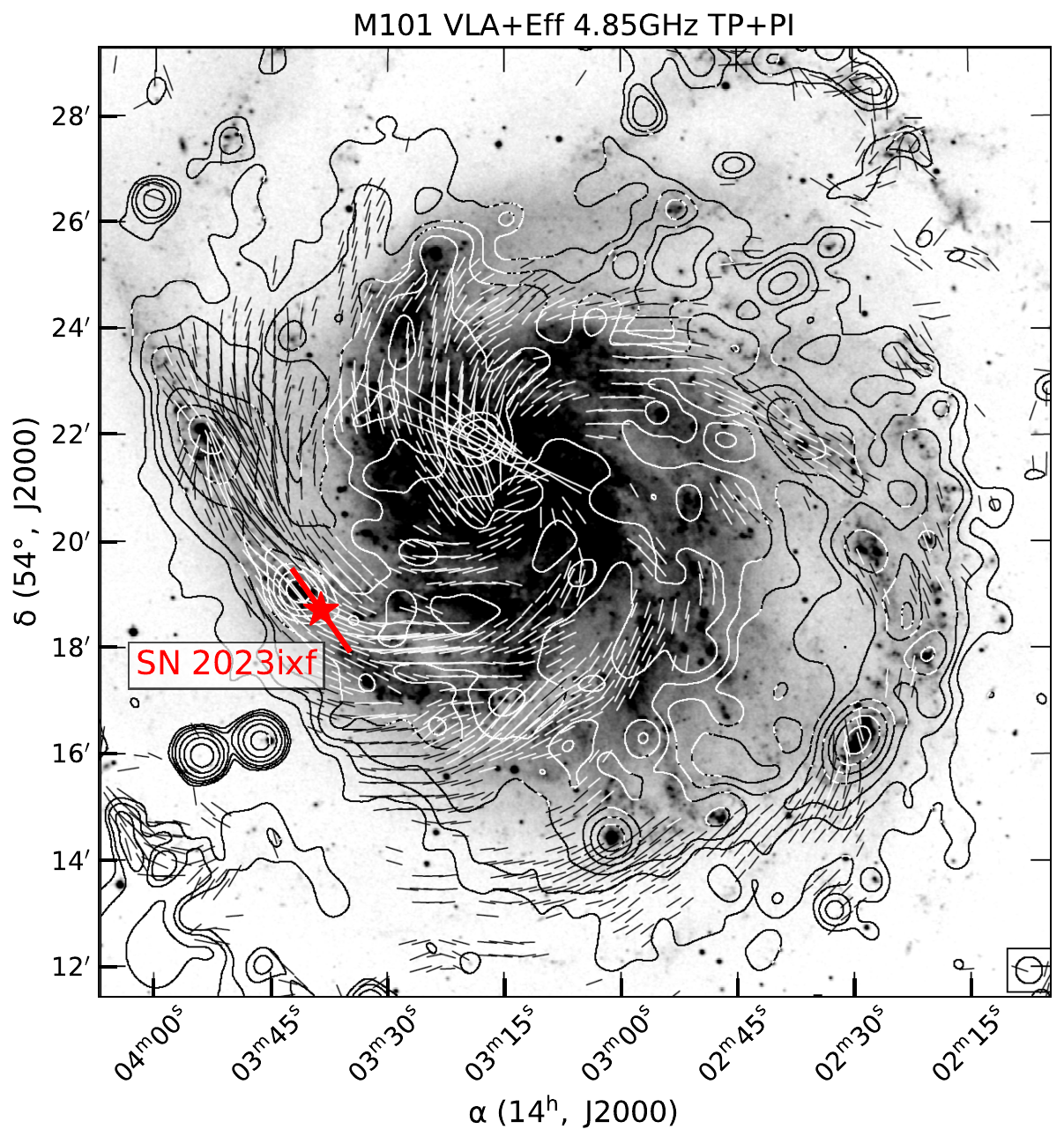}
    \caption{VLA+Effelsberg 4.85\,GHz radio continuum and polarization intensity map of M101, adapted from \citet{berkhuijsen_radio_2016}. The grayscale and contours show the total-intensity radio emission, while the white line segments show the magnetic field orientation as traced by polarized emission from the hot ISM, with their lengths proportional to the polarized intensity. The location of SN\,2023ixf is marked with a red star, and the red line indicates the 
    PA of the ISP derived from our spectropolarimetric observations (Section \ref{sec:isp}). The red line is of arbitrary length. The determined PA$_{\text{ISP}}$ is aligned with the local magnetic field in the spiral arm. 
}~\label{fig:m101}
\end{figure}

\begin{figure*}
    \centering
\includegraphics[width=0.85\textwidth]{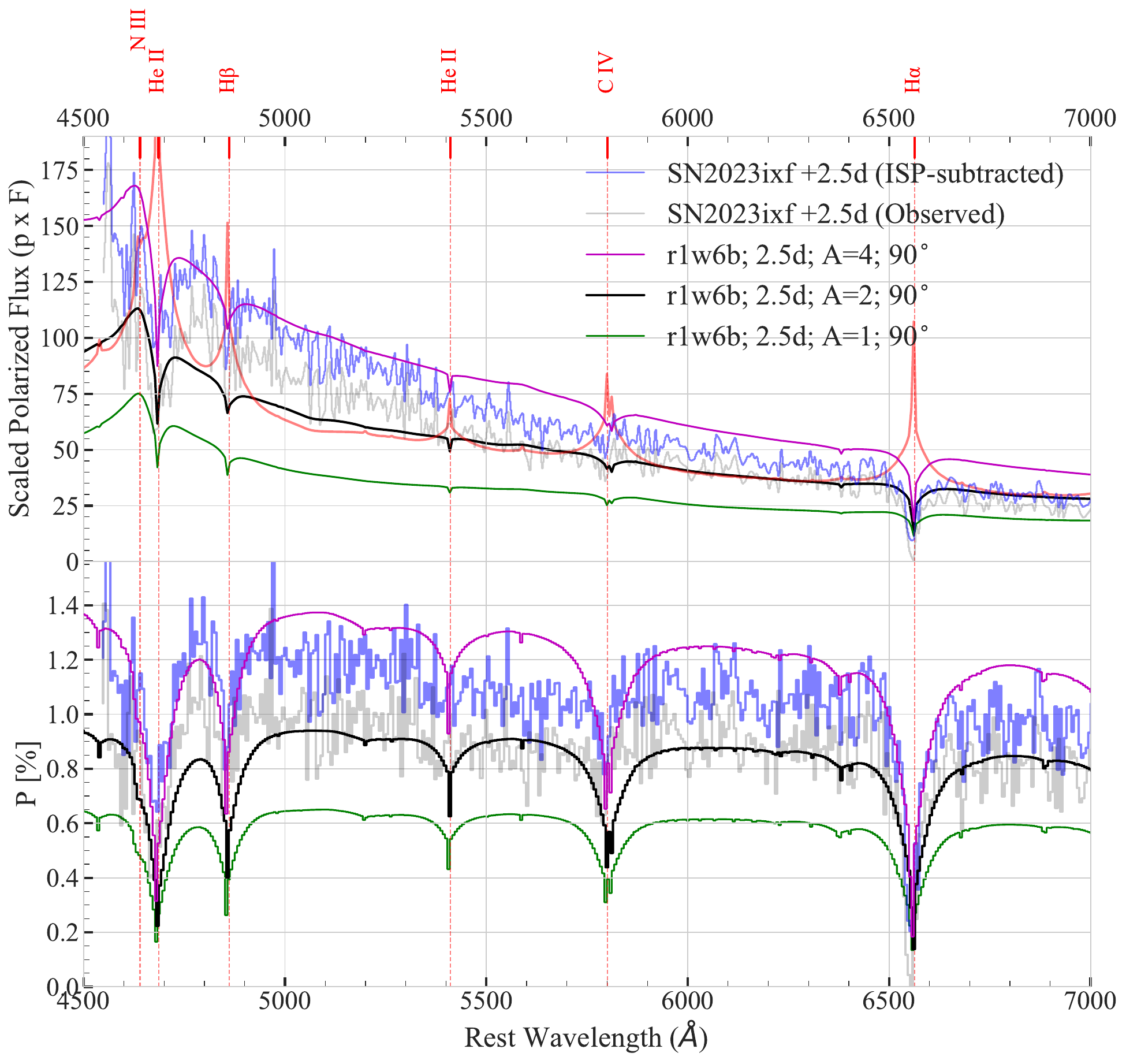}
    \caption{Observed (gray) and ISP-subtracted (blue) polarized flux density (top panel) and fractional polarization (bottom panel) for SN\,2023ixf on day +2.5 compared to the r1w6b model with $A = 1.0$ (green), 2.0 (black), and 4.0 (magenta) with fixed inclination angle $i = 90^{\circ}$. The model flux density is marked as a solid red line.
}~\label{fig:pflux_2p5}
\end{figure*}

\section{Results of Observations and Modeling}~\label{sec:results}
In this section, we present the complete spectropolarimetric sample of SN\,2023ixf obtained with the Kast double spectrograph during the early flash phase, the plateau/photospheric phase, and the onset of the nebular phase. The first two weeks of evolution (days 1.4 to 14.5 after explosion) were presented in Paper I \citep{vasylyev_early_2023}. We expand on that study by comparing the observations to 2D polarized radiative-transfer models and other well-studied SNe with high-quality spectropolarimetric data. See Figure~\ref{fig:story} for a summary of the temporal evolution of continuum polarization and PA, and  Appendix A for the spectropolarimetric observations including the individual Stokes parameters.

\subsection{Interstellar Polarization}~\label{sec:isp}
As discussed in Paper I, the observed polarization of SN\,2023ixf includes a contribution from interstellar polarization (ISP). This ISP is caused by the dichroism of magnetically aligned aspherical dust grains in the Milky Way and host galaxy along the line of sight to the SN. However, the total 
reddening, $E(B-V)_{\text{tot}}$, is low for SN\,2023ixf owing to its high Galactic latitude ($b \approx 59.8^{\circ}$) and its location in a face-on host galaxy, suggesting  $P_{\text{ISP}} \lesssim 0.35$\%. 

With access to photospheric-phase spectropolarimetry, we estimate the ISP by independently fitting Serkowski profiles (\citealt{serkowski_wavelength_1975}; $K = 1.15
$) to the $q$ and $u$ Stokes parameters across the H$\alpha$ and H$\beta$ emission peaks within $\pm$ 100 \AA\ and $\pm$ 50 \AA\ of their respective rest wavelengths (6563 \AA\ and 4861 \AA), while constraining $\lambda_{\text{max}}$ to be identical for both fits. The resulting mean ISP value, measured from days $+32$ to $+62$ (see Figures \ref{fig:specpol_quppa_32}--\ref{fig:specpol_quppa_62}), is  $P_{\text{ISP2}} \approx 0.25$\% at 6000\,\AA, with $q_{\text{ISP2}} = 0.1 \pm 0.05\%$ and $u_{\text{ISP2}} = 0.27\pm 0.1\%$. This ISP determination (marked as ISP~2 in Figure \ref{fig:story}) falls within the expected upper limit based on reddening and is adopted for all subsequent analysis in this work.

\citet{singh_unravelling_2024} find a similarly low contribution to the total polarization from the ISP, $q_{\text{ISP1}} = -0.07 \pm 0.02\%$ and $u_{\text{ISP1}} = 0.22 \pm 0.06\%$, which we will denote ISP~1. They adopt Serkowski-law parameters with $P_{\text{max}} \approx 0.37\%$, $\lambda_{\text{max}} \approx 3650$\,\AA, and $K = 1.15$. \citet{shrestha_spectropolarimetry_2024} find a slightly lower ISP contribution $q_{\text{ISP}} = -0.08 \pm 0.05\%$ and $u_{\text{ISP}} = 0.10 \pm 0.05\%$, denoted as ISP~3. Although this small ISP contribution does not significantly alter our interpretations of the global geometry, we demonstrate that precise ISP determination is crucial for accurately analyzing line-polarization behavior and PA variations. The importance of correcting for the ISP has been discussed in detail by \citet{leonard_evidence_2000} and \citet{stevance_spectropolarimetry_2016} (see also Appendix A of \citealt{leonard_is_2001}). The ISP is time-invariant; therefore, determining it during any particular epoch is sufficient for correcting the data across all observed phases.

We determine the ISP PA 
on the sky and probe the local magnetic-field structure in M101. Maintaining a north-south slit orientation throughout our observations simplifies this procedure. Through the Davis-Greenstein mechanism \citep{davis_polarization_1951,roberge_davisgreenstein_1999}, aspherical dust grains align their short axes (and spin axis) parallel to the magnetic field. Dichroic extinction then preferentially removes light polarized along the grains' long axes, causing the observed polarization PA to trace the magnetic-field direction. In Figure \ref{fig:m101}, we show the inferred ISP polarization angle (PA$_{\text{ISP}} = 35^{\circ} \pm 5^{\circ}$) overlaid on a radio polarization intensity map of M101. The polarization PA of the ISP is aligned with the direction of the local magnetic field in the spiral arm.

\subsection{The Early Flash Phase}

During the first three days after first light, the total-flux spectrum is dominated by narrow lines (FWHM $\approx 100$\,km\,s$^{-1}$) with intermediate-width (FWHM $\approx 1000$\,km\,s$^{-1}$) Lorentzian wings superposed on an otherwise featureless continuum. These features indicate the presence of optically thick CSM, where the electron-scattering photosphere resides at such early times \citep{smith_high-resolution_2023,jacobson-galan_sn_2023}. During this phase, both the total-flux and polarization measurements probe only the geometry of the CSM, as it completely reprocesses the underlying emission from the expanding ejecta. The confined CSM is adopted to have a density profile following $\rho(r) \propto r^{-2}$, as this structure can explain both the shape and rapid evolution of these ionization features, consistent with wind-driven mass loss prior to explosion. However, the physical mechanism driving this wind-like mass loss remains uncertain.

As discussed in Paper I, the continuum polarization (5600--6400\,\AA) initially at $\sim$1\% on days $+1.4$ and $+2.5$ drops to $\sim$0.5\% by day $+3.5$. In this work, we used a smaller bin size of 6\,\AA\ for day $+2.5$ to analyze narrow polarization features. We present the ISP-corrected spectropolarimetry on days $+1.4$, $+2.5$, and $+3.5$ in Figures \ref{fig:specpol_quppa_1p4}, \ref{fig:specpol_quppa_2p5}, and \ref{fig:specpol_quppa_3p5}, respectively. During these three epochs, the polarization is smooth across the wavelength range except for the narrow line cores of He\,\textsc{ii} $\lambda$4686/N\,\textsc{iii} $\lambda$4640, H$\beta$, and H$\alpha$, where there is significant depolarization. As suggested by \citet{leonard_evidence_2000} and \citet{patat_asymmetries_2011}, and later shown analytically by \citet{dessart_synthetic_2011}, the depolarization in the narrow line cores is owing to the dilution of polarized photons by the recombination process. 

In Figure~\ref{fig:r1w6b}, we present the ejecta and CSM structure for the r1w6b model evolved to day $+2.5$ using the \texttt{HERACLES} radiation-hydrodynamics code. At this epoch, the spectrum forms at optical depths $\tau \lesssim 10$. The upper-left panel shows that the velocity profile $v(R)$ output by \texttt{HERACLES} is nonmonotonic. As mentioned in Section~\ref{sec:2dpol} and discussed by \citet{dessart_spectropolarimetric_2024}, \texttt{LONG\_POL} is not currently compatible with a nonmonotonic velocity profile. Therefore, the velocity profile output by \texttt{HERACLES} is modified to follow homologous expansion. 

The results of the 2D polarized radiative-transfer modeling and comparison to the observed polarized flux density ($p \times F$) and percent polarization ($p$) are presented in Figure \ref{fig:pflux_2p5}. In this comparison, the models assume a fixed inclination angle $i = 90^\circ$, corresponding to an edge-on view of the SN. For a given axially symmetric configuration, an edge-on viewing angle maximizes the observed continuum polarization. The asymmetry parameter ``$A$" (see Eq.~\ref{eq1}) is the free parameter in these models. In our analysis, the ``best-fit model'' gives a lower limit on the asymmetry parameter. 

The 2D r1w6b model with $A = 2.0$ (pole-equator density contrast $\rho_{\text{pole}}/\rho_{\text{eq}} =3$) best reproduces the \textit{observed} continuum polarization in SN\,2023ixf while also matching the relative strength of depolarization across the aforementioned lines. The ISP-subtracted polarization requires models with an asymmetry parameter $A > 2$, suggesting a more pronounced pole-to-equator density contrast in the CSM. Our estimation of the intrinsic polarization agrees with that of \citet{singh_unravelling_2024}, who find a continuum polarization level of 1.1\% assuming a slightly different $q_{\text{ISP}}$ and $u_{\text{ISP}}$. The lack of a clearly defined depolarization feature across less-prominent lines such as He\,\textsc{ii} $\lambda$5411 and C\,\textsc{iv} $\lambda\lambda$5801, 5812 is also consistent with the observations. These weaker lines, while present in the total-flux spectrum, do not significantly affect the polarization profile, likely owing to their lower optical depths compared to the stronger lines of H and He\,\textsc{ii} $\lambda$4686. Furthermore, the instrumental resolution (18\,\AA), corresponding to a velocity resolution of $\sim 800$\,km\,s$^{-1}$ at H$\alpha$, may be insufficient to fully resolve any narrow polarization features associated with these weaker lines. This instrumental limitation could potentially smooth out subtle polarization signatures in less-prominent spectral features.

Modeling of the polarized-flux spectrum ($p \times F$) further supports this interpretation of the narrow line core depolarization. In the top panel of Figure \ref{fig:pflux_2p5}, the polarized flux shows clear deficits at the rest wavelengths of narrow line core features, confirming that the recombination process is removing polarized photons from the underlying continuum. These deficits exhibit a characteristic funnel-like shape, reflecting the statistical nature of the electron-scattering process where the probability for photons to be absorbed by the line decreases with increasing distance from the rest wavelength \citep{dessart_interacting_2024}. The model also predicts that electron scattering in the CSM should produce enhancements or subtle ``bumps'' in the polarization at the wavelengths corresponding to the Lorentzian wings; however, the resolution and signal-to-noise ratio (S/N) of our observations are insufficient to definitively detect such signatures. 

This model also predicts the polarized flux $F_Q > 0$, which, as discussed in Section \ref{sec:2dpol}, corresponds to a polarization vector \textit{parallel} to the long symmetry axis of a prolate geometry (asymmetry parameter $A > 0$). This behavior differs from the nebular \citet{brown_polarisation_1977} 
approximation, where the polarization vector would be perpendicular to the long symmetry axis and therefore $F_Q < 0$. Such a difference emphasizes the importance of properly treating optical-depth effects in the prenebular phase, including the polarization's dependence on the distribution of scatterers and of the emergent intensity on the plane of the sky. In this regime, more flux emerges from directions with less CSM, producing a net polarization with the electric field parallel to the long axis in the prolate geometry. 

By day $+9.5$, the flash features are no longer distinguishable in our observations. SN\,2023ixf's spectrum evolves to a smooth, quasiblackbody continuum, marking the start of the cold dense shell (CDS) phase. The CDS forms as the SN ejecta sweep up and accelerate the CSM, while simultaneously decelerating, resulting in a dense shell of piled-up material moving at several thousand km\,s$^{-1}$. The CDS cools efficiently owing to its high density, yet maintains a speed close to that of the outer ejecta. Model spectra by \cite{dessart_modeling_2022} predict narrow absorption components during this CDS phase, which are absent in the observations of SN\,2023ixf (as also noted by \citealt{jacobson-galan_sn_2023}). 

On day $+14.5$ (Fig. \ref{fig:specpol_quppa_14p5}), the quasiblackbody develops only weak broad features associated with the high-velocity ejecta. The observed polarization and PA remain constant across the entire wavelength range at roughly 0.4\% and $230^\circ$, respectively. After subtracting the ISP, the continuum polarization drops to 0.2\%, while the PA stays roughly the same within the uncertainties. The uncertainty in PA becomes large ($>10^{\circ}$) as $p$ approaches 0 because $\sigma_{\rm PA} \propto 1/p$.

\begin{figure*}
    \centering
\includegraphics[width=0.85\textwidth]{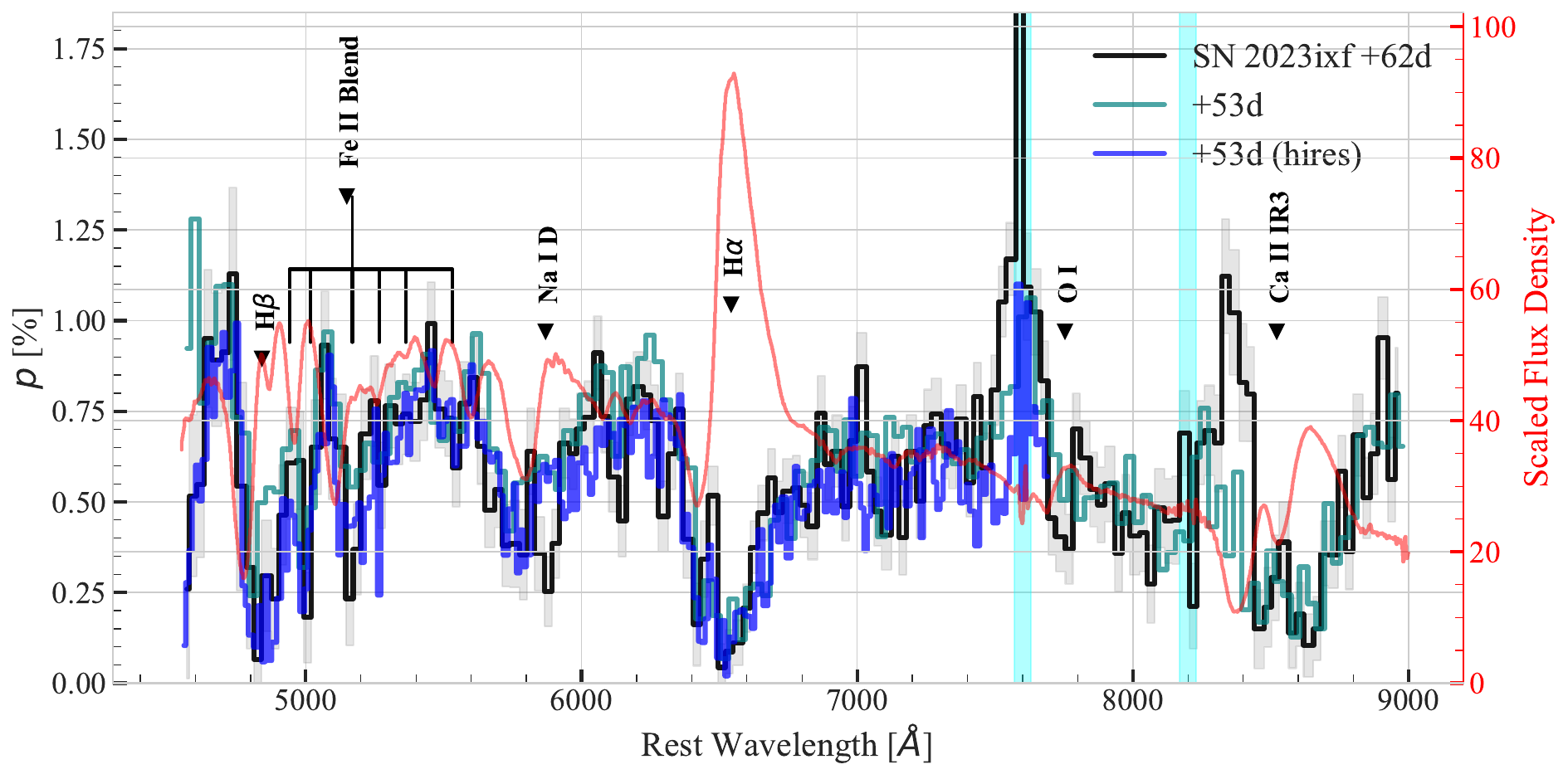}
    \caption{ Spectropolarimetry of SN\,2023ixf on days $+53$ and $+62$. The polarization level (black histogram) is shown for day $+62$, while measurements using both the 300\,lines\,mm$^{-1}$ (teal) and 600\,lines\,mm$^{-1}$ (blue) gratings are shown for day $+53$. The relative-flux spectrum at day $+62$ is plotted in red. Prominent spectral features are labeled at their rest wavelengths, and gray shaded regions indicate 1$\sigma$ statistical uncertainties in the polarization measurements. Cyan vertical bands mark regions affected by imperfect telluric removal. 
}~\label{fig:specpol_53_62}
\end{figure*}
\subsection{The Ejecta Phase}
By approximately three weeks after the explosion, the total-flux spectrum of SN\,2023ixf begins to show broad P~Cygni Balmer lines, indicating that the spectrum is now forming in the fast-moving ejecta that have swept out the CSM. We define the continuum polarization here as the weighted mean polarization calculated in the interval 6800--7200\,\AA, avoiding the prominent lines. The wavelength range adopted to determine the continuum in this study differs from that of Paper I, which used 5600--6400\,\AA. As discussed in that work, at early times, the wavelength choice for the continuum does not change the calculation within the systematic uncertainties. Therefore, it is sufficient to construct the continuum polarization evolution using 5600--6400\,\AA\ for early times and 6800--7200\,\AA\ throughout the plateau phase.

On day $+23$, the observed polarization spectrum remains featureless within statistical uncertainties, with an observed and ISP-subtracted mean continuum polarization of $p_\mathrm{cont} \approx 0.3\%$. As illustrated in Figure~\ref{fig:specpol_quppa_23}, the PA remains consistent with the CDS phase value ($220\degr$), although the uncertainties become large when subtracting the ISP. However, by day $+32$, although the polarization level remains steady, the continuum PA drops back down to the value measured during days $+1.4$ to $+3.5$ and is well constrained. Multiple polarization features emerge above the noise level, as shown in Figure~\ref{fig:specpol_quppa_32}. A polarization trough is observed blueward of the emission peak of H$\alpha$, approximately $-4200$\,km\,s$^{-1}$ from the rest wavelength. In the non-ISP-corrected data, this trough exhibits a distinctive notch-like feature. This notch is also present in the subsequent epochs. However, when accounting for the ISP, two significant changes occur. The notch becomes less prominent or disappears, revealing a more typical depolarization signature commonly observed in many other SNe II (specifically IIP/L), and the wavelength-dependent PA across the H$\alpha$ line becomes statistically insignificant on day $+32$. These changes highlight the critical importance of proper ISP correction in interpreting polarization features, particularly when analyzing fine structures in $p$ and assessing the significance of PA variations. 

Similar behavior is observed on day $+35$ as on day $+32$. The variation in the continuum level between these two epochs for the $q$ and $u$ Stokes parameters is within the systematic uncertainties of $\sim$0.1\%. While poor data quality on day $+35$ produces spurious features adjacent to the blue wing of H$\alpha$, the dominant spectropolarimetric features remain consistent with those observed on day $+32$, including the polarization signatures across H$\beta$, He\,\textsc{i}\,$\lambda$5876/Na\,\textsc{i}\,D, H$\alpha$, and the Ca\,\textsc{ii} near-infrared triplet (NIR3).

We obtained spectropolarimetry on day $+53$ with two instrumental configurations of the Kast spectrograph: the standard low-resolution mode (Fig.~\ref{fig:specpol_quppa_53_lowres}) using the 300\,lines\,mm$^{-1}$ grating, and a higher resolution mode (Fig.~\ref{fig:specpol_quppa_53_hires}) using the 600\,lines\,mm$^{-1}$ grating blazed at 5000\,\AA. The higher resolution mode provides better sampling of spectral features, particularly useful for resolving the wavelength-dependent behavior across the H$\alpha$ line. Data reduction was performed independently on the two configurations. The excellent agreement between the low- and high-resolution modes within the statistical uncertainties serves as an additional confirmation of the robustness of our reduction procedure and its stability against variations in observing conditions.

The observed continuum polarization level on day $+53$ increases to $\sim$ 0.6\%, up from $\sim$ 0.3\% on day $+35$. Both configurations reveal several polarization troughs associated with Fe\,\textsc{ii} lines at 4924, 5018, 5169\,\AA, as well as at 5276\,\AA\ and 5316\,\AA. The consistent location of these features between the high- and low-resolution data provides additional confidence in their detection (see Figure \ref{fig:specpol_53_62}). After applying our ISP correction, the polarization level near the emission peaks of the Balmer lines approaches zero, consistent with all previous epochs showing resolvable P~Cygni profiles. However, we observe wavelength-dependent PA variations across H$\alpha$, and possibly H$\beta$ (though the resolution is insufficient for definitive measurement of the latter). Notably, the PA near the H$\alpha$ emission peak rotates to the values measured for the continuum PA during the CDS and earliest photospheric phases. While an alternative ISP correction using a Serkowski profile that passes through H$\alpha$ at day $+53$ could eliminate this PA feature, such a fit would require an ISP level exceeding the upper limit $P_{\text{ISP}} < 0.35$\% from the reddening law. Since the ISP should remain constant over time, we apply the same average Serkowski parameters to this epoch, even though this results in a sign change in $u$ and consequently a PA rotation of $\sim 50^\circ$. Further discussion of the wavelength-dependent line polarization behavior is presented in Section~\ref{Sec:quplane}.

\subsection{End of the Plateau and Nebular Phases}
As the ejecta continue to expand, the photosphere recedes into deeper layers of the SN. Eventually the photosphere reaches the inner core and the ejecta become optically thin --- for typical Type IIP SNe, this occurs $\sim 100$ days after the explosion. However, the $V$-band light curve of SN\,2023ixf begins to drop off onto the radioactive tail at $\sim 75$ days (\citealt{zimmerman_complex_2024, hsu_one_2024,zheng_sn_2025}), consistent with a Type IIL or short-plateau classification \citep{hiramatsu_luminous_2021}. We use the midpoint of the fall from the plateau to mark the end of the plateau ($t_p = 80$ days) for SN\,2023ixf. Thus, our spectropolarimetric observations on day $+62$ correspond to a fractional phase of 0.8 relative to the end of the plateau, capturing the beginning of the transition toward the nebular phase. 

The continuum polarization reaches its highest post-CSM phase value of $p_\mathrm{cont} \approx 0.7\%$ on day $+62$. This increase in polarization before the onset of the nebular phase --- when the ejecta transition from the multiple- to single-scattering regime and the electron-density profile flattens \citep{dessart_polarization_2021} --- 
is generally consistent with observations of many SNe II \citep{leonard_non-spherical_2006,chornock_large_2010,kumar_broad-band_2014,nagao_aspherical_2019,nagao_spectropolarimetry_2024}. While this increase in polarization during the late-plateau phase is commonly observed, the timing, magnitude, and duration of the rise varies significantly among different SNe II; see Section~\ref{sec:comparison} for further discussion.

\begin{figure}
    \centering
\includegraphics[width=0.5\textwidth]{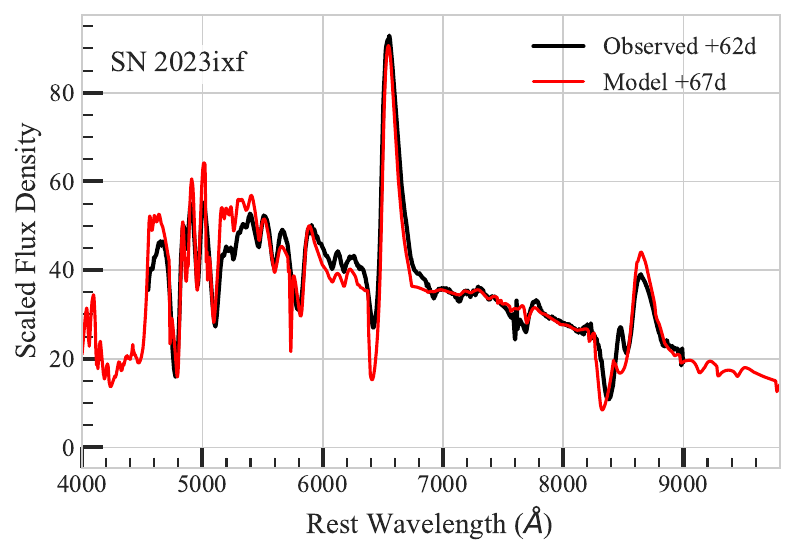}
    \caption{Observed SN\,2023ixf total-flux spectrum on day $+62$ (black) vs. a modified version of the x5p0 model from \citet{hillier_photometric_2019} (red) computed for day $+67$. The modified model assumes an interaction power of $3\times 10^{40}$\,erg\,s$^{-1}$ and a dense shell at $\sim 8000$\,km\,s$^{-1}$.    
}~\label{fig:xp50_67}
\end{figure}

\begin{figure*}
    \centering
\includegraphics[width=\textwidth]{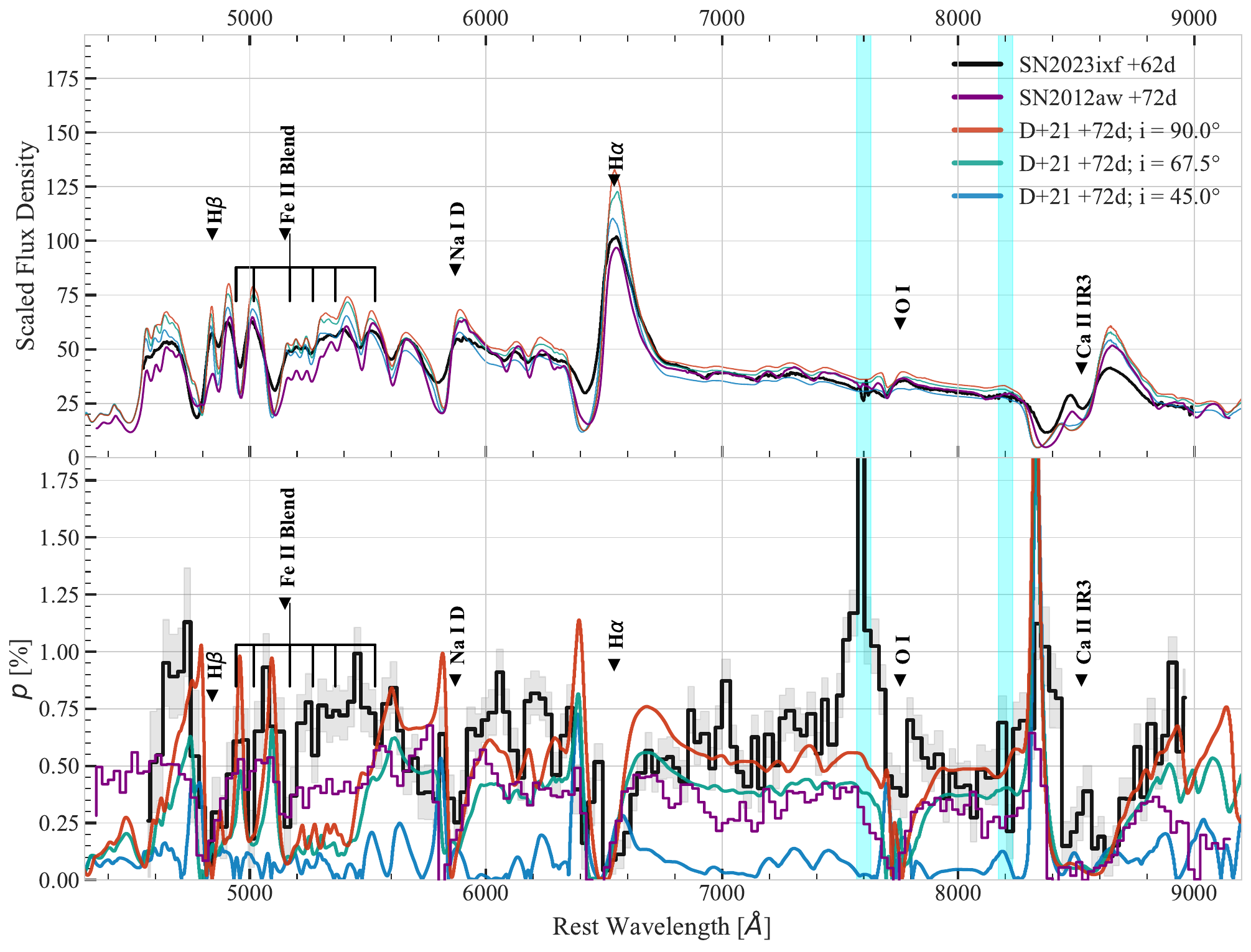}
    \caption{Comparison of SN\,2023ixf observations on day $+62$ (black) and SN\,2012aw on day $+72$ with the bipolar 2D-X1-X2b model from \citet{dessart_multiepoch_2021} (D+21) viewed at different inclination angles (colored lines). {\it Top panel:} Total-flux spectra with prominent spectral features labeled. {\it Bottom panel:} Level of polarization as a function of wavelength. The model was computed for a 2D geometry constructed by combining two different 1D \texttt{CMFGEN} models: one with enhanced $^{56}$Ni mixing (1D-X2b) containing a $^{56}$Ni-rich shell at 4000\,km\,s$^{-1}$ assigned to polar angles $0^{\circ}$ to $22.5^{\circ}$, and 1D-X1 covering polar angles $22.5^{\circ}$ to $90^{\circ}$; see Section \ref{sec:2dpol} for model description. Different viewing (inclination) angles from $45^{\circ}$ to edge-on ($90^{\circ}$) are shown. Gray shaded regions in the polarization panel indicate 1$\sigma$ uncertainties in the observed data, while cyan bands mark regions affected by telluric subtraction.}~\label{fig:pol_comp_62_ts_22}
\end{figure*}

\begin{figure*}
    \centering
\includegraphics[width=\textwidth]{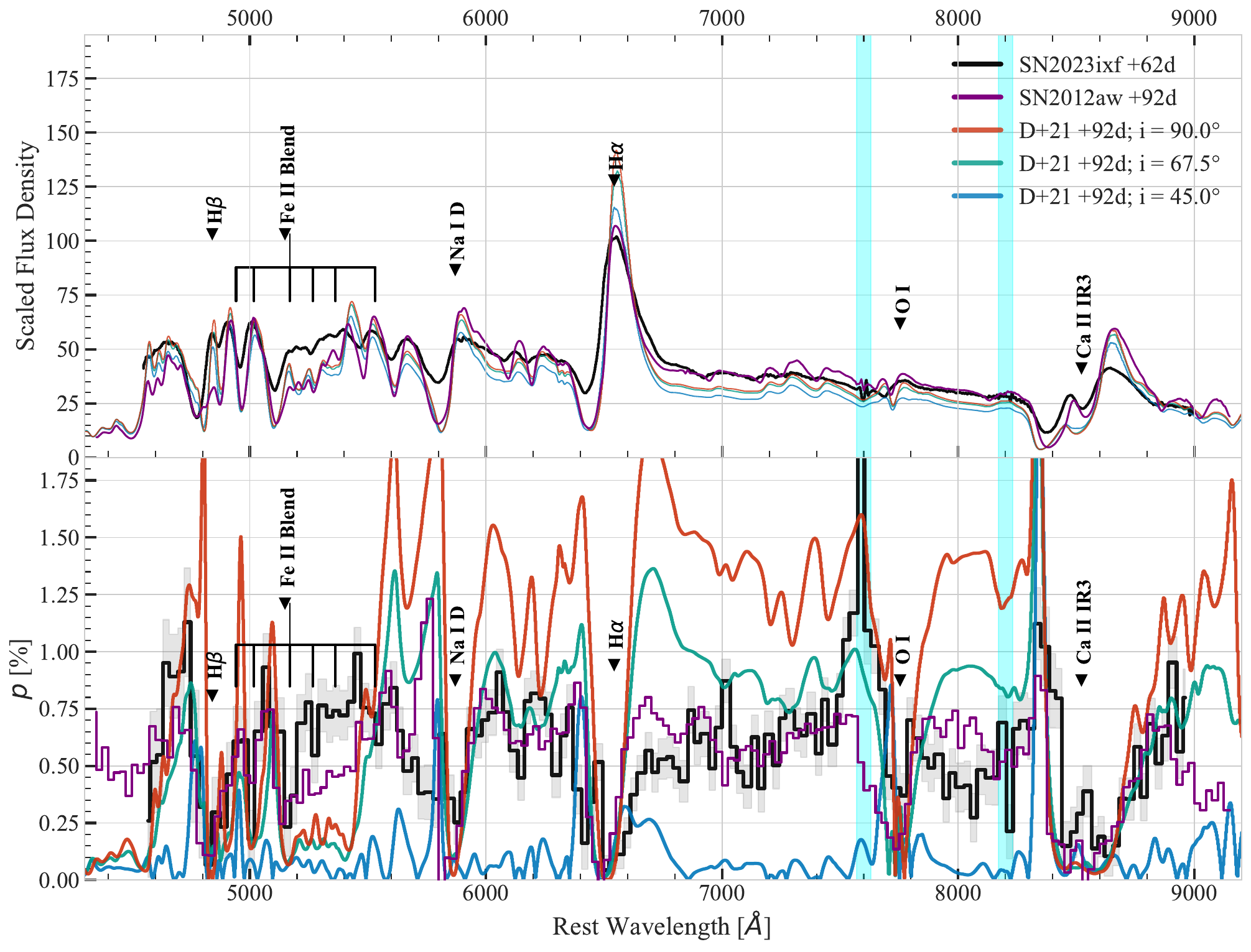}
    \caption{Same as Figure \ref{fig:pol_comp_62_ts_22} but comparing similar phases relative to the end of the plateau (corresponding to SN\,2012aw on day $+92$)}~\label{fig:pol_comp_62_ts_25}.
\end{figure*}

\begin{figure}
    \centering
\includegraphics[width=0.5\textwidth]{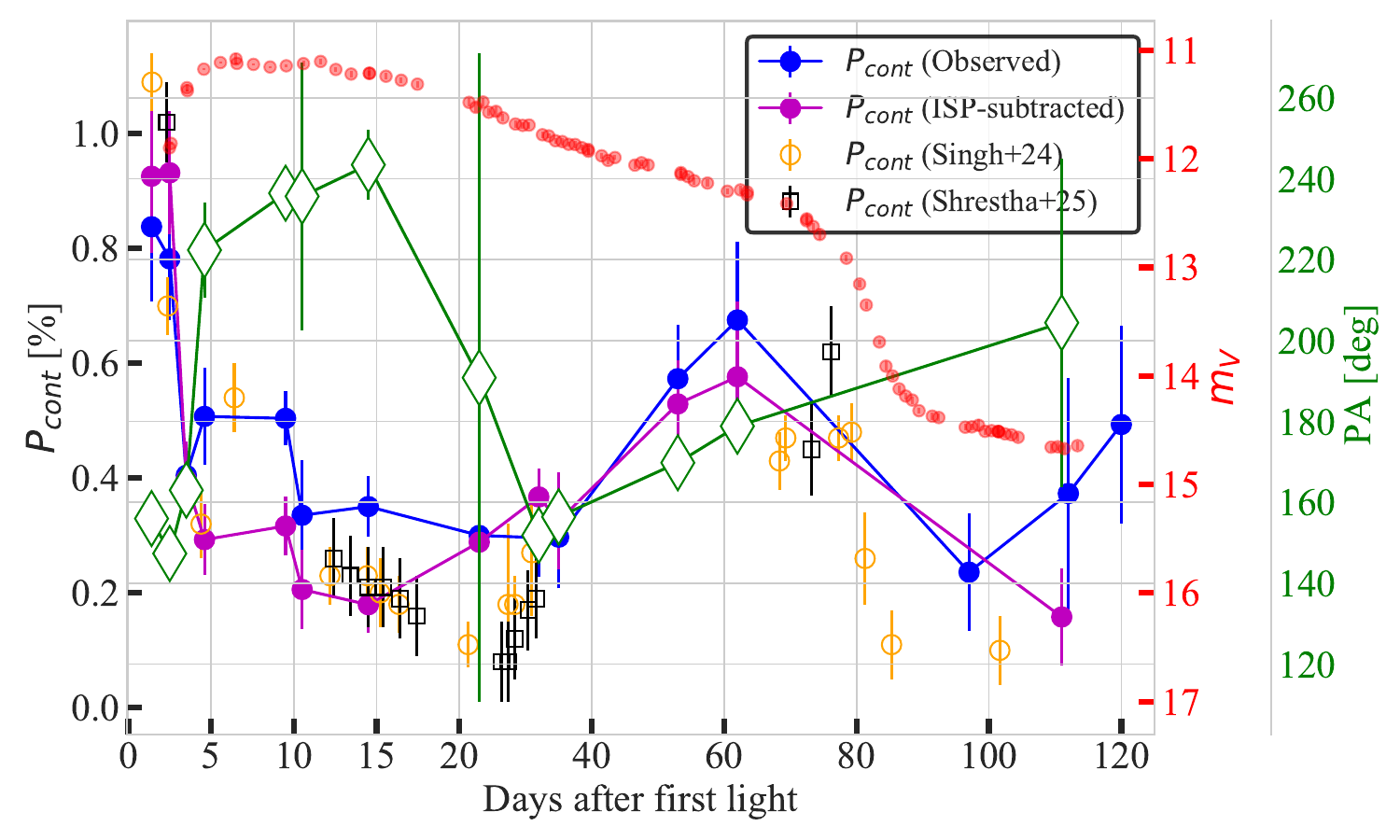}
    \caption{Temporal evolution of the observed (blue points) and ISP-subtracted (magenta points) continuum polarization, and ISP-subtracted PA (green diamonds) for SN\,2023ixf. Yellow (black) open circles (squares) correspond to the ISP-subtracted continuum polarization data from \citet{singh_unravelling_2024}\citep{shrestha_spectropolarimetry_2024}.
     The red points show the apparent $V$-band magnitude ($m_V$) from \citet{zheng_sn_2025}. 
     The magenta point on day $+111$ is the mean ISP-subtracted polarization value between days +97, 112, and 120.
}~\label{fig:tempol}
\end{figure}

\begin{figure*}
    \centering
\includegraphics[width=0.99\textwidth]{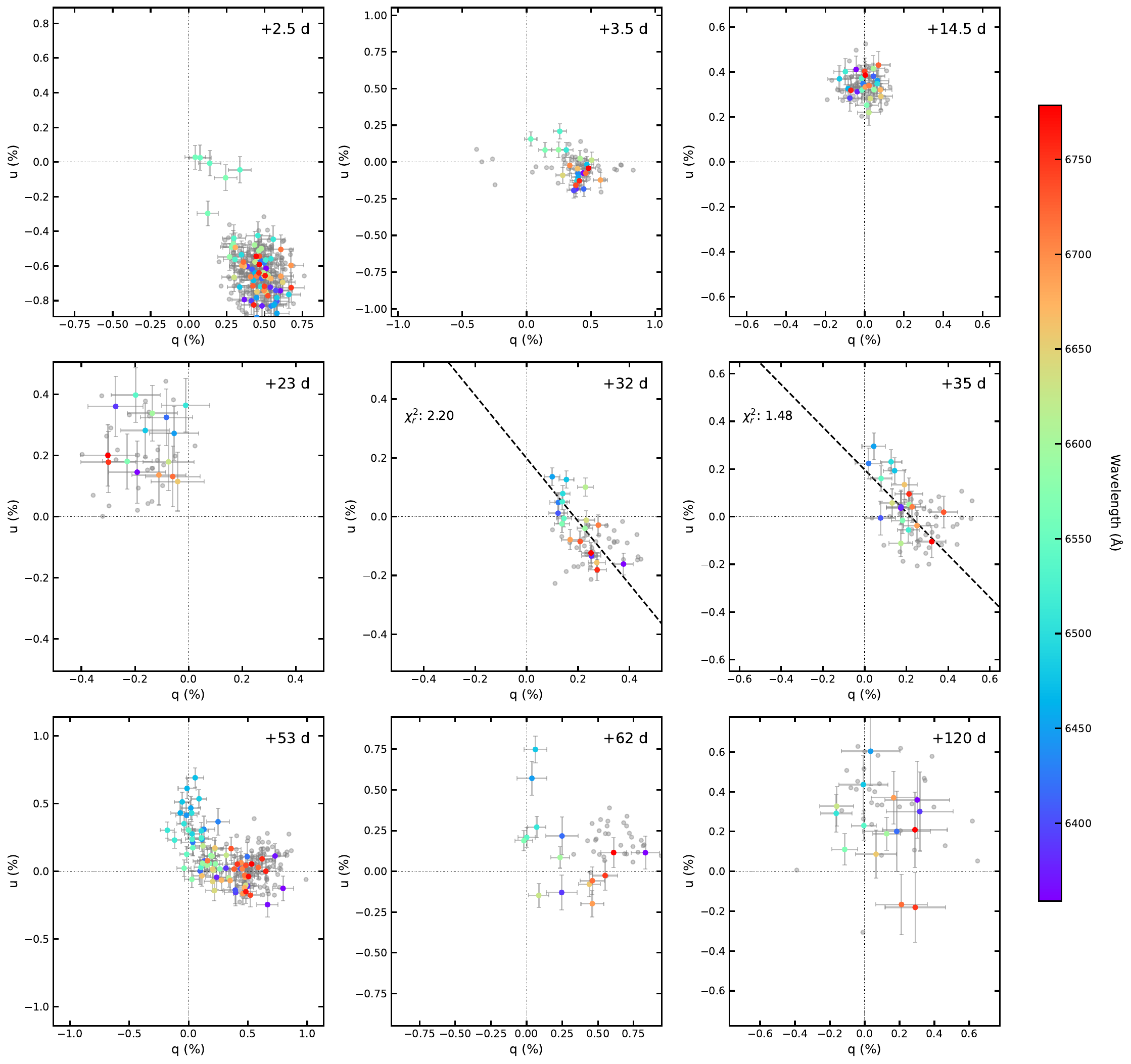}
    \caption{Temporal evolution of the polarization of SN\,2023ixf in the $q$--$u$ plane for selected epochs. Points are colored according to wavelength as shown in the color bar (corresponding to H$\alpha$), with gray points representing continuum-dominated wavelengths. Error bars indicate 1$\sigma$ statistical uncertainties. The black dashed lines in the panels corresponding to days +32 and +35 represents the best fits for the dominant axis. The reduced chi-squared is labeled as $\chi^2_r$
}~\label{fig:specpol_qu}
\end{figure*}

\begin{figure*}
    \centering
\includegraphics[width=0.95\textwidth]{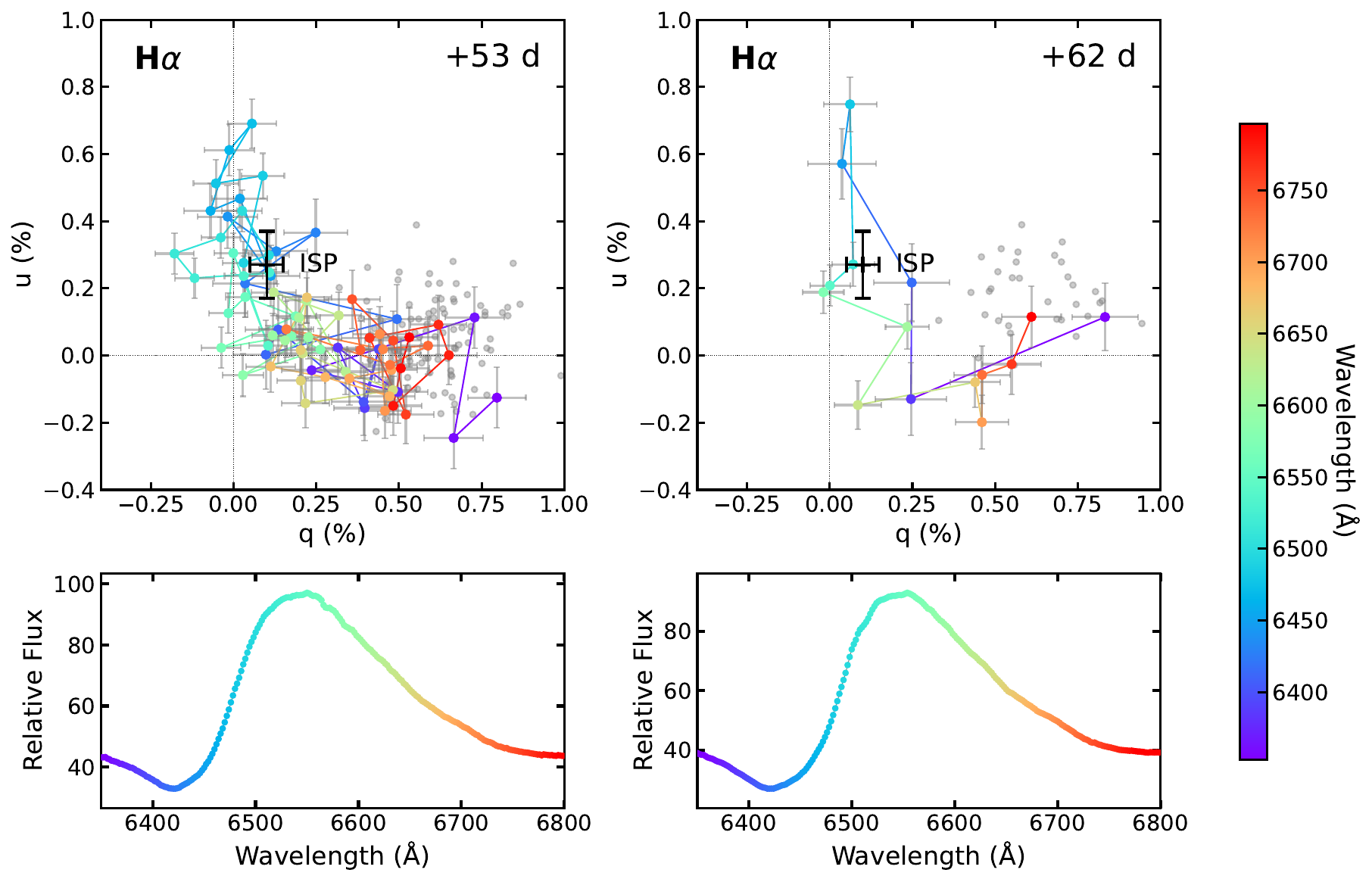}
    \caption{Spectropolarimetry of SN\,2023ixf on days $+53$ (left panels) and $+62$ (right panels). Top panels show the wavelength-dependent Stokes parameters in the $q$--$u$ plane across the H$\alpha$ region (6350--6800\,\AA). Points are colored according to wavelength as shown in Figure \ref{fig:specpol_qu}. Bottom panels show the corresponding total-flux spectra over the same wavelength range on an arbitrary relative flux scale, with colors matching their positions in the $q$--$u$ plane above. Black cross labeled ISP is the estimated interstellar polarization level (ISP~2; see Section \ref{sec:isp}).
}~\label{fig:qu_ha}
\end{figure*}

\begin{figure}
    \centering
\includegraphics[width=0.5\textwidth]{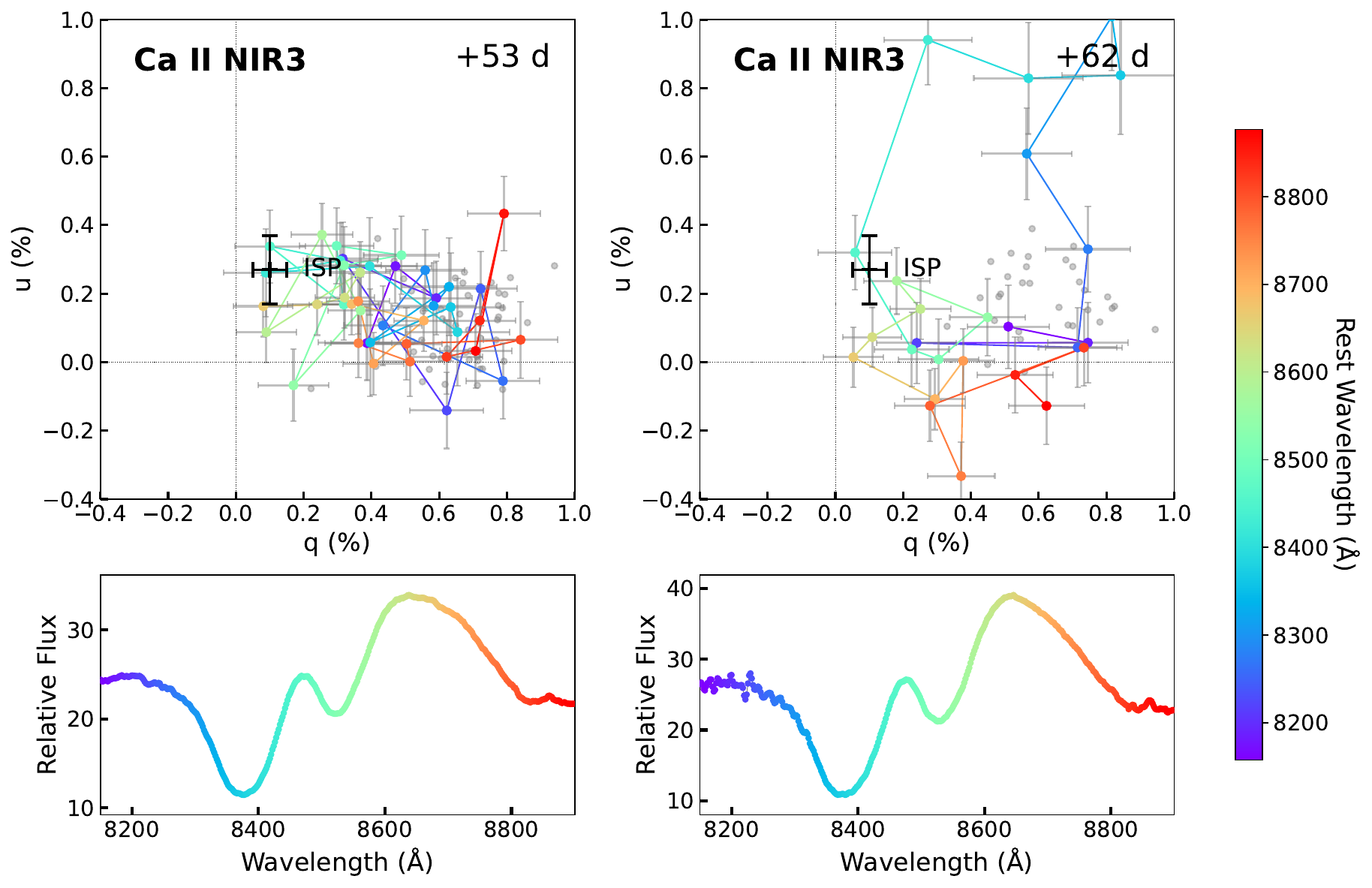}
    \caption{Same as Figure \ref{fig:qu_ha} but for the Ca~II NIR3 region (8250--8900\,\AA). 
}~\label{fig:qu_ca}
\end{figure}

We have computed a model flux spectrum for this epoch that matches the observations well (Fig.~\ref{fig:xp50_67}) using a modified version of the x5p0 model from \citet{hillier_photometric_2019} with an interaction power of $3\times10^{40}$\,erg\,s$^{-1}$ and a dense shell of CSM at $\sim$8000\,km\,s$^{-1}$. While latitudinal density scaling (using a single model) was used for modeling the early-time CSM geometry, this approach becomes physically inappropriate for the ejecta phase. A simple latitudinal density contrast would imply unrealistic mass variations with polar angle rather than the physically meaningful variations in kinetic energy or $^{56}$Ni mixing that are likely responsible for the observed polarization at this phase. A more physically motivated approach, as demonstrated for SN\,2012aw \citep{dessart_multiepoch_2021}, involves combining ejecta from the same progenitor but with different explosion energies or degrees of $^{56}$Ni mixing at different latitudes. While we have not performed detailed polarized radiative-transfer calculations specifically for SN\,2023ixf at this epoch, such modeling would be necessary to properly understand the geometry during this transition phase.

Given that SN\,2023ixf exhibits a light curve and spectroscopic evolution that resemble an SN~IIP/L, and since we have not performed polarized radiative-transfer calculations specifically tailored to SN\,2023ixf, we use the bipolar model geometry (specifically 2D-X1-X2b from \citealt{dessart_multiepoch_2021}) described in Section \ref{sec:2dpol}. In Figure \ref{fig:pol_comp_62_ts_22}, we compare  the model predictions at the closest available spectroscopic phase to our day $+62$ observations for a range of inclinations. That is, we compare the epoch of SN\,2012aw for which it is the closest match spectroscopically to SN\,2023ixf on day $+62$. In the comparison we also include the SN\,2012aw total-flux and ISP-subtracted polarization data from \citet{dessart_multiepoch_2021} corresponding to day $+72$. Many features of the ISP-subtracted polarization in SN\,2023ixf are reproduced by the models favoring an edge-on orientation in the plane of the sky. 

In Figure \ref{fig:pol_comp_62_ts_25}, we instead compare the data at similar phases relative to the end of the plateau ($t/t_p$, corresponding to SN\,2012aw at day $+92$; see Section \ref{sec:comparison}). While edge-on inclinations are still favored, the required viewing angle is less extreme than in the spectroscopic-phase comparison. Notably, the polarization signatures of SN\,2012aw and SN\,2023ixf are remarkably similar when compared at the same fractional plateau phase, despite the longer plateau duration of SN\,2012aw. Both SNe are well described by a subset of models that assume bipolar geometry. However, the models underpredict the polarization level in the blue range (5150--5450\,\AA), where line blanketing from Fe~II and Ti~II should destroy polarized photons from the asymmetric inner ejecta. \citet{dessart_multiepoch_2021}, in the case of SN\,2012aw, suggest this discrepancy might be resolved by introducing a high-velocity $^{56}$Ni enhancement that would act as an external scattering source, producing polarization that scales with the incoming photon flux while preserving some wavelength dependence due to line opacity at large velocities. In the case of SN\,2023ixf, the ejecta ionization is higher as a result of the continued interaction between ejecta and CSM evidenced by UV observations at comparable epochs \citep{bostroem_circumstellar_2024}. This higher ionization reduces line blanketing \citep{dessart_modeling_2022} in SN\,2023ixf, leading to less depolarization in the blue wavelength region.

Following day $+62$ and the plateau dropoff of the light curve, the continuum polarization level decreases to $p_\mathrm{cont} \approx 0.2\%$. To improve the S/N during this late phase, we combined the spectropolarimetry obtained on days $+97$, $+112$, and $+120$, as the polarization remains constant within statistical uncertainties across these three epochs. The rise in polarization before the end of the plateau and subsequent decrease during the nebular phase follows behavior commonly seen in other SNe~II, with the decline in $p$ driven by decreasing optical depth $\tau$. In the single-scattering limit which is applicable during the nebular phase, $p \propto \tau \propto t^{-2}$ owing to geometric dilution as a result of expansion of the ejecta.
A summary of the observed and ISP-subtracted polarization and PA is given in Table \ref{tbl:contpol} and illustrated in Figure \ref{fig:tempol} relative to the light curve in apparent $V$-band magnitude. In this figure, we also compare our results to those of \citet{singh_unravelling_2024} (using HONIR $R$-band imaging polarimetry and SPOL spectropolarimetry on the Bok 2.3\,m telescope) and \citet{shrestha_spectropolarimetry_2024} (using MOPTOP $R$-band imaging polarimetry and SPOL spectropolarimetry). Although both studies utilized different instruments and assumed comparable but distinct ISP values, our ISP-subtracted continuum polarization measurements show broad agreement with their findings, particularly the overall trends of initial decline and later rise in polarization. However, some deviations exist, potentially attributable to the differences in instrumentation, observing conditions, the passbands considered for the continuum measurement, and adopted ISP corrections.

\begin{table*}
\centering
\caption{Observed and ISP-subtracted continuum polarization of SN\,2023ixf$^a$} 
\begin{tabular}{lccccccc}
   \hline 
   \hline
    Phase$^b$ & \multicolumn{4}{c}{Observed} & \multicolumn{2}{c}{ISP-subtracted} \\
    \cline{2-5} \cline{6-7}
    (days) & $q_{\texttt{cont}}$ & $u_{\texttt{cont}}$ & $p_{\texttt{cont}}$ & PA$_{\texttt{cont}}$ & $p_{\texttt{cont,sub}}$ & PA$_{\texttt{cont,sub}}$ \\ 
    & (\%) & (\%) & (\%) & (deg) & (\%) & (deg) \\ 
   \hline 
   1.4 & 0.69(12) & $-$0.47(15) & 0.84(13) & 163(5) & 0.93(13) & 156(4) \\
   2.5 & 0.47(11) & $-$0.63(11) & 0.78(11) & 153(4) & 0.93(11) & 147(3) \\ 
   3.5 & 0.40(05) & $-$0.01(07) & 0.40(05) & 179(5) & 0.40(07) & 163(4) \\ 
   4.6 & 0.10(08) & 0.49(10) & 0.51(08) & 219(6) & 0.29(06) & 222(12) \\
   9.5 & $-$0.04(06) & 0.50(05) & 0.50(05) & 227(3) & 0.32(05) & 237(5) \\
   10.5 & -0.05(12) & 0.33(10) & 0.34(10) & 229(12) & 0.21(07) & 236(33) \\
   14.5 & $-$0.02(05) & 0.35(05) & 0.35(05) & 227(4) & 0.18(05) & 244(9) \\
   23 & $-$0.19(11) & 0.24(10) & 0.30(12) & 244(10) & 0.29(11) & 191(80) \\
   32 & 0.28(07) & $-$0.08(04) & 0.30(07) & 172(5) & 0.37(05) & 152(5) \\
   35 & 0.30(09) & $-$0.01(08) & 0.30(09) & 179(8) & 0.33(09) & 156(7) \\
   53 & 0.57(09) & 0.04(08) & 0.57(09) & 182(4) & 0.53(08) & 170(5) \\
   62 & 0.65(14) & 0.20(09) & 0.68(14) & 189(4) & 0.58(13) & 179(5) \\
   97 & 0.01(15) & 0.23(11) & 0.24(10) & 224(20) & 0.17(10) & 172(58) \\
   112 & 0.22(15) & 0.31(20) & 0.37(20) & 202(24) & 0.26(15) & 198(33) \\
   120 & 0.21(22) & 0.43(19) & 0.49(17) & 213(14) & 0.35(15) & 214(25) \\
   \hline 
\end{tabular}\\
{$^a$}The continuum polarization is calculated over the range 6800--7200\,\AA. \\
{$^b$}{Days after the estimated time of first light on MJD~60082.75 (18 May 2023 UTC).} \\
{$^c$}{Values in parentheses represent the uncertainty in the last two digits of the measurement.}~\label{tbl:contpol}
\end{table*}

\section{Discussion}~\label{sec:discussion}
To summarize, the rapid evolution of the flash features in SN\,2023ixf during the first $\sim 3$ days after the explosion suggests the presence of an optically thick CSM confined to a radius of $\sim 5\times10^{14}$\,cm \citep{bostroem_early_2023} or $\sim 10^{15}$\,cm \citep{2023arXiv230604721J}. X-ray and radio signatures were observed from SN\,2023ixf by \citet{grefenstette_nustar_2023} and \citet{berger_millimeter_2023}, respectively, suggestive of a truncated CSM. An analysis of X-ray and radio observations by \citet{nayana_dinosaur_2024} revealed different density profiles at various radii around SN\,2023ixf, indicating an asymmetric, possibly clumpy CSM structure (along the line of sight) in the immediate environment ($R < 10^{15}$ cm) -- consistent with our spectropolarimetric findings. The unusually red color of the SN in early photometric observations, along with the duration of dust sublimation, also points to such an asymmetry. There is also evidence for continued interaction with an extended wind well into the nebular phase, which traces mass-loss decades prior to the explosion \citep{bostroem_circumstellar_2024,hsu_one_2024,kumar_signatures_2025}.

\subsection{The q--u Plane}
\label{Sec:quplane}
A useful method for characterizing the global geometry and identifying deviations from axial symmetry across individual spectral lines is by plotting the polarization as a function of wavelength in the $q$--$u$ plane. In this representation, the ISP manifests as a translation of the origin to ($q_{\text{ISP}}$, $u_{\text{ISP}}$) in the plane. For nonpolarized configurations, the data will cluster about the origin or ISP. For axially symmetric but aspherical configurations, the data across a spectral line will lie along a line (``dominant axis") that passes through the origin or ISP. In such cases, the wavelength-independent continuum polarization will remain clustered around a single point offset from the origin (or ISP). More complex configurations, particularly those exhibiting departures from global symmetry or multiple symmetry axes for different elements, will show scatter from this dominant axis. As was the case in Figure \ref{fig:story}, a $\theta$ degree rotation in the $q$--$u$ plane  corresponds to $\Delta$PA $= \theta/2$.

Figure~\ref{fig:specpol_qu} illustrates the temporal evolution of the spectropolarimetry in the $q$--$u$ plane from day $+2.5$ to $+120$.
Here, we analyze the polarization behavior across the H$\alpha$ line relative to the continuum wavelengths. 
On day $+2.5$, the continuum and the vicinity of the H$\alpha$ line are clustered. A handful of points capture the depolarization of the narrow H$\alpha$ profile. Owing to the resolution limits, there are not many data points sampled within 1000\,km\,s$^{-1}$ of the rest wavelength of the emission lines. Although there is not a clearly defined dominant axis, the data fall within a cone drawn from the assumed ISP level. There is no evidence for two distinct axes of symmetry during the CSM phase. See Section \ref{sec:comparison} for further discussion and comparison to the $q$--$u$ behavior of SN\,1998S. 

Prior to the ejecta phase ($< 35$ days), the polarization behavior across the continuum and the H$\alpha$ line regions in the $q$--$u$ plane is identical within statistical uncertainties. However, beginning on day $+32$, a distinct pattern emerges: the continuum wavelengths remain clustered, while the wavelengths corresponding to the H$\alpha$ region (6300--6800\,\AA) align along the dominant axis. This behavior is commonly observed during the photospheric or ejecta phase in CCSNe \citep{wang_spectropolarimetry_2008}. The alignment of data along a dominant axis across a spectral line is indicative of asphericity with axial symmetry \citep{tanaka_three-dimensional_2017}. 

The polarization behavior across H$\alpha$ shows distinct evolution in the $q$--$u$ plane. On days $+32$ and $+35$, the data points trace a dominant axis without significant deviations, suggesting axial symmetry. However, by days $+53$ and $+62$ (shown in Fig. \ref{fig:qu_ha}), the polarization can no longer be described by a single dominant axis. The wavelength region between the absorption minimum and emission peak of H$\alpha$ primarily follows $q=0$, while wavelengths outside this region tend to align along $u=0$, indicating a deviation from global axial symmetry.
A similar dual-component structure in the $q$--$u$ plane across H$\alpha$ was observed in SN\,1998S \citep{dessart_evolution_2024} but during the CSM phase ($+5$\,d) instead.  

The Ca~II NIR3 exhibits even more complex behavior in the $q$--$u$ plane. Unlike  H$\alpha$, which can be described by two 
dominant axes, the Ca~II data trace a continuous loop pattern in the $q$--$u$ plane (Fig. \ref{fig:qu_ca}). Such loop structures occur when both the polarization percent and angle vary smoothly across the line profile, indicating that the geometry changes continuously with velocity. This behavior cannot be decomposed into distinct geometric components or explained by a simple departure from axial symmetry. Rather, it suggests that the Ca~II NIR3 line-forming regions may have a more complex spatial 
distribution. 

Unlike asymmetries in SNe Ia and Ib/c, which are often attributed to compositional inhomogeneities \citep{kasen_analysis_2003,cikota_linear_2019}, those in SNe~II are typically linked to nonuniform ionization, particularly in the inner ejecta \citep{chugai_polarization_1992}. Thus, the Ca~II loop might indicate an inhomogeneous calcium ionization structure resulting from uneven decay heating by a clumpy distribution of $^{56}$Ni \citep{chugai_polarization_1992,kasen_time-dependent_2006,hole_spectropolarimetric_2010}. The presence of this loop structure in Ca~II NIR3 but not H$\alpha$ is unexpected, as both lines form within the H-rich envelopes of SNe~II \citep{dessart_synthetic_2011,dessart_type_2013}. Similar loops have been seen in other SNe \citep{kasen_analysis_2003, maund_spectropolarimetry_2007,chornock_deviations_2008, tanaka_three-dimensional_2017,patat_introduction_2017}, but fully replicating them proves difficult for current 2D analytical models. Placing constraints on the scale of these  $^{56}$Ni clumps requires 3D modeling \citep{tanaka_three-dimensional_2017} and is left for future studies.


\subsection{Bipolar Geometry in SNe II}
\label{sec:comparison}
There is mounting evidence that some SNe~II, particularly those exhibiting strong CSM interaction, are bipolar explosions \citep{wang_bipolar_2001,Wang1987A2002ApJ...579..671W,dessart_multiepoch_2021,vasylyev_spectropolarimetry_2024,bilinski_multi-epoch_2024,nagao_evidence_2024}. The spectropolarimetric evolution of SN\,2023ixf provides some of the strongest observational evidence yet for this interpretation, with significant polarization at both early and late times, and a dual axial symmetry. While our modeling indicates equatorial enhancement of the CSM, the geometry does not require extreme configurations such as jets or disk-like structures.

\begin{figure}
    \centering
\includegraphics[width=0.5\textwidth]{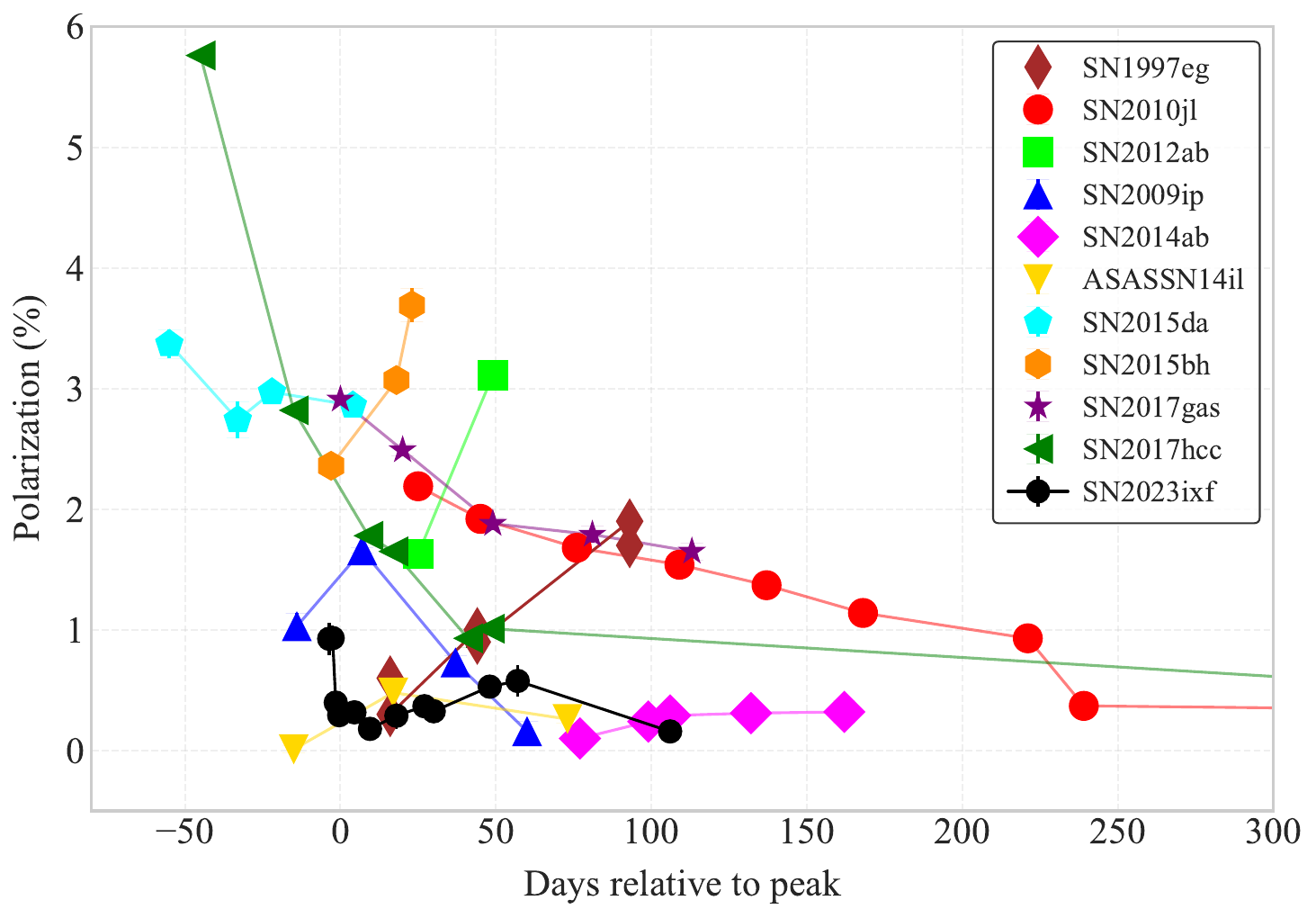}
    \caption{Temporal evolution of the observed continuum polarization of SN\,2023ixf (black points) compared compared to that of other SNe~IIn from the \citet{bilinski_multi-epoch_2024} sample and SN\,1997eg from \citet{hoffman_dual-axis_2008}. 
}~\label{fig:bilinski}
\end{figure}

The polarimetric evolution varies significantly among SNe~II, with distinct patterns emerging when comparing objects with long-lasting SN~IIn signatures to those exhibiting typical SN~IIP spectroscopic features. Complicating this picture, the duration of SN~IIn features varies considerably --- from fleeting flash features lasting only a few days to persistent signatures that endure for months and prevent the formation of a true nebular phase. A recent study of 14 SNe~IIn by \citet{bilinski_multi-epoch_2024} found that more than half of their sample reached or exceeded 1\% polarization, with optically thick CSM persisting for more than 3 months as shown in Figure \ref{fig:bilinski}. Line-core depolarization similar to that of SN\,2023ixf was observed in a subset of this sample (e.g., SN\,1997eg, \citealt{hoffman_dual-axis_2008}; SN\,2010jl, \citealt{patat_asymmetries_2011}).

\begin{figure}
    \centering
\includegraphics[width=0.5\textwidth]{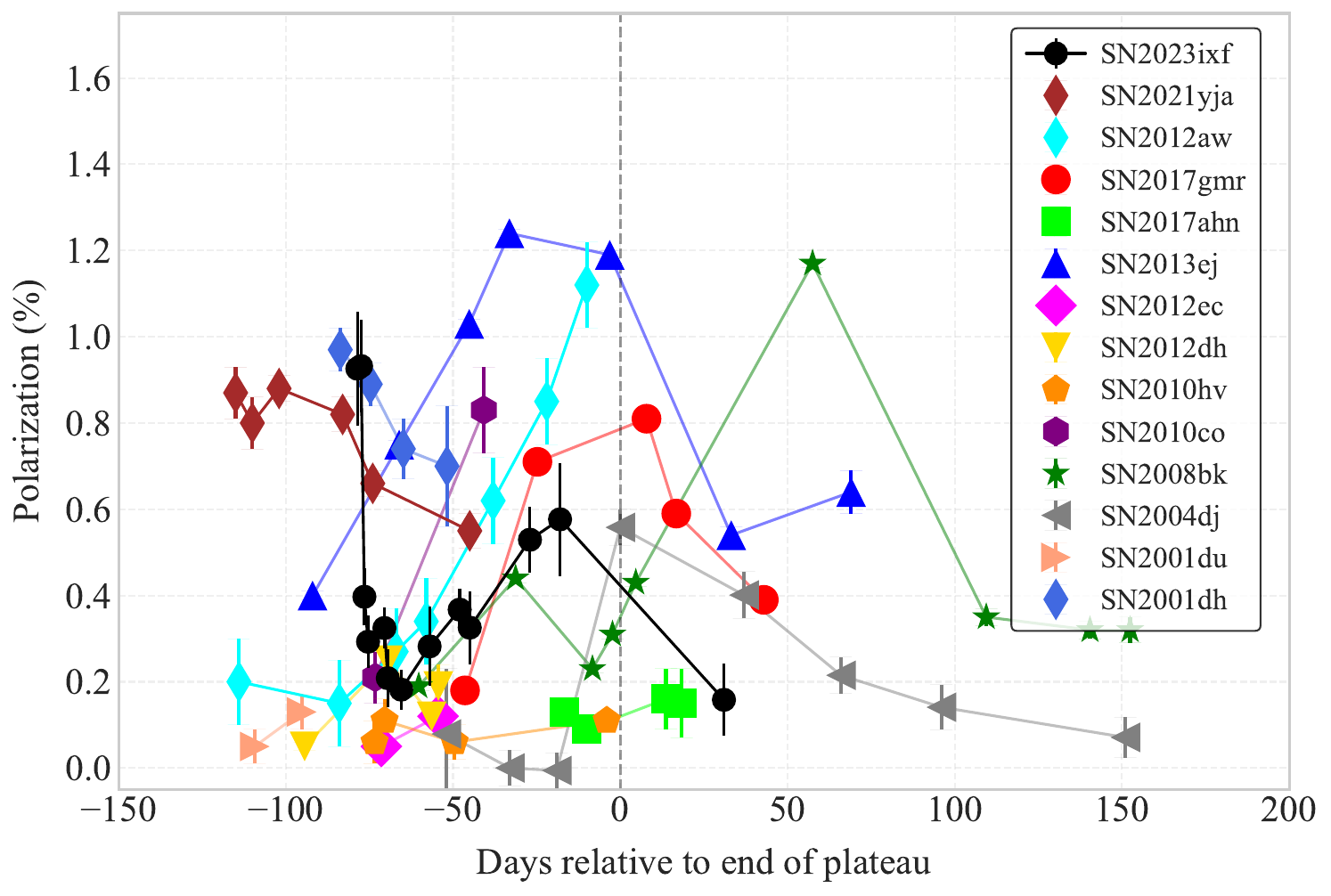}
    \caption{Temporal evolution of the continuum polarization for SN\,2023ixf (black circles) compared with that of other Type IIP/L SNe from \citet{nagao_spectropolarimetry_2024}. Time is shown relative to the end of the plateau phase (vertical dashed line at 0 days) to account for different plateau durations among the sample. Error bars represent 1$\sigma$ uncertainties in the polarization measurements. SN\,2023ixf follows the general trend seen in other SNe~II, with polarization increasing as the SN approaches the end of its plateau phase. 
}~\label{fig:nagao}
\end{figure}

\begin{figure}
    \centering
\includegraphics[width=0.5\textwidth]{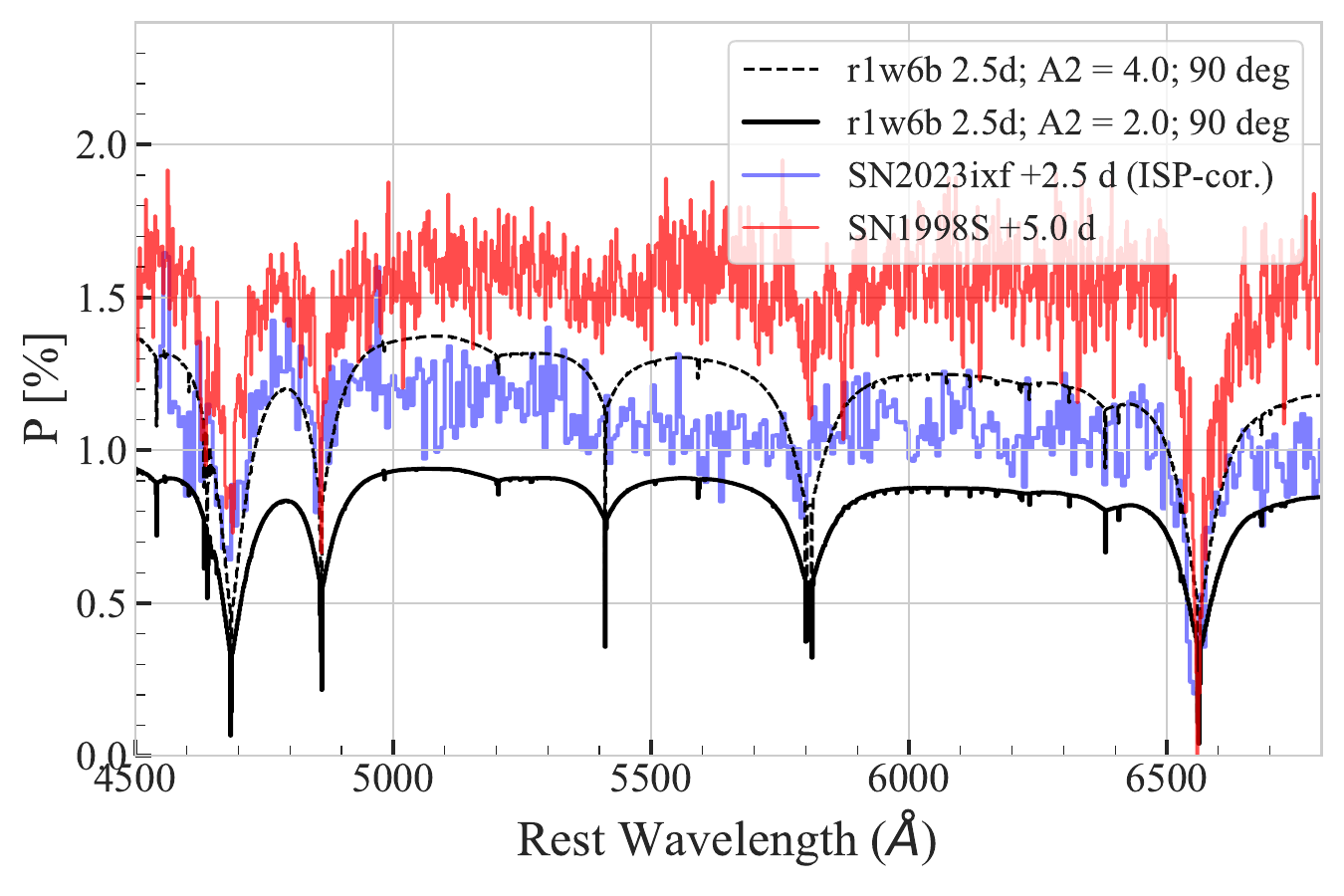}
    \caption{Early-time spectropolarimetry of SN\,2023ixf on day $+2.5$ (blue) compared with the r1w6b bipolar model viewed edge-on with asymmetry parameter $A=2.0$ (black) and ISP-corrected observations of SN\,1998S at day $+5$ (red) from \citet{leonard_evidence_2000}. The model, which has a pole-to-equator density contrast of 3, reproduces both the continuum polarization level and the depolarization features across strong emission lines in SN\,2023ixf. SN\,1998S shows a higher polarization level, requiring a higher degree of asphericity along our line of sight compared with SN\,2023ixf.
}~\label{fig:specpol_23ixf_98s}
\end{figure}

\begin{figure*}
    \centering
    \includegraphics[width=0.7\textwidth]{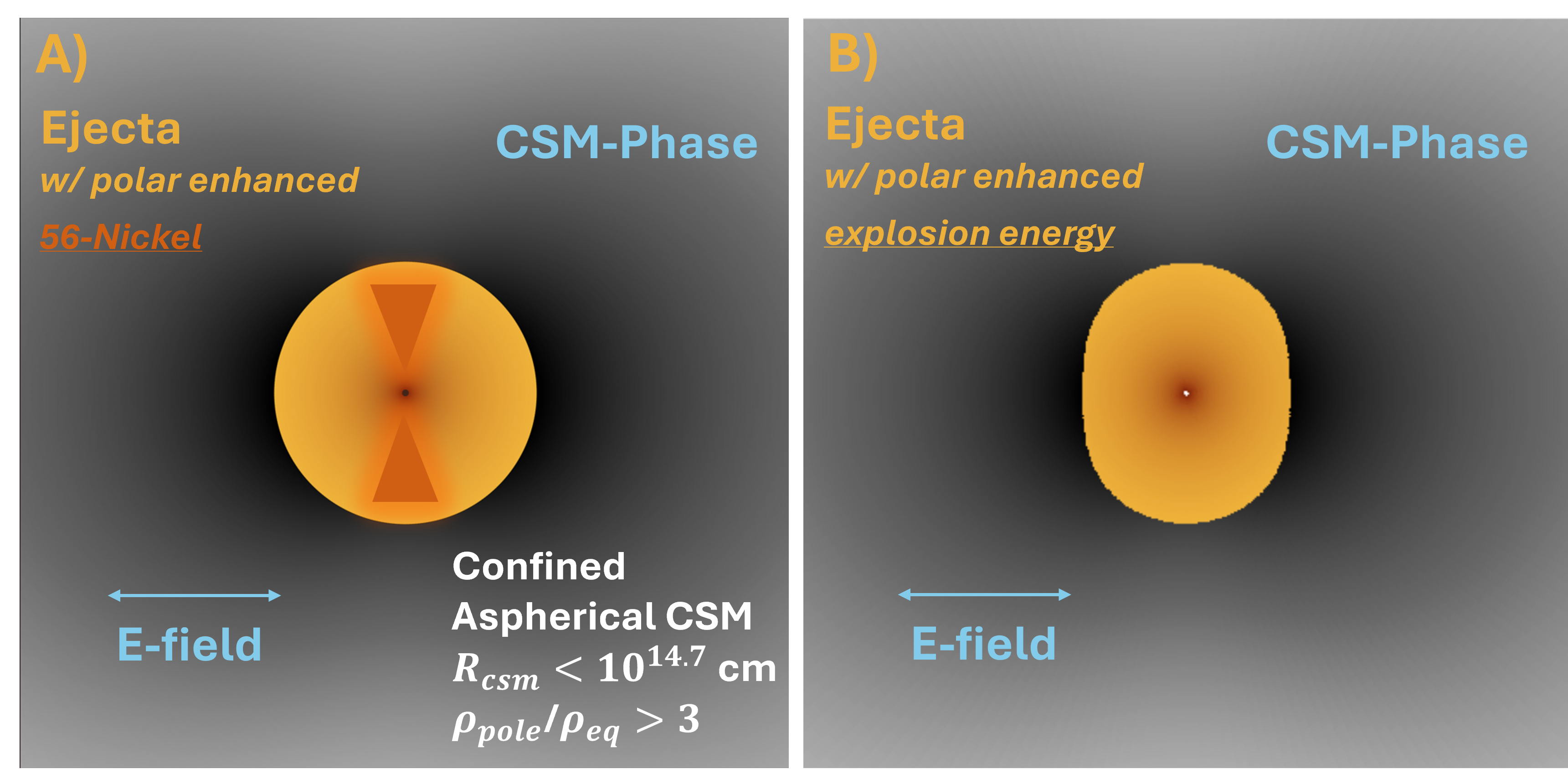}
    \caption{Simplified schematic representations of the density structure of SN\,2023ixf during the early CSM interaction phase, using a latitudinally scaled r1w6b model. The photosphere during this phase is located in the CSM. Scenario A (B) assumes the asymmetry in the inner ejecta is induced by polar enhanced $^{56}$Ni distribution (kinetic energy).  The central orange/red region shows the high-velocity ejecta [$\rho(r) \propto r^{-8}$], surrounded by a wind-like CSM [$\rho(r) \propto r^{-2}(1 + A\cos^2\theta)$] shown in grayscale. The CSM density is enhanced along the poles (assuming a prolate configuration) and extends beyond the field of view, with no sharp outer boundary. The net polarization vector is \textit{parallel} to the long symmetry axis of the CSM. This CSM morphology suggests an aspherical mass-loss history from the progenitor prior to explosion. We note that this picture is idealized and does not capture clumps/instabilities in the CSM or ejecta.}
    \label{fig:story_schematic}
\end{figure*}
In contrast, SNe~IIP/L typically show different behavior --- the recent VLT-FORS2 sample study by \citet{nagao_spectropolarimetry_2024} found that these objects tend to exhibit low polarization (only a few tenths of a percent) at early times, increasing rapidly to a peak within weeks of the plateau dropoff in many cases, as shown for a subset of these SNe alongside SN\,2023ixf in Figure \ref{fig:nagao}. There are, however, notable exceptions to this pattern. SN\,2021yja \citep{vasylyev_spectropolarimetry_2024} and SN\,2001dh \citep{nagao_spectropolarimetry_2024} exhibited high early-time polarization that decreased steadily over weeks, on much longer timescales than observed in SN\,2023ixf. Despite the uncharacteristically high polarization observed for SN\,2021yja during its early photospheric phase, neither optical flux \citep{hosseinzadeh_weak_2022} nor UV observations \citep{vasylyev_early-time_2022} provided significant evidence that this polarization was induced by interaction with dense CSM during the observed epochs. Indeed, the UV spectrum of SN\,2021yja was remarkably similar to that of SN\,2022wsp \citep{vasylyev_early_2023}, a Type IIP SN exhibiting standard evolution without strong CSM interaction signatures. For SN\,2001dh, however, the potential role of CSM interaction remains ambiguous, as no high-resolution spectra confirming such interaction have been published to our knowledge. At the other extreme, SN\,2001du and SN\,2017ahn maintained consistently low polarization ($< 0.2$\%) throughout their evolution (though these two SNe may have been viewed pole-on). 

SN\,2023ixf represents an intermediate case between these two populations. The transformation from Type IIn-like to Type IIP/L-like behavior probably occurs as the CSM is swept up by the expanding ejecta and any additional material in the extended wind is optically thin \citep{zimmerman_complex_2024}. At early times, its spectroscopic and polarimetric behavior closely resembled that of SN\,1998S. However, SN\,1998S exhibited more prolonged SN~IIn features ($\delta t> 5$ days) and higher intrinsic continuum polarization ($p \approx 1.5\%$; \citealt{leonard_evidence_2000}), suggesting a more extended CSM. As illustrated in Figure \ref{fig:specpol_23ixf_98s}, a higher density contrast ($\rho_{\rm pole}/\rho_{\rm eq} >5$, $A = A_2 > 4$; \citealt{dessart_spectropolarimetric_2024}) is required to match the polarization level of SN\,1998S for an edge-on view ($i = 90^{\circ}$), with even greater contrast needed for smaller inclination angles.
During its later evolution, the behavior of SN\,2023ixf more closely matched that of SNe~IIP/L like SN\,2012aw.  While SN\,2012aw exhibited low polarization ($<0.2\%$) for most of its evolution, the continuum polarization increased to $\sim 1.2\%$ by day $+108$, coinciding with the plateau dropoff \citep{dessart_multiepoch_2021}. Unlike the prototypical Type IIP SN\,2004dj \citep{leonard_non-spherical_2006}, this rise in polarization began weeks before the end of the plateau.

While disk-like configurations have been proposed to explain rare events such as SN\,2009ip \citep{mauerhan_multi-epoch_2014,reilly_spectropolarimetry_2017}, our analysis suggests that most Type II SNe can be explained by more modest aspherical geometries. The early-time polarization can be reproduced by an optically thick CSM shell/cocoon with equatorial enhancement and a pole-to-equator density contrast of only a factor of a few. During the later phases, when the inner ejecta become visible, the observed geometry can be explained by enhanced $^{56}$Ni mixing along the polar directions (scenario ``A" in Figure \ref{fig:story_schematic}) or by a a moderate (factor of a few) pole-to-equator contrast in kinetic energy (scenario ``B" in Figure \ref{fig:story_schematic}). Scenario ``A" simulates the effect of polar $^{56}$Ni blobs on the ionization and therefore the shape of the electron-scattering photosphere in the inner ejecta. However, the true distribution is likely more complex, deviating from a simple bipolar geometry (e.g., featuring fingers or clumps) while maintaining an overall preferential axis (See Section \ref{sec:results}). These two scenarios produce similar polarization signatures in the computed models, complicating the determination of whether the observed asymmetry results from enhanced $^{56}$Ni mixing, aspherical kinetic energy distribution, or a combination of both \citep{dessart_multiepoch_2021}. This degeneracy extends to other parameters such as inclination angle, density contrast, half-opening angle $\beta_{1/2}$, and whether the configuration is unipolar or bipolar, further challenging our interpretation. 

The polar-enhanced $^{56}$Ni and kinetic energy models from \citet{dessart_multiepoch_2021} provide a satisfactory match to SN\,2023ixf during the late photospheric phase for inclinations $i > 56^{\circ}$. The PA during this late phase was similar (within tens of degrees) to that observed during the earlier CSM-dominated phase (days +1.4 and +2.5). These models predict a polarized flux $F_Q < 0$ for days +72 and 92, meaning the net polarization vector is \textit{perpendicular} to the long axis of symmetry. Since the late-photospheric-phase PA is roughly aligned with the CSM-phase PA (which itself was parallel to the CSM long axis; see Section~\ref{sec:2dpol}), this suggests that the axis of the $^{56}$Ni or kinetic energy enhancement is oriented \textit{perpendicularly} to the CSM long axis.

A key outstanding question is whether the CSM and ejecta asymmetries share a common origin. Recent theoretical work has explored several mechanisms that could produce bipolar structures in both the explosion and surrounding medium. These include binary interaction shaping the CSM distribution \citep{smith_observed_2011}, asymmetric $^{56}$Ni deposition (plumes/blobs) during core collapse (\citealt{chugai_asymmetry_2006, wongwathanarat_three-dimensional_2015}), and large-scale convective instabilities in the progenitor envelope \citep{goldberg_shock_2022}, including the recently proposed effervescent zone \citep{soker_pre-explosion_2023} and boil-off \citep{fuller_boil-off_2024} mechanisms.

We did not detect significant evolution of the dominant axis between the early phase when the ejecta-CSM interaction is strong and the photosphere resides largely inside the CSM (see top panels of Figure \ref{fig:specpol_qu}) and the later phase when the photosphere is clearly inside the hydrogen envelope of the ejecta (see middle panels of Figure \ref{fig:specpol_qu}). Such correlation between the dominant axes of the CSM and the ejecta asymmetry has been observed in other SNe \citep[e.g.,][]{wang_spectropolarimetry_2008}, and very recently for SN~2024ggi (Yang et al., in prep.). In these observations, the dominant axis of the CSM is typically a few degrees off from the dominant axis of the ejecta, a behavior first observed in SN~1987A \citep{wang_axisymmetric_2002}. The recent spectropolarimetric observations of SN~2024ggi within 1.5 days of first light showed similar alignment (Yang et al., in prep.). These observations strongly suggest that the geometry of core-collapse SNe is intrinsically correlated with their progenitor systems, and that the initial angular momentum of the progenitors plays a critical role in models of CCSNe.

\section{Conclusions}\label{sec:conclusion}
We present a spectropolarimetric sequence of the nearby Type II SN\,2023ixf spanning 1.4 to 120 days after explosion. The first epochs of these observations (days $+1.4$ and $+2.5$), initially presented in Paper I \citep{vasylyev_early_2023}, represent some of the earliest spectropolarimetric measurements ever obtained for an SN and were acquired before the disappearance of narrow, highly ionized emission features. This unique timing and high observational cadence provided an unprecedented view of such ``flash" features, offering direct constraints on the geometry of both the outermost layers of the explosion and its immediate circumstellar environment.

The spectropolarimetric evolution traces two distinct physical regimes. During the first five days, when the spectrum showed SN~IIn-like features, 2D polarized radiative-transfer modeling indicates an optically thick, aspherical CSM with a modest pole-to-equator density contrast of $\sim 3$. This geometry successfully reproduces both the observed polarization level and many of the spectral features without requiring more extreme configurations like disks or tori that have been suggested for other SNe~II. Our analysis supports the visualization of the geometry in Figure 3 of \citet{vasylyev_early_2023}. After this early CSM interaction phase, SN\,2023ixf evolved like a typical SN~II through its $\sim 80$\,day plateau, with polarization increasing weeks before the plateau end.

From the theoretical modeling perspective, much work remains to overcome degeneracies in inclination, density contrast, and half-opening angle $\beta_{1/2}$. Similar polarized fluxes can be produced through different physical mechanisms creating asymmetry in the ejecta, complicating the interpretation of the origin of these asymmetries. The models presented in this work can only provide limits on the intrinsic asphericity rather than the geometry projected in the plane of the sky, since the true inclination angle of the SN cannot be determined. Breaking these degeneracies will require a larger statistical sample of SNe~II. Future models should incorporate 3D geometries to better capture the effects of clumpy ejecta, and an extensive exploration of the parameter space is necessary to explain the diverse polarization signatures observed even among SNe of the same class. The physical mechanism responsible for creating such an aspherical CSM also remains unclear.

Future early-time spectropolarimetric observations of similar events, particularly when combined with multiwavelength data probing the CSM structure (e.g., UV spectroscopy; \citealt{vasylyev_early-time_2022,bostroem_circumstellar_2024}), will be essential for disentangling these scenarios. While comprehensive coverage throughout the first few months is important, high-cadence observations during the weeks surrounding the plateau dropoff are particularly critical, as the polarization peak during this phase provides a direct probe of the ejecta asphericity when models are most sensitive to geometric parameters. Continued monitoring of SN\,2023ixf over the coming decade will help characterize its circumstellar environment, constrain its progenitor nature, and detect any ongoing interaction.



\section{acknowledgments} 
A major upgrade of the Kast spectrograph on the Shane 3\,m telescope at Lick Observatory, led by Brad Holden, was made possible through gifts from the Heising-Simons Foundation, William and Marina Kast, and the University of California Observatories.
We appreciate the expert assistance of the staff at Lick Observatory.  
Research at Lick Observatory is partially supported by a gift from Google. 

This research was funded in part by {\it HST} grant GO-16656 from the Space Telescope Science Institute, which is operated by the Association of Universities for Research in Astronomy (AURA), Inc., under NASA contract NAS5-26555. Additional generous financial support was provided to A.V.F.’s supernova group at U.C. Berkeley by the Christopher R. Redlich Fund, Steven Nelson, Sunil Nagaraj, Landon Noll, Sandra Otellini, Gary and Cynthia Bengier, Clark and Sharon Winslow, Alan Eustace, William Draper, Timothy and Melissa Draper, Briggs and Kathleen Wood, and Sanford Robertson (S.S.V. is a Steven Nelson Graduate Fellow in Astronomy, K.C.P. was a Nagaraj-Noll-Otellini Graduate Fellow in Astronomy, W.Z. is a Bengier-Winslow-Eustace Specialist in Astronomy, T.G.B. is a Draper-Wood-Robertson Specialist in Astronomy, Y.Y. was a Bengier-Winslow-Robertson Fellow in Astronomy).
Numerous other donors to his group and/or research at Lick Observatory include Michael and Evelyn Antin, Charlie Baxter and Jinee Tao, Duncan Beardsley, Jim Connelly and Anne Mackenzie, Curtis and Shelley Covey, Byron and Allison Deeter, Paul and Silvia Edwards, Arthur and Cindy Folker, Thomas and Dana Grogan, Heidi Gerster, Harvey Glasser, Charles and Gretchen Gooding, Judith and Timothy Hachman, Alan and Gladys Hoefer, the Hugh Stuart Center Charitable Trust, George Hume, Stephen Imbler, Michael Kast and Rebecca Lyon, Lata Krishnan and Ajay Shah, Walter and Karen Loewenstern, Rand Morimoto and Anna Henderson, Edward Oates, Doug and Emilie Ogden, Jon and Susan Reiter, Paul Robinson, Catherine Rondeau, Laura Sawczuk and Luke Ellis, Stanley and Miriam Schiffman, Richard and Betsey Sesler, Hans Spiller, Justin and Seana Stephens, Ilya Strebulaev and Anna Dvornikova, Diane Tokugawa and Alan Gould, David Turner III and Joanne Turner, David and Malin Walrod, Gerry and Virginia Weiss, Janet Westin and Mike McCaw, David and Angie Yancey, and others.
A.V.F. is grateful for the hospitality of the Hagler Institute for Advanced Study as well as the Department of Physics and Astronomy at Texas A\&M University during part of this investigation.

R.M. acknowledges support by the National Science Foundation (NSF) under awards AST-2221789 and AST-2224255. L.W. acknowledges the NSF for support through award AST-1817099. The TReX team at U.C. Berkeley is partially funded by the Heising-Simons Foundation under grant \#2021-3248 (PI R. Margutti).
\bigskip


\software{The KastShiv, {Astropy \citep{astropy:2013, astropy:2018}, 
IDL Astronomy user's library \citep{landsman_idl_1993}}} 
\bigskip
\newpage
\appendix
\section{Appendix: Supplementary Spectropolarimetry of SN\,2023ixf~\label{sec:app}}
\restartappendixnumbering



\begin{table*}[!h]
\begin{center}
\caption{Complete Journal of Spectropolarimetric Observations of SN\,2023ixf.} 
    \begin{tabular}{cccccccc}
	\hline 
	\hline
	UTC Date & MJD$^b$ & Phase$^a$ & Airmass & Avg. Seeing &  Exp. Time$^c$   \\ 
	(MM-DD-YYYY)&   & (days) & & (arcsec) & (s)   \\ 
	\hline 
    05-20-2023 & 60084.18 & 1.4 & 1.12--1.44 & 1.2 & $600 \times 4 \times 4$ \\
    05-21-2023 &60085.23& 2.5 & 1.06& 1.2 & $600 \times 4 \times 3$ \\
    05-22-2023 &60086.24& 3.5 & 1.05 & 1.2 &  $600 \times 4 \times 1$\\
    05-23-2023 &60087.35& 4.6 & 1.15 & 1.2 &  $600 \times 4 \times 1$ \\
    05-28-2023 &60092.20& 9.5 & 1.05 & 1.2 &  $600 \times 4 \times 4$ \\
   06-02-2023 &60097.26& 14.5 & 1.06 & 1.2 &  $200 \times 4 \times 2$\\
   06-10-2023 &60105.26& 22.5 & 1.10 & 1.2 &  $200 \times 4 \times 3^d$\\
   06-19-2023 &60114.25& 31.5 & 1.17 & 1.2 &  $600 \times 4 \times 3$\\
   06-22-2023 &60117.27& 34.5 & 1.22 & 1.4 &  $600 \times 4 \times 2$\\
   07-10-2023$^e$ &60135.25& 52.5 & 1.33 & 1.8 &  $550 \times 4 \times 3$\\
   07-10-2023 &60135.32& 52.5 & 1.64 & 1.8 &  $200 \times 4 \times 2$\\
   07-19-2023 &60144.30& 61.5 & 1.62 & 1.5 &  $200 \times 4 \times 2$\\
   08-24-2023 &60180.21& 97.4 & 1.65 & 1.0 &  $360 \times 4 \times 1$\\
   09-08-2023 &60195.16& 112.4 & 1.68 & 1.2 &  $600 \times 4 \times 1$\\
   09-16-2023 &60203.15& 120.4 & 1.73 & 1.0 &  $540 \times 4 \times 1$\\
	\hline 
\end{tabular}\\
{$^a$}{Days after the estimated time of first light on MJD~60082.75 (UTC 18 May 2023).} \\
{$^b$}{MJD given as the start time of the CCD exposure.}\\
{$^c$}{Exposure time of a single exposure $\times$ 4 retarder-plate angles $\times$ number of loops.\\ Wavelength range for Kast is 4550--9800\,\AA.}\\
{$^d$}{Only the first loop is used in the analysis.} 
{$^e$}{Higher resolution 600\,lines\,mm$^{-1}$ \\ grating configuration blazed at 5000\,\AA\ used for this epoch.} 
\label{tbl:specpol_log}
\end{center}
\end{table*}

\clearpage
\twocolumngrid

\begin{figure}
    \centering
    \includegraphics[width=0.5\textwidth]{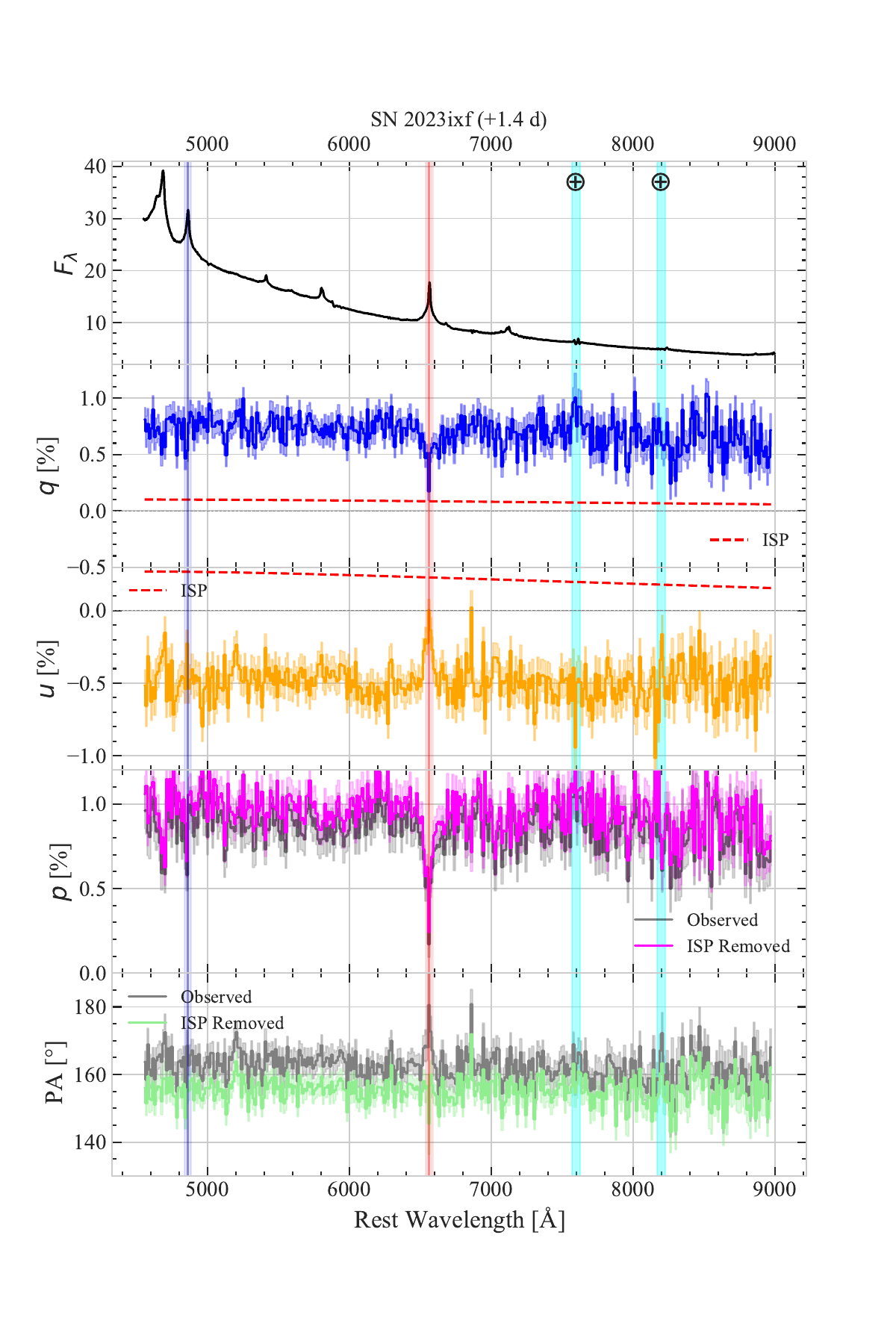}
    \caption{Spectropolarimetric observations of SN\,2023ixf at $+1.4$ days after first light. Plotted from top to bottom: total-flux spectrum (F$_{\lambda}$), Stokes $q$ parameter, Stokes $u$ parameter, total fractional polarization $p$, and polarization position angle (PA). The ISP-subtracted (observed) polarization is shown in magenta (gray) and the mean ISP best fit to the individual Stokes parameters (see Sec. \ref{sec:isp}) is indicated by red dashed lines.}~\label{fig:specpol_quppa_1p4}
\end{figure}

\begin{figure}
    \centering
    \includegraphics[width=0.5\textwidth]{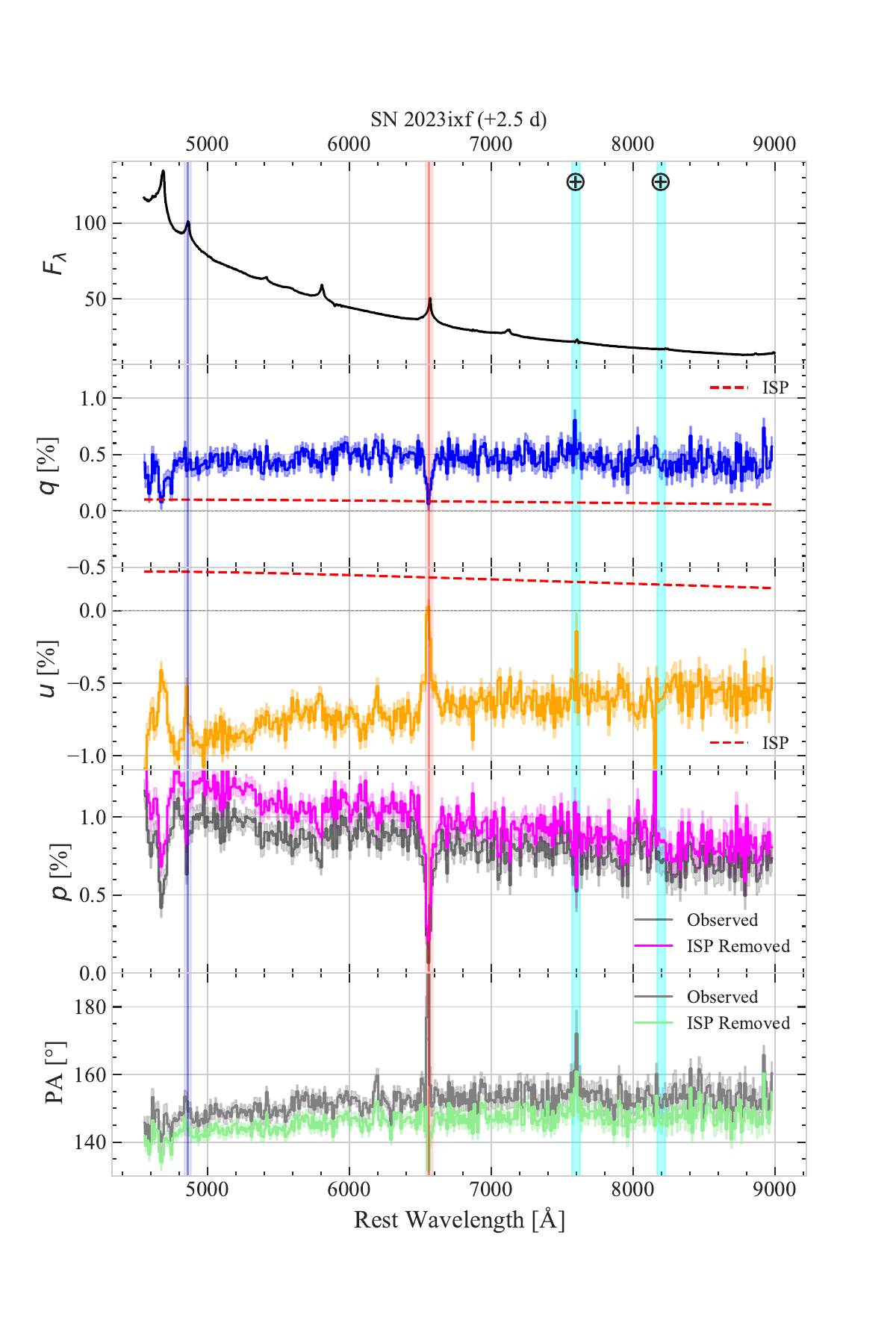}
    \caption{Same as Figure \ref{fig:specpol_quppa_1p4}, but on day $+2.5$.
}~\label{fig:specpol_quppa_2p5}
\end{figure}

\begin{figure}
    \centering
    \includegraphics[width=0.5\textwidth]{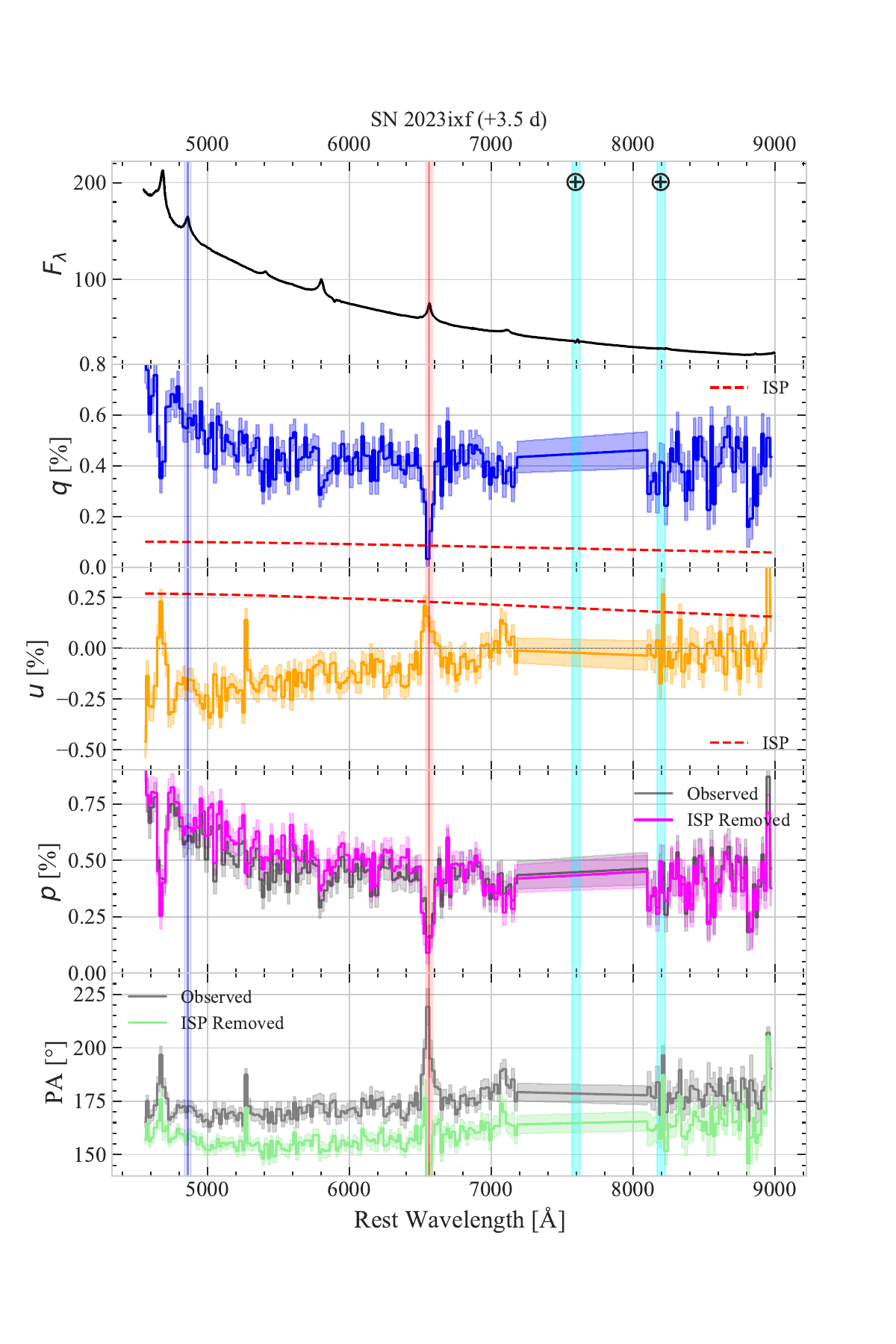}
    \caption{Same as Figure \ref{fig:specpol_quppa_1p4}, but on day $+3.5$.
}~\label{fig:specpol_quppa_3p5}
\end{figure}

\begin{figure}
    \centering
    \includegraphics[width=0.5\textwidth]{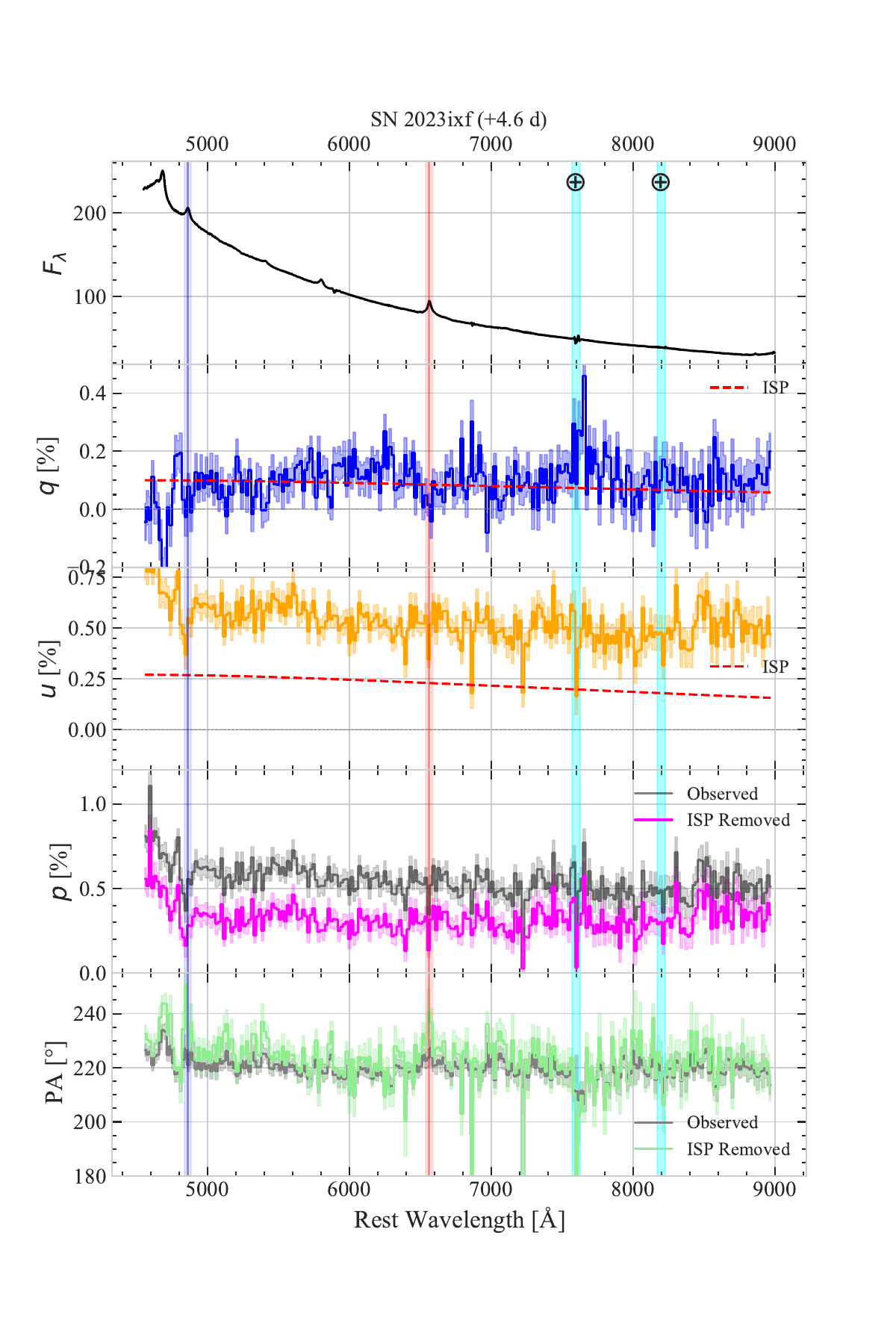}
    \caption{Same as Figure \ref{fig:specpol_quppa_1p4}, but on day $+4.6$.
}~\label{fig:specpol_quppa_4p6}
\end{figure}

\begin{figure}
    \centering
    \includegraphics[width=0.5\textwidth]{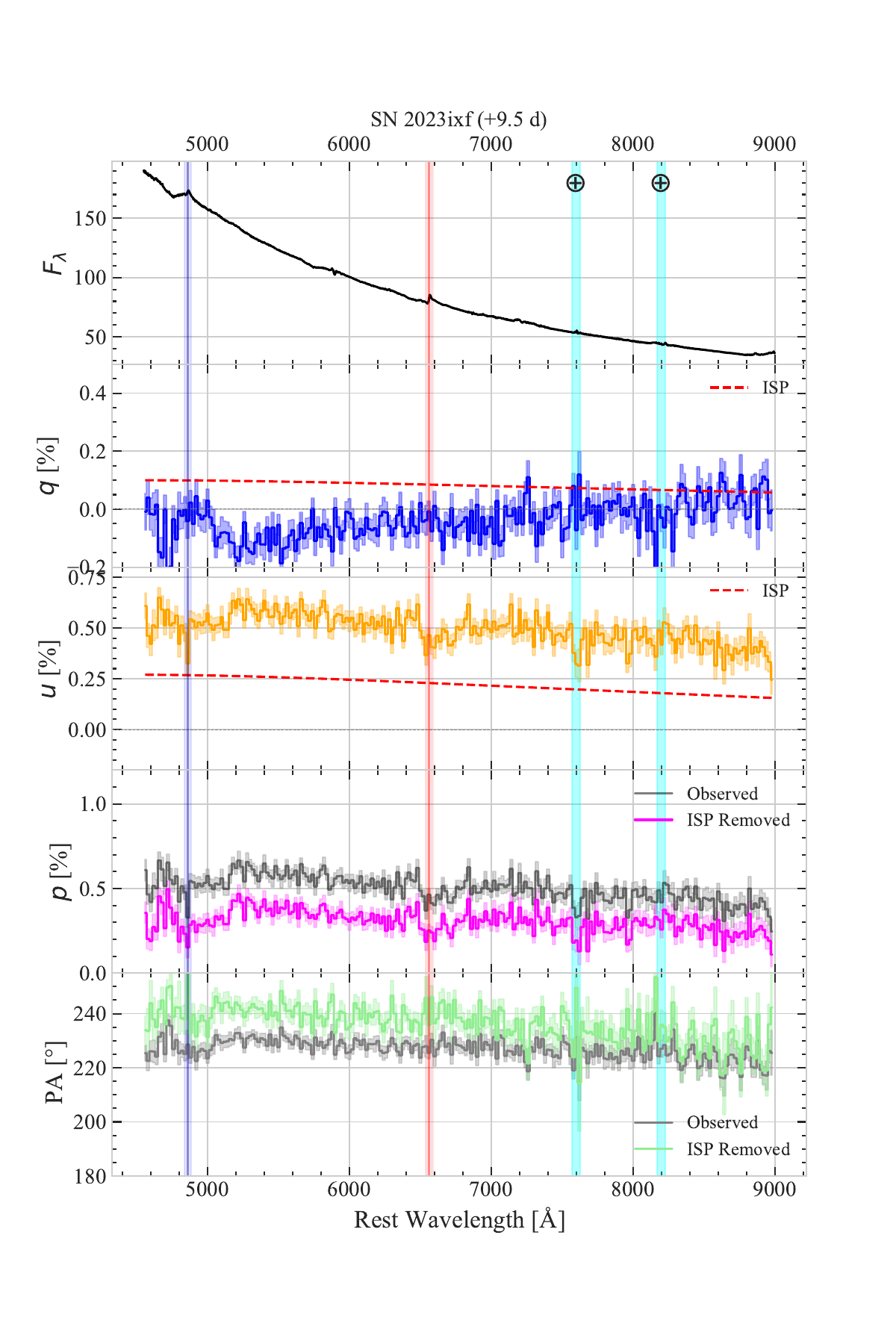}
    \caption{Same as Figure \ref{fig:specpol_quppa_1p4}, but on day $+9.5$.
}~\label{fig:specpol_quppa_9p5}
\end{figure}

\begin{figure}
    \centering
    \includegraphics[width=0.5\textwidth]{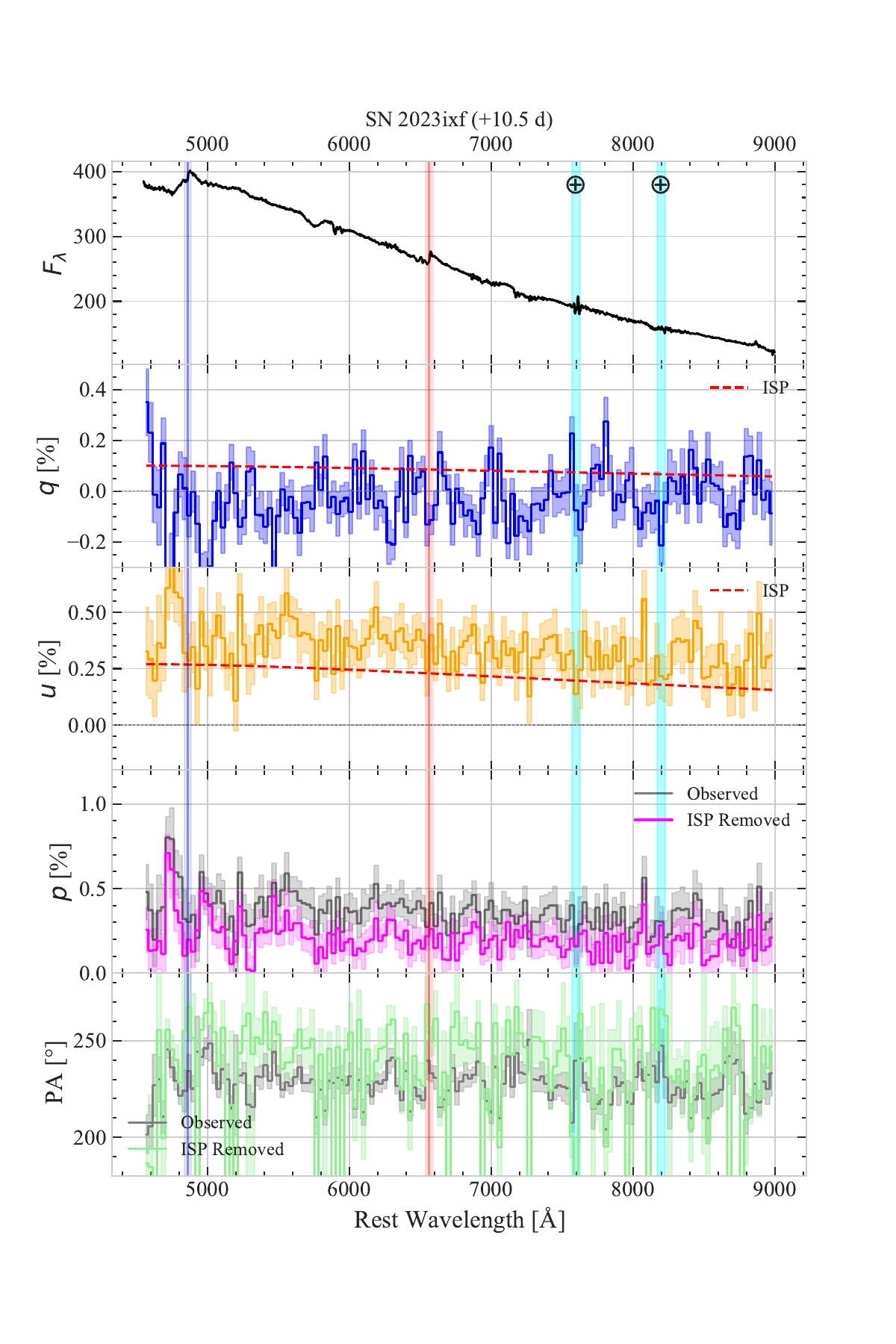}
    \caption{Same as Figure \ref{fig:specpol_quppa_1p4}, but on day $+10.5$.
}~\label{fig:specpol_quppa_10p5}
\end{figure}

\begin{figure}
    \centering
    \includegraphics[width=0.5\textwidth]{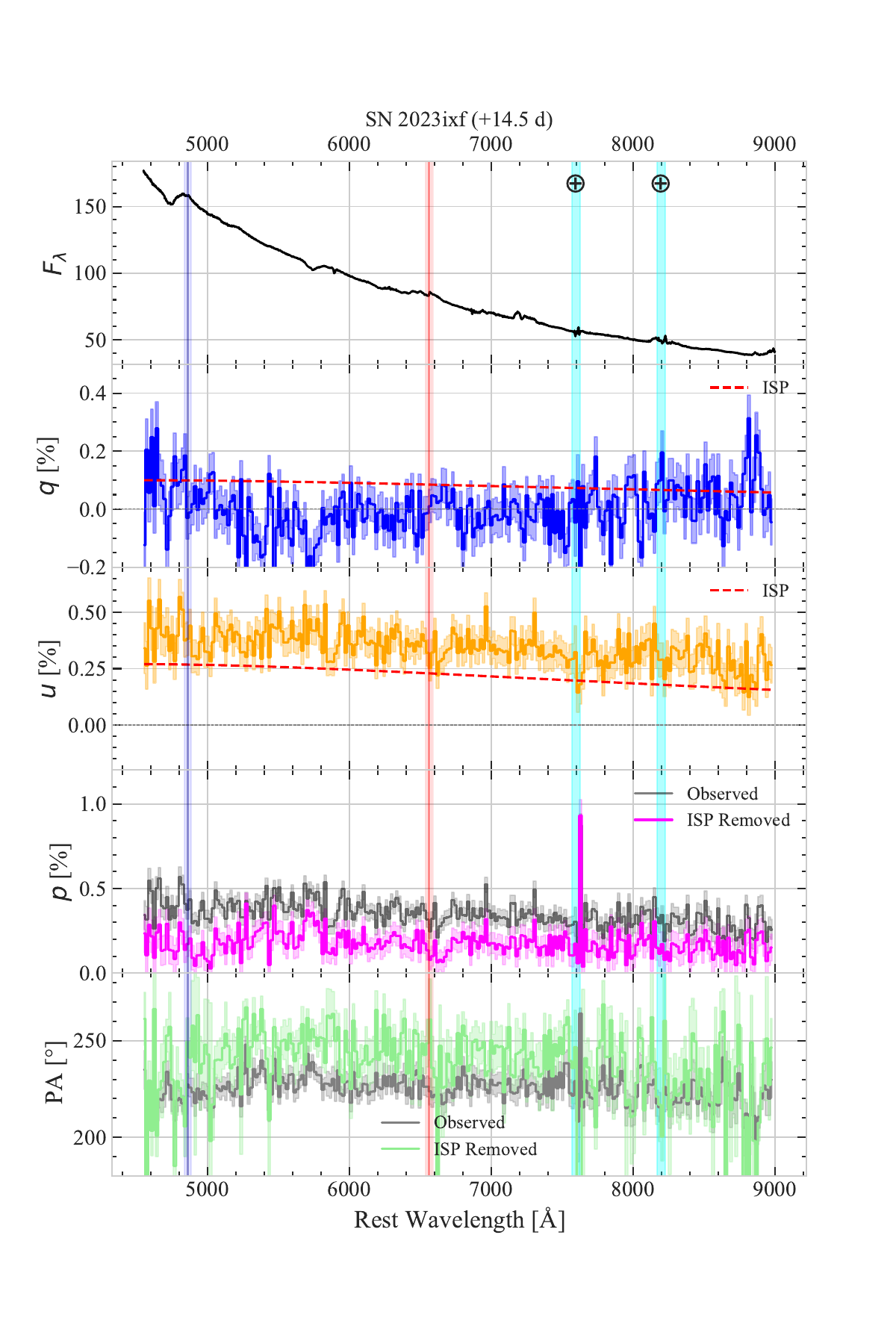}
    \caption{Same as Figure \ref{fig:specpol_quppa_1p4}, but on day $+14.5$.
}~\label{fig:specpol_quppa_14p5}
\end{figure}

\begin{figure}
    \centering
    \includegraphics[width=0.5\textwidth]{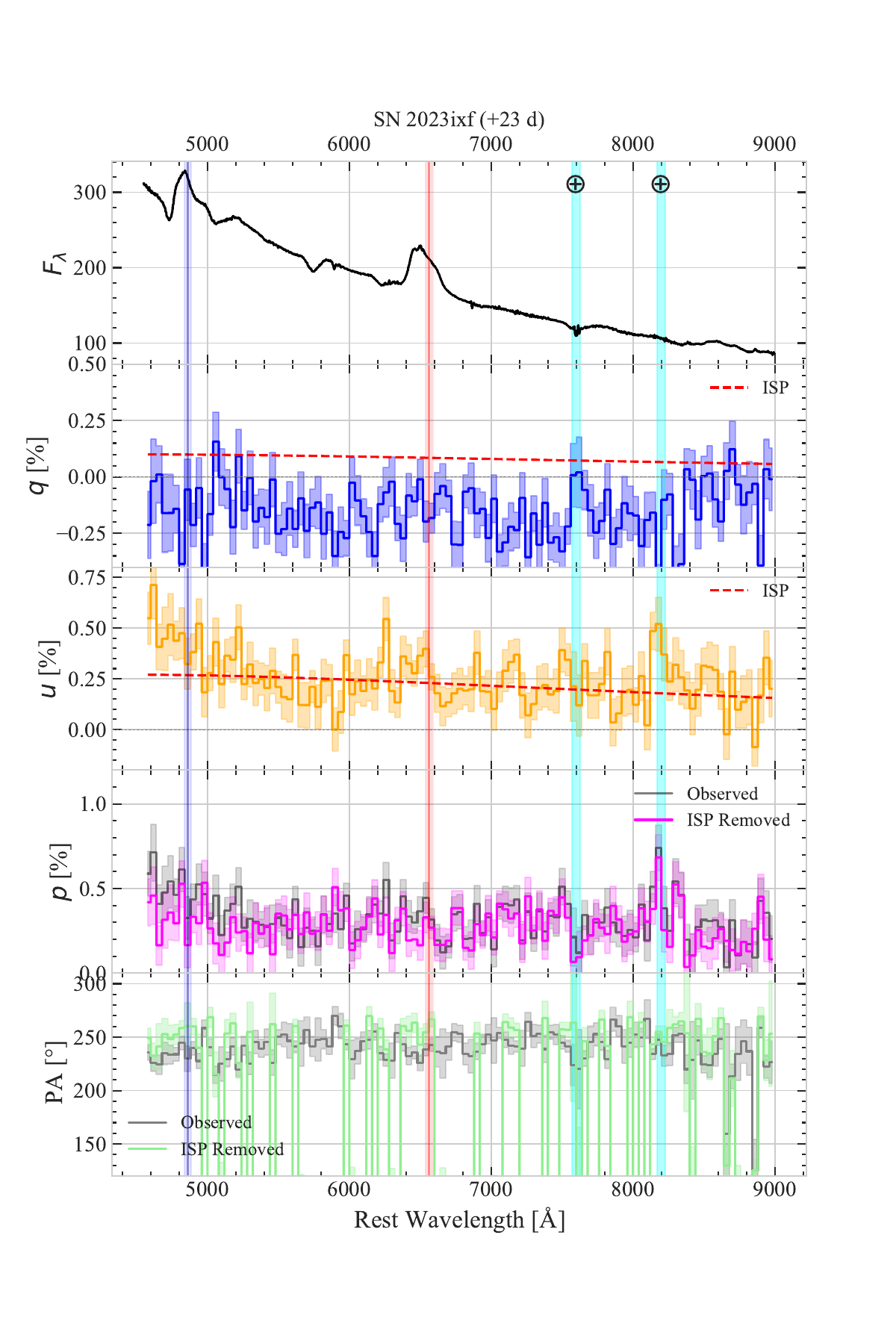}
    \caption{Same as Figure \ref{fig:specpol_quppa_1p4}, but on day $+23$.
}~\label{fig:specpol_quppa_23}
\end{figure}

\begin{figure}
    \centering
    \includegraphics[width=0.5\textwidth]{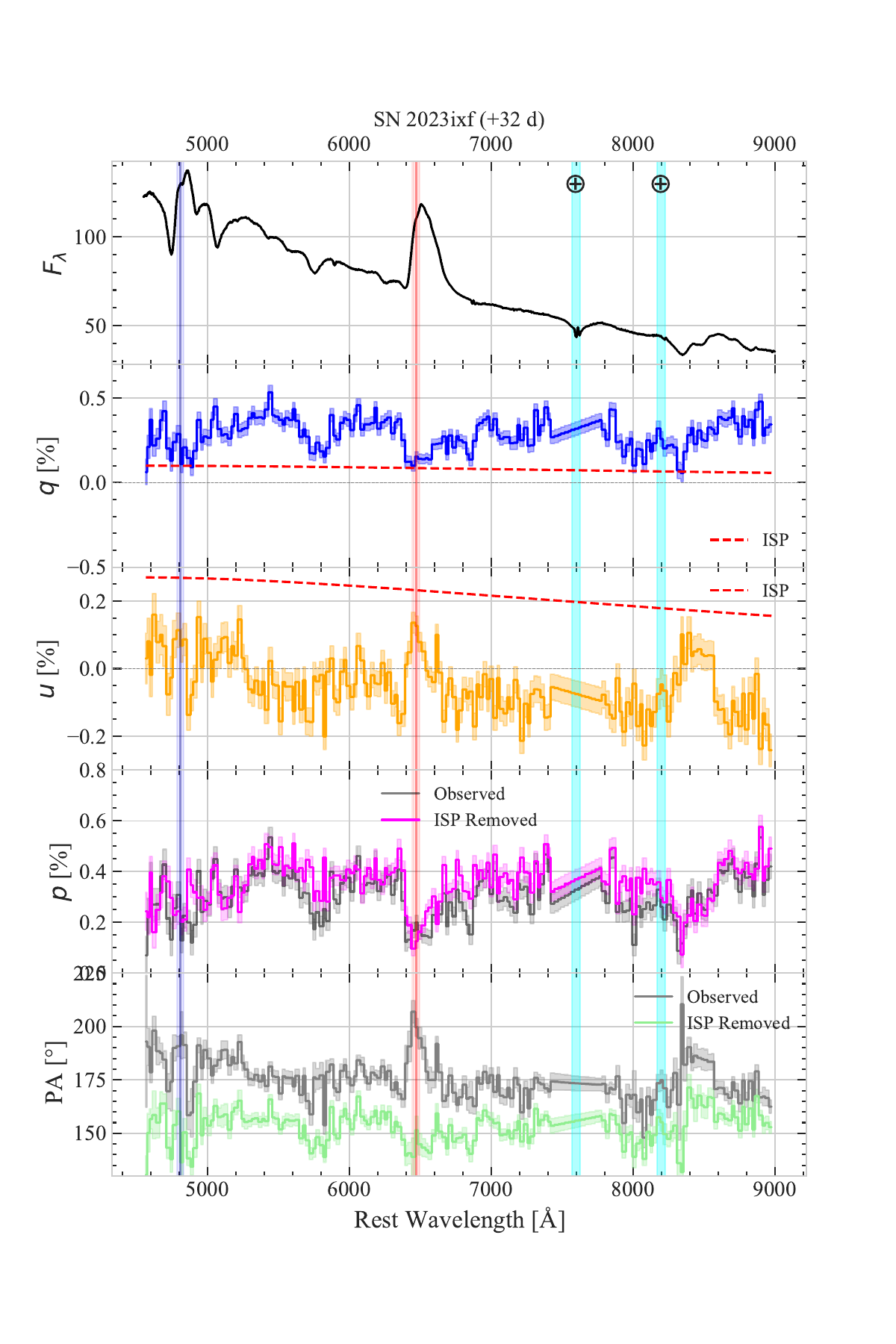}
    \caption{Same as Figure \ref{fig:specpol_quppa_1p4}, but on day $+32$.
}~\label{fig:specpol_quppa_32}
\end{figure}

\begin{figure}
    \centering
    \includegraphics[width=0.5\textwidth]{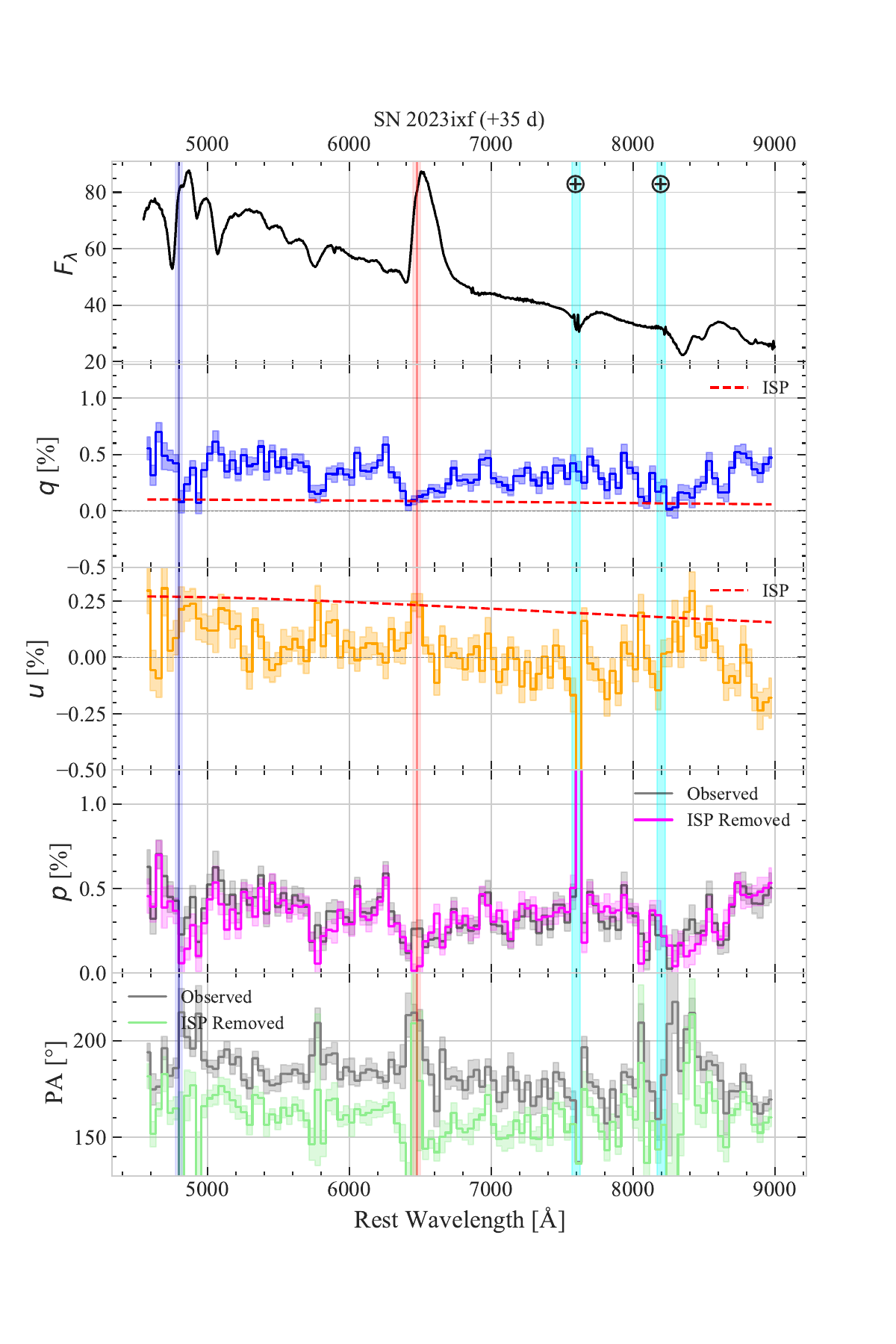}
    \caption{Same as Figure \ref{fig:specpol_quppa_1p4}, but on day $+35$.
}~\label{fig:specpol_quppa_35}
\end{figure}

\begin{figure}
    \centering
    \includegraphics[width=0.5\textwidth]{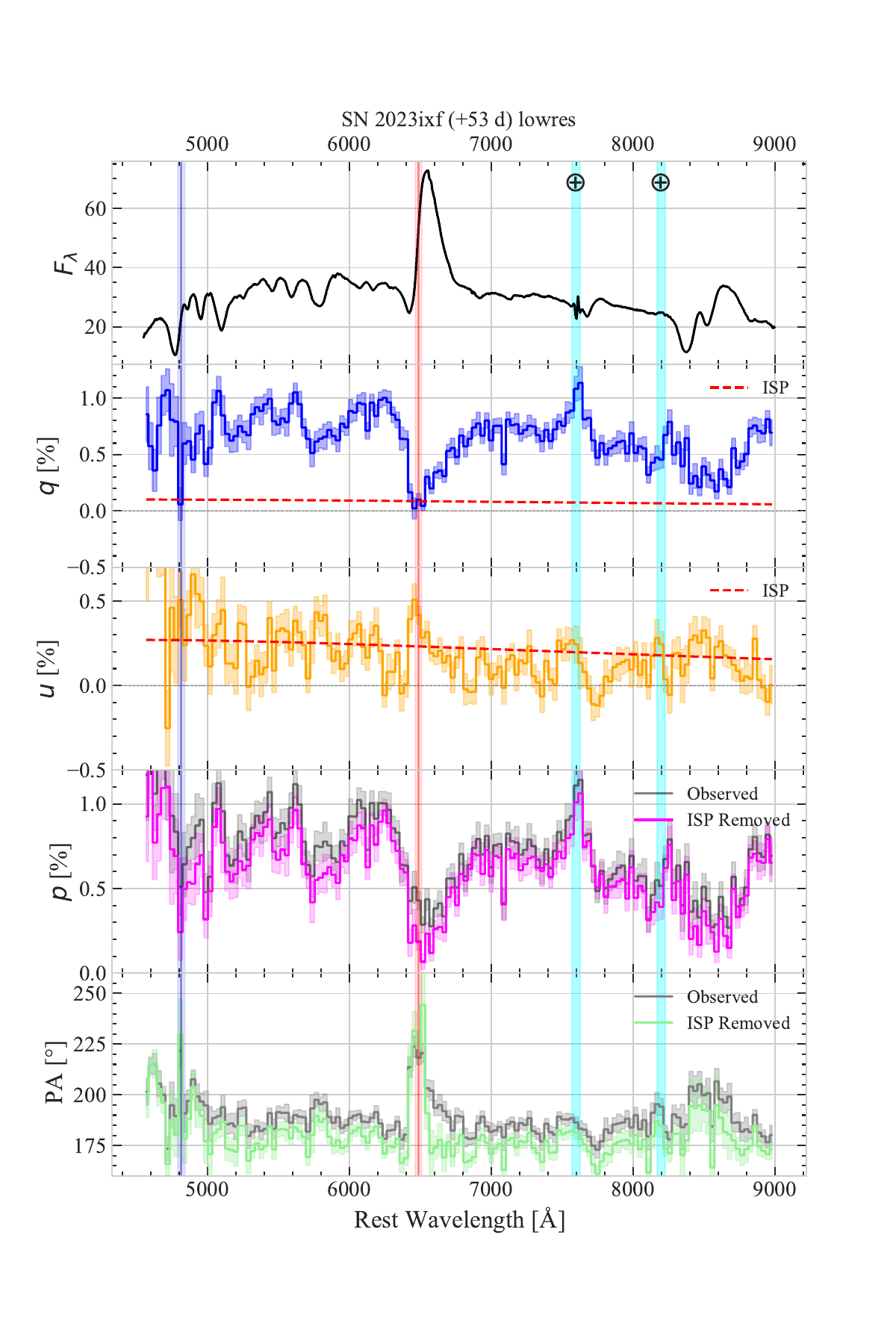}
    \caption{Same as Figure \ref{fig:specpol_quppa_1p4}, but on day $+53$ using the 300\,lines\,mm$^{-1}$ grating configuration.
}~\label{fig:specpol_quppa_53_lowres}
\end{figure}

\begin{figure}
    \centering
    \includegraphics[width=0.5\textwidth]{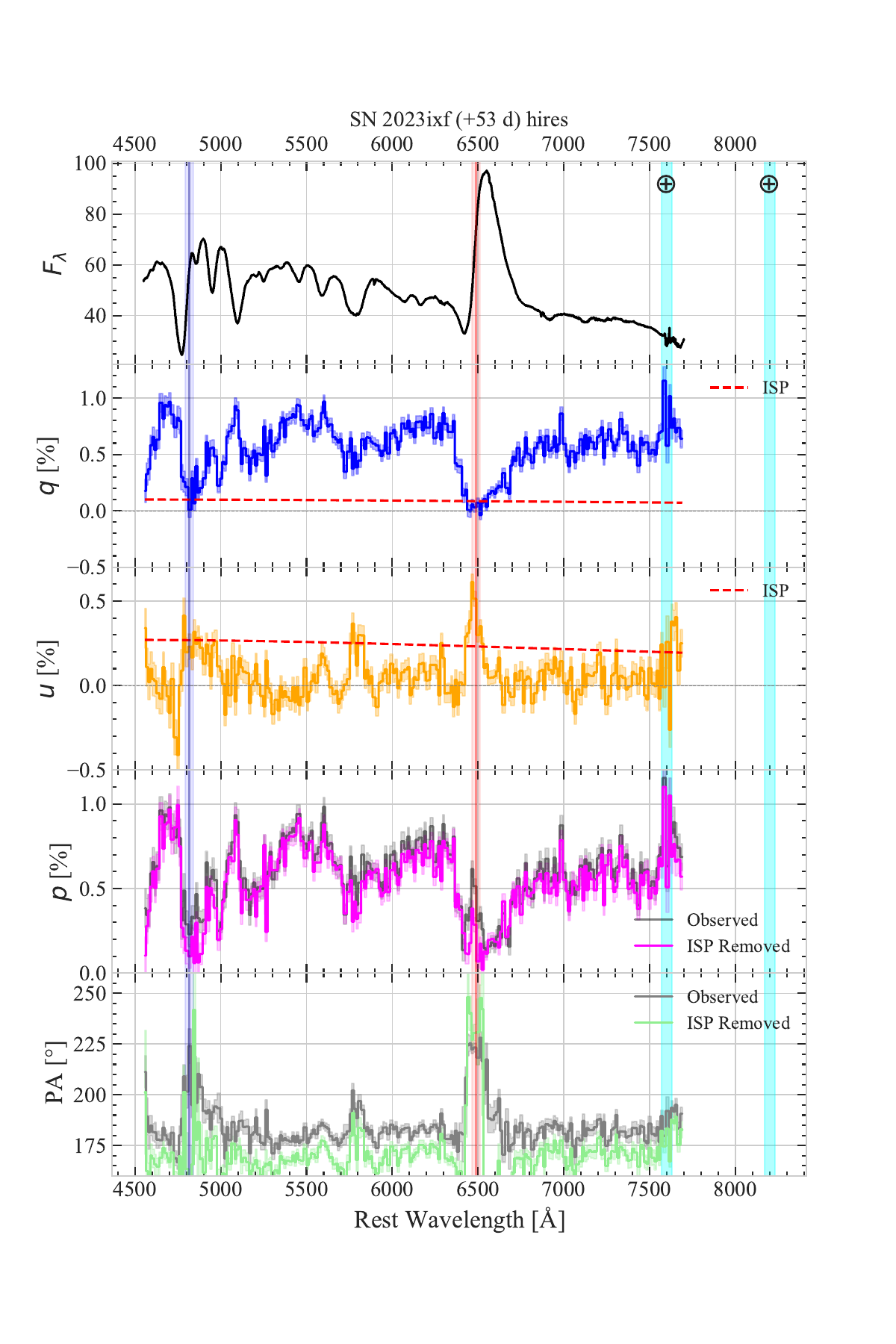}
    \caption{Same as Figure \ref{fig:specpol_quppa_1p4}, but on day $+53$ using the 600\,lines\,mm$^{-1}$ grating configuration.
}~\label{fig:specpol_quppa_53_hires}
\end{figure}

\begin{figure}
    \centering
    \includegraphics[width=0.5\textwidth]{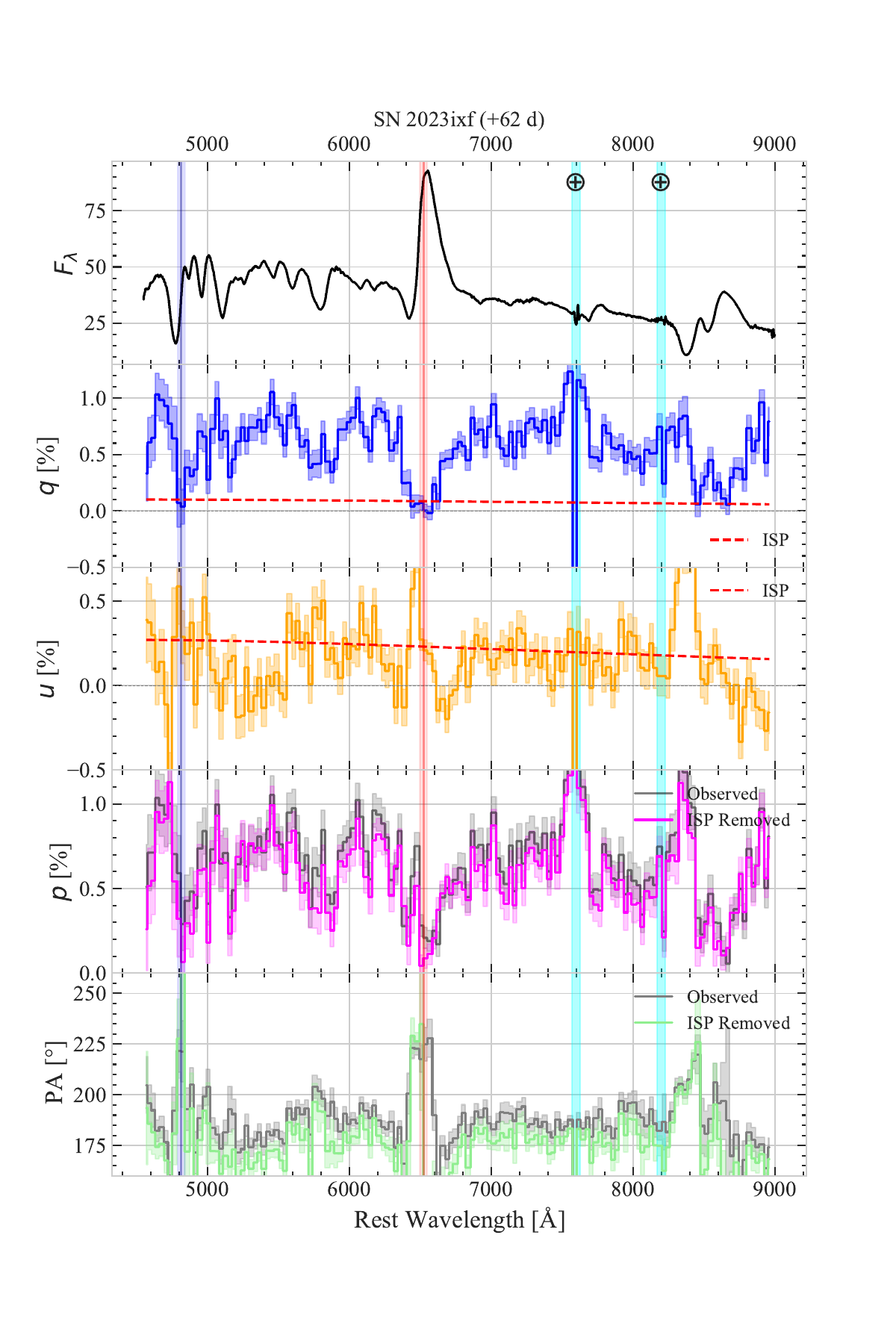}
    \caption{Same as Figure \ref{fig:specpol_quppa_1p4}, but on day $+62$.
}~\label{fig:specpol_quppa_62}
\end{figure}

\begin{figure}
    \centering
    \includegraphics[width=0.5\textwidth]{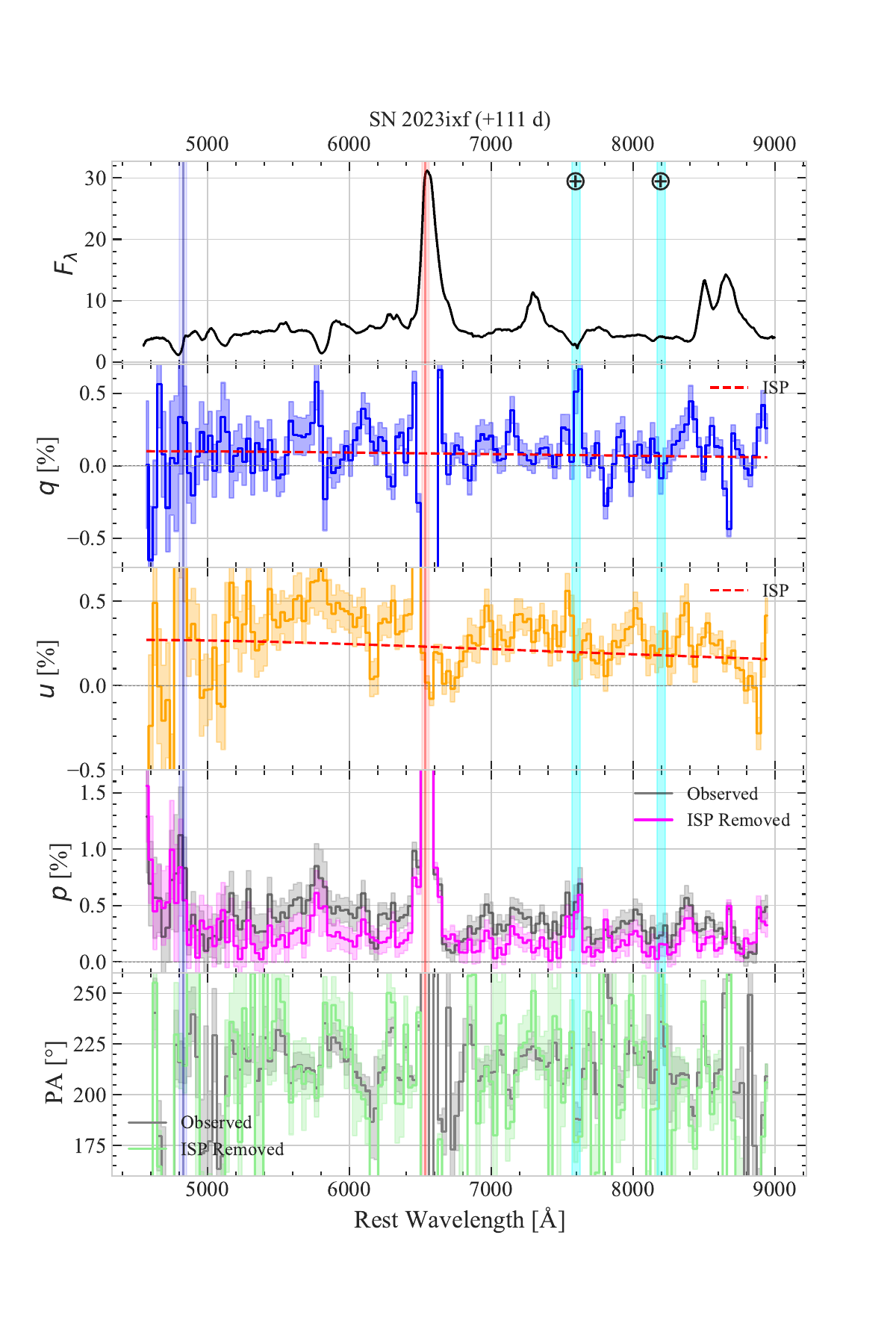}
    \caption{Same as Figure \ref{fig:specpol_quppa_1p4}, but averaged across days $+97$, $+112$, and $+120$ day.
}~\label{fig:specpol_quppa_111}
\end{figure}

\clearpage
\bibliographystyle{aasjournal}
\bibliography{2023ixf}

\begin{thebibliography}{}
\expandafter\ifx\csname natexlab\endcsname\relax\def\natexlab#1{#1}\fi
\providecommand{\url}[1]{\href{#1}{#1}}
\providecommand{\dodoi}[1]{doi:~\href{http://doi.org/#1}{\nolinkurl{#1}}}
\providecommand{\doeprint}[1]{\href{http://ascl.net/#1}{\nolinkurl{http://ascl.net/#1}}}
\providecommand{\doarXiv}[1]{\href{https://arxiv.org/abs/#1}{\nolinkurl{https://arxiv.org/abs/#1}}}

\bibitem[{{Astropy Collaboration} {et~al.}(2013){Astropy Collaboration}, {Robitaille}, {Tollerud}, {Greenfield}, {Droettboom}, {Bray}, {Aldcroft}, {Davis}, {Ginsburg}, {Price-Whelan}, {Kerzendorf}, {Conley}, {Crighton}, {Barbary}, {Muna}, {Ferguson}, {Grollier}, {Parikh}, {Nair}, {Unther}, {Deil}, {Woillez}, {Conseil}, {Kramer}, {Turner}, {Singer}, {Fox}, {Weaver}, {Zabalza}, {Edwards}, {Azalee Bostroem}, {Burke}, {Casey}, {Crawford}, {Dencheva}, {Ely}, {Jenness}, {Labrie}, {Lim}, {Pierfederici}, {Pontzen}, {Ptak}, {Refsdal}, {Servillat}, \& {Streicher}}]{astropy:2013}
{Astropy Collaboration}, {Robitaille}, T.~P., {Tollerud}, E.~J., {et~al.} 2013, \aap, 558, A33, \dodoi{10.1051/0004-6361/201322068}

\bibitem[{{Astropy Collaboration} {et~al.}(2018){Astropy Collaboration}, {Price-Whelan}, {Sip{\H{o}}cz}, {G{\"u}nther}, {Lim}, {Crawford}, {Conseil}, {Shupe}, {Craig}, {Dencheva}, {Ginsburg}, {Vand erPlas}, {Bradley}, {P{\'e}rez-Su{\'a}rez}, {de Val-Borro}, {Aldcroft}, {Cruz}, {Robitaille}, {Tollerud}, {Ardelean}, {Babej}, {Bach}, {Bachetti}, {Bakanov}, {Bamford}, {Barentsen}, {Barmby}, {Baumbach}, {Berry}, {Biscani}, {Boquien}, {Bostroem}, {Bouma}, {Brammer}, {Bray}, {Breytenbach}, {Buddelmeijer}, {Burke}, {Calderone}, {Cano Rodr{\'\i}guez}, {Cara}, {Cardoso}, {Cheedella}, {Copin}, {Corrales}, {Crichton}, {D'Avella}, {Deil}, {Depagne}, {Dietrich}, {Donath}, {Droettboom}, {Earl}, {Erben}, {Fabbro}, {Ferreira}, {Finethy}, {Fox}, {Garrison}, {Gibbons}, {Goldstein}, {Gommers}, {Greco}, {Greenfield}, {Groener}, {Grollier}, {Hagen}, {Hirst}, {Homeier}, {Horton}, {Hosseinzadeh}, {Hu}, {Hunkeler}, {Ivezi{\'c}}, {Jain}, {Jenness}, {Kanarek}, {Kendrew}, {Kern}, {Kerzendorf}, {Khvalko}, {King}, {Kirkby}, {Kulkarni},
  {Kumar}, {Lee}, {Lenz}, {Littlefair}, {Ma}, {Macleod}, {Mastropietro}, {McCully}, {Montagnac}, {Morris}, {Mueller}, {Mumford}, {Muna}, {Murphy}, {Nelson}, {Nguyen}, {Ninan}, {N{\"o}the}, {Ogaz}, {Oh}, {Parejko}, {Parley}, {Pascual}, {Patil}, {Patil}, {Plunkett}, {Prochaska}, {Rastogi}, {Reddy Janga}, {Sabater}, {Sakurikar}, {Seifert}, {Sherbert}, {Sherwood-Taylor}, {Shih}, {Sick}, {Silbiger}, {Singanamalla}, {Singer}, {Sladen}, {Sooley}, {Sornarajah}, {Streicher}, {Teuben}, {Thomas}, {Tremblay}, {Turner}, {Terr{\'o}n}, {van Kerkwijk}, {de la Vega}, {Watkins}, {Weaver}, {Whitmore}, {Woillez}, {Zabalza}, \& {Astropy Contributors}}]{astropy:2018}
{Astropy Collaboration}, {Price-Whelan}, A.~M., {Sip{\H{o}}cz}, B.~M., {et~al.} 2018, \aj, 156, 123, \dodoi{10.3847/1538-3881/aabc4f}

\bibitem[{Ben-Ami {et~al.}(2015)Ben-Ami, Hachinger, Gal-Yam, Mazzali, Filippenko, Horesh, Matheson, Modjaz, Sauer, Silverman, Smith, \& Yaron}]{ben-ami_ultraviolet_2015}
Ben-Ami, S., Hachinger, S., Gal-Yam, A., {et~al.} 2015, ApJ, 803, 40, \dodoi{10.1088/0004-637X/803/1/40}

\bibitem[{Berger {et~al.}(2023)Berger, Keating, Margutti, Maeda, Alexander, Cendes, Eftekhari, Gurwell, Hiramatsu, Ho, Laskar, Rao, \& Williams}]{berger_millimeter_2023}
Berger, E., Keating, G.~K., Margutti, R., {et~al.} 2023, Millimeter {Observations} of the {Type} {II} {SN2023ixf}: {Constraints} on the {Proximate} {Circumstellar} {Medium}, \dodoi{10.48550/arXiv.2306.09311}

\bibitem[{Berkhuijsen {et~al.}(2016)Berkhuijsen, Urbanik, Beck, \& Han}]{berkhuijsen_radio_2016}
Berkhuijsen, E.~M., Urbanik, M., Beck, R., \& Han, J.~L. 2016, A \& A, 588, A114, \dodoi{10.1051/0004-6361/201527322}

\bibitem[{Bilinski {et~al.}(2024)Bilinski, Smith, Williams, Smith, Leonard, Hoffman, Andrews, \& Milne}]{bilinski_multi-epoch_2024}
Bilinski, C., Smith, N., Williams, G.~G., {et~al.} 2024, Monthly Notices of the Royal Astronomical Society, 529, 1104, \dodoi{10.1093/mnras/stae380}

\bibitem[{Bilinski {et~al.}(2018)Bilinski, Smith, Williams, Smith, Zheng, Graham, Mauerhan, Andrews, Filippenko, Akerlof, Chatzopoulos, Hoffman, Huk, Leonard, Marion, Milne, Quimby, Silverman, Vinkó, Wheeler, \& Yuan}]{bilinski_sn2012ab_2018}
---. 2018, Monthly Notices of the Royal Astronomical Society, 475, 1104, \dodoi{10.1093/mnras/stx3214}

\bibitem[{Bilinski {et~al.}(2020)Bilinski, Smith, Williams, Smith, Andrews, Clubb, Zheng, Filippenko, Fox, Hosseinzadeh, Howell, Kelly, Milne, Sand, Hoffman, Leonard, Cargill, Casper, Halevy, Kim, Kumar, Pina, \& Yuk}]{bilinski_sn_2020}
---. 2020, Monthly Notices of the Royal Astronomical Society, 498, 3835, \dodoi{10.1093/mnras/staa2617}

\bibitem[{Bose {et~al.}(2013)Bose, Kumar, Sutaria, Kumar, Roy, Bhatt, Pandey, Chandola, Sagar, Misra, \& Chakraborti}]{bose_supernova_2013}
Bose, S., Kumar, B., Sutaria, F., {et~al.} 2013, Monthly Notices of the Royal Astronomical Society, 433, 1871, \dodoi{10.1093/mnras/stt864}

\bibitem[{Bostroem {et~al.}(2023)Bostroem, Pearson, Shrestha, Sand, Valenti, Jha, Andrews, Smith, Terreran, Green, Dong, Lundquist, Haislip, Hoang, Hosseinzadeh, Janzen, Jencson, Kouprianov, Paraskeva, Retamal, Reichart, Arcavi, Bonanos, Coughlin, Farah, Hawley, Hebb, Hiramatsu, Howell, Iijima, Ilyin, McCully, Moran, Morris, Mura, Newsome, Pabst, Ochner, Gonzalez, Pastorello, Pellegrino, Ravi, Reguitti, Salo, Vinko, Wheeler, Williams, \& Wyatt}]{bostroem_early_2023}
Bostroem, K.~A., Pearson, J., Shrestha, M., {et~al.} 2023, Early {Spectroscopy} and {Dense} {Circumstellar} {Medium} {Interaction} in {SN}{\textasciitilde}2023ixf,  arXiv.
\newblock \url{http://arxiv.org/abs/2306.10119}

\bibitem[{Bostroem {et~al.}(2024)Bostroem, Sand, Dessart, Smith, Jha, Valenti, Andrews, Dong, Filippenko, Gomez, Hiramatsu, Hoang, Hosseinzadeh, Howell, Jencson, Lundquist, McCully, Mehta, Retamal, Pearson, Ravi, Shrestha, \& Wyatt}]{bostroem_circumstellar_2024}
Bostroem, K.~A., Sand, D.~J., Dessart, L., {et~al.} 2024, Circumstellar {Interaction} in the {Ultraviolet} {Spectra} of {SN} 2023ixf 14-66 {Days} {After} {Explosion},  arXiv.
\newblock \url{http://arxiv.org/abs/2408.03993}

\bibitem[{Brown \& McLean(1977)}]{brown_polarisation_1977}
Brown, J.~C., \& McLean, I.~S. 1977, A \& A, 57, 141.
\newblock \url{https://ui.adsabs.harvard.edu/abs/1977A&A....57..141B}

\bibitem[{Bruch {et~al.}(2021)Bruch, Gal-Yam, Schulze, Yaron, Yang, Soumagnac, Rigault, Strotjohann, Ofek, Sollerman, Masci, Barbarino, Ho, Fremling, Perley, Nordin, Cenko, Adams, Adreoni, Bellm, Blagorodnova, Bulla, Burdge, De, Dhawan, Drake, Duev, Dugas, Graham, Graham, Irani, Jencson, Karamehmetoglu, Kasliwal, Kim, Kulkarni, Kupfer, Liang, Mahabal, Miller, Prince, Riddle, Sharma, Smith, Taddia, Taggart, Walters, \& Yan}]{bruch_large_2021}
Bruch, R.~J., Gal-Yam, A., Schulze, S., {et~al.} 2021, The Astrophysical Journal, 912, 46, \dodoi{10.3847/1538-4357/abef05}

\bibitem[{Bruch {et~al.}(2023)Bruch, Gal-Yam, Yaron, Chen, Strotjohann, Irani, Zimmerman, Schulze, Yang, Kim, Bulla, Sollerman, Rigault, Ofek, Soumagnac, Masci, Fremling, Perley, Nordin, Cenko, Ho, Adams, Adreoni, Bellm, Blagorodnova, Burdge, De, Dekany, Dhawan, Drake, Duev, Graham, Graham, Jencson, Karamehmetoglu, Kasliwal, Kulkarni, Miller, Neill, Prince, Riddle, Rusholme, Sharma, Smith, Sravan, Taggart, Walters, \& Yan}]{bruch_prevalence_2023}
Bruch, R.~J., Gal-Yam, A., Yaron, O., {et~al.} 2023, The Astrophysical Journal, 952, 119, \dodoi{10.3847/1538-4357/acd8be}

\bibitem[{Bullivant {et~al.}(2018)Bullivant, Smith, Williams, Mauerhan, Andrews, Fong, Bilinski, Kilpatrick, Milne, Fox, Cenko, Filippenko, Zheng, Kelly, \& Clubb}]{bullivant_sn_2018}
Bullivant, C., Smith, N., Williams, G.~G., {et~al.} 2018, Monthly Notices of the Royal Astronomical Society, 476, 1497, \dodoi{10.1093/mnras/sty045}

\bibitem[{Chandra {et~al.}(2023)Chandra, Chevalier, Maeda, Ray, \& Nayana}]{chandra_chandras_2023}
Chandra, P., Chevalier, R.~A., Maeda, K., Ray, A.~K., \& Nayana, A.~J. 2023, Chandra's insight into {SN} 2023ixf,  arXiv.
\newblock \url{http://arxiv.org/abs/2311.04384}

\bibitem[{Chevalier \& Fransson(1994)}]{chevalier_emission_1994}
Chevalier, R.~A., \& Fransson, C. 1994, The Astrophysical Journal, 420, 268, \dodoi{10.1086/173557}

\bibitem[{Chornock \& Filippenko(2008)}]{chornock_deviations_2008}
Chornock, R., \& Filippenko, A.~V. 2008, The Astronomical Journal, 136, 2227, \dodoi{10.1088/0004-6256/136/6/2227}

\bibitem[{Chornock {et~al.}(2010)Chornock, Filippenko, Li, \& Silverman}]{chornock_large_2010}
Chornock, R., Filippenko, A.~V., Li, W., \& Silverman, J.~M. 2010, ApJ, 713, 1363, \dodoi{10.1088/0004-637X/713/2/1363}

\bibitem[{Chugai(2001)}]{chugai_broad_2001}
Chugai, N. 2001, MNRAS, 326, 1448, \dodoi{10.1111/j.1365-2966.2001.04717.x}

\bibitem[{Chugai(1992)}]{chugai_polarization_1992}
Chugai, N.~N. 1992, Soviet Astronomy Letters, 18, 168.
\newblock \url{https://ui.adsabs.harvard.edu/abs/1992SvAL...18..168C}

\bibitem[{Chugai(2006)}]{chugai_asymmetry_2006}
---. 2006, Astronomy Letters, 32, 739, \dodoi{10.1134/S1063773706110041}

\bibitem[{Cikota {et~al.}(2019)Cikota, Patat, Wang, Wheeler, Bulla, Baade, Höflich, Cikota, Clocchiatti, Maund, Stevance, \& Yang}]{cikota_linear_2019}
Cikota, A., Patat, F., Wang, L., {et~al.} 2019, Monthly Notices of the Royal Astronomical Society, 490, 578, \dodoi{10.1093/mnras/stz2322}

\bibitem[{Davis \& Greenstein(1951)}]{davis_polarization_1951}
Davis, Jr., L., \& Greenstein, J.~L. 1951, The Astrophysical Journal, 114, 206, \dodoi{10.1086/145464}

\bibitem[{Dessart(2024)}]{dessart_interacting_2024}
Dessart, L. 2024, Interacting supernovae,  arXiv.
\newblock \url{http://arxiv.org/abs/2405.04259}

\bibitem[{Dessart \& Hillier(2011)}]{dessart_synthetic_2011}
Dessart, L., \& Hillier, D.~J. 2011, MNRAS, 415, 3497, \dodoi{10.1111/j.1365-2966.2011.18967.x}

\bibitem[{Dessart \& Hillier(2022)}]{dessart_modeling_2022}
---. 2022, A \& A, 660, L9, \dodoi{10.1051/0004-6361/202243372}

\bibitem[{Dessart {et~al.}(2017)Dessart, Hillier, \& Audit}]{dessart_explosion_2017}
Dessart, L., Hillier, D.~J., \& Audit, E. 2017, Astronomy \& Astrophysics, 605, A83, \dodoi{10.1051/0004-6361/201730942}

\bibitem[{Dessart {et~al.}(2021{\natexlab{a}})Dessart, Hillier, \& Leonard}]{dessart_polarization_2021}
Dessart, L., Hillier, D.~J., \& Leonard, D.~C. 2021{\natexlab{a}}, A\&A, 651, A10, \dodoi{10.1051/0004-6361/202140409}

\bibitem[{Dessart {et~al.}(2024{\natexlab{a}})Dessart, Hillier, \& Leonard}]{dessart_evolution_2024}
---. 2024{\natexlab{a}}, Astronomy \& Astrophysics, 684, A16, \dodoi{10.1051/0004-6361/202347808}

\bibitem[{Dessart {et~al.}(2013)Dessart, Hillier, Waldman, \& Livne}]{dessart_type_2013}
Dessart, L., Hillier, D.~J., Waldman, R., \& Livne, E. 2013, Monthly Notices of the Royal Astronomical Society, 433, 1745, \dodoi{10.1093/mnras/stt861}

\bibitem[{Dessart {et~al.}(2021{\natexlab{b}})Dessart, Leonard, Hillier, \& Pignata}]{dessart_multiepoch_2021}
Dessart, L., Leonard, D.~C., Hillier, D.~J., \& Pignata, G. 2021{\natexlab{b}}, A\&A, 651, A19, \dodoi{10.1051/0004-6361/202140281}

\bibitem[{Dessart {et~al.}(2024{\natexlab{b}})Dessart, Leonard, Vasylyev, \& Hillier}]{dessart_spectropolarimetric_2024}
Dessart, L., Leonard, D.~C., Vasylyev, S.~S., \& Hillier, D.~J. 2024{\natexlab{b}}, Spectropolarimetric modeling of interacting {Type} {II} supernovae. {Application} to early-time observations of {SN1998S},  arXiv.
\newblock \url{http://arxiv.org/abs/2409.13562}

\bibitem[{Dickinson {et~al.}(2024)Dickinson, Milisavljevic, Garretson, Dessart, Margutti, Chornock, Subrayan, Hillier, Golub, Li, Logsdon, Rajagopal, Ridgway, Smith, \& Cynamon}]{dickinson_immediate_2024}
Dickinson, D., Milisavljevic, D., Garretson, B., {et~al.} 2024, The {Immediate}, {eXemplary}, and {Fleeting} echelle spectroscopy of {SN} 2023ixf: {Monitoring} acceleration of slow progenitor circumstellar material, driven by shock interaction,  arXiv, \dodoi{10.48550/arXiv.2412.14406}

\bibitem[{Fassia {et~al.}(2001)Fassia, Meikle, Chugai, Geballe, Lundqvist, Walton, Pollacco, Veilleux, Wright, Pettini, Kerr, Puchnarewicz, Puxley, Irwin, Packham, Smartt, \& Harmer}]{fassia_optical_2001}
Fassia, A., Meikle, W. P.~S., Chugai, N., {et~al.} 2001, MNRAS, 325, 907, \dodoi{10.1046/j.1365-8711.2001.04282.x}

\bibitem[{Filippenko(1982)}]{filippenko_importance_1982}
Filippenko, A.~V. 1982, Publications of the Astronomical Society of the Pacific, 94, 715, \dodoi{10.1086/131052}

\bibitem[{Filippenko(1997)}]{filippenko_optical_1997}
---. 1997, Annual Review of A \& A, 35, 309, \dodoi{10.1146/annurev.astro.35.1.309}

\bibitem[{Fuller \& Tsuna(2024)}]{fuller_boil-off_2024}
Fuller, J., \& Tsuna, D. 2024, The Open Journal of Astrophysics, 7, \dodoi{10.33232/001c.120130}

\bibitem[{Gal-Yam(2017)}]{gal-yam_observational_2017}
Gal-Yam, A. 2017, in Handbook of {Supernovae}, ed. A.~W. Alsabti \& P.~Murdin (Cham: Springer International Publishing), 1--43, \dodoi{10.1007/978-3-319-20794-0_35-1}

\bibitem[{Georgy {et~al.}(2013)Georgy, Walder, Folini, Bykov, Marcowith, \& Favre}]{georgy_circumstellar_2013}
Georgy, C., Walder, R., Folini, D., {et~al.} 2013, Astronomy \& Astrophysics, 559, A69, \dodoi{10.1051/0004-6361/201321226}

\bibitem[{Goldberg {et~al.}(2022)Goldberg, Jiang, \& Bildsten}]{goldberg_shock_2022}
Goldberg, J.~A., Jiang, Y.-F., \& Bildsten, L. 2022, The Astrophysical Journal, 933, 164, \dodoi{10.3847/1538-4357/ac75e3}

\bibitem[{González {et~al.}(2007)González, Audit, \& Huynh}]{gonzalez_heracles_2007}
González, M., Audit, E., \& Huynh, P. 2007, Astronomy \& Astrophysics, 464, 429, \dodoi{10.1051/0004-6361:20065486}

\bibitem[{Grefenstette(2023)}]{grefenstette_nustar_2023}
Grefenstette, B. 2023, The Astronomer's Telegram, 16049, 1.
\newblock \url{https://ui.adsabs.harvard.edu/abs/2023ATel16049....1G}

\bibitem[{Gutiérrez {et~al.}(2017)Gutiérrez, Anderson, Hamuy, Morrell, González-Gaitan, Stritzinger, Phillips, Galbany, Folatelli, Dessart, Contreras, Valle, Freedman, Hsiao, Krisciunas, Madore, Maza, Suntzeff, Prieto, González, Cappellaro, Navarrete, Pizzella, Ruiz, Smith, \& Turatto}]{gutierrez_type_2017}
Gutiérrez, C.~P., Anderson, J.~P., Hamuy, M., {et~al.} 2017, ApJ, 850, 89, \dodoi{10.3847/1538-4357/aa8f52}

\bibitem[{Hillier(1994)}]{hillier_calculation_1994}
Hillier, D.~J. 1994, Astronomy and Astrophysics, 289, 492.
\newblock \url{https://ui.adsabs.harvard.edu/abs/1994A&A...289..492H}

\bibitem[{Hillier(1996)}]{hillier_calculation_1996}
---. 1996, Astronomy and Astrophysics, 308, 521.
\newblock \url{https://ui.adsabs.harvard.edu/abs/1996A&A...308..521H}

\bibitem[{Hillier \& Dessart(2012)}]{hillier_time-dependent_2012}
Hillier, D.~J., \& Dessart, L. 2012, Monthly Notices of the Royal Astronomical Society, 424, 252, \dodoi{10.1111/j.1365-2966.2012.21192.x}

\bibitem[{Hillier \& Dessart(2019)}]{hillier_photometric_2019}
---. 2019, A\&A, 631, A8, \dodoi{10.1051/0004-6361/201935100}

\bibitem[{Hillier \& Lanz(2001)}]{hillier_cmfgen_2001}
Hillier, D.~J., \& Lanz, T. 2001, 247, 343.
\newblock \url{https://ui.adsabs.harvard.edu/abs/2001ASPC..247..343H}

\bibitem[{Hiramatsu {et~al.}(2021)Hiramatsu, Howell, Moriya, Goldberg, Hosseinzadeh, Arcavi, Anderson, Gutiérrez, Burke, McCully, Valenti, Galbany, Fang, Maeda, Folatelli, Hsiao, Morrell, Phillips, Stritzinger, Suntzeff, Gromadzki, Maguire, Müller-Bravo, \& Young}]{hiramatsu_luminous_2021}
Hiramatsu, D., Howell, D.~A., Moriya, T.~J., {et~al.} 2021, The Astrophysical Journal, 913, 55, \dodoi{10.3847/1538-4357/abf6d6}

\bibitem[{Hiramatsu {et~al.}(2023)Hiramatsu, Tsuna, Berger, Itagaki, Goldberg, Gomez, De, Hosseinzadeh, Bostroem, Brown, Arcavi, Bieryla, Blanchard, Esquerdo, Farah, Howell, Matsumoto, McCully, Newsome, Gonzalez, Pellegrino, Rhee, Terreran, Vinkó, \& Wheeler}]{hiramatsu_discovery_2023}
Hiramatsu, D., Tsuna, D., Berger, E., {et~al.} 2023, The Astrophysical Journal Letters, 955, L8, \dodoi{10.3847/2041-8213/acf299}

\bibitem[{Hoeflich(1991)}]{hoflich_asphericity_1991}
Hoeflich, P. 1991, A \& A, 246, 481

\bibitem[{Hoeflich {et~al.}(1996)Hoeflich, Wheeler, Hines, \& Trammell}]{hoeflich_analysis_1996}
Hoeflich, P., Wheeler, J.~C., Hines, D.~C., \& Trammell, S.~R. 1996, The Astrophysical Journal, 459, 307, \dodoi{10.1086/176894}

\bibitem[{Hoffman {et~al.}(2008)Hoffman, Leonard, Chornock, Filippenko, Barth, \& Matheson}]{hoffman_dual-axis_2008}
Hoffman, J.~L., Leonard, D.~C., Chornock, R., {et~al.} 2008, ApJ, 688, 1186, \dodoi{10.1086/592261}

\bibitem[{Hole {et~al.}(2010)Hole, Kasen, \& Nordsieck}]{hole_spectropolarimetric_2010}
Hole, K.~T., Kasen, D., \& Nordsieck, K.~H. 2010, The Astrophysical Journal, 720, 1500, \dodoi{10.1088/0004-637X/720/2/1500}

\bibitem[{Hosseinzadeh {et~al.}(2022)Hosseinzadeh, Kilpatrick, Dong, Sand, Andrews, Bostroem, Janzen, Jencson, Lundquist, Meza, Pearson, Valenti, Wyatt, Burke, Hiramatsu, Howell, McCully, Newsome, Gonzalez, Pellegrino, Terreran, Auchettl, Davis, Foley, Miao, Pan, Rest, Siebert, Taggart, Tucker, Leung, Swift, Yang, Anderson, Ashall, Benetti, Brown, Cartier, Chen, Della~Valle, Galbany, Gomez, Gromadzki, Haislip, Hsiao, Inserra, Jha, Killestein, Kouprianov, Kozyreva, Müller-Bravo, Nicholl, Paraskeva, Reichart, Ryder, Shahbandeh, Shappee, Smith, \& Young}]{hosseinzadeh_weak_2022}
Hosseinzadeh, G., Kilpatrick, C.~D., Dong, Y., {et~al.} 2022, arXiv:2203.08155 [astro-ph].
\newblock \url{http://arxiv.org/abs/2203.08155}

\bibitem[{{Hosseinzadeh} {et~al.}(2023){Hosseinzadeh}, {Farah}, {Shrestha}, {Sand}, {Dong}, {Brown}, {Bostroem}, {Valenti}, {Jha}, {Andrews}, {Arcavi}, {Haislip}, {Hiramatsu}, {Hoang}, {Howell}, {Janzen}, {Jencson}, {Kouprianov}, {Lundquist}, {McCully}, {Meza Retamal}, {Modjaz}, {Newsome}, {Padilla Gonzalez}, {Pearson}, {Pellegrino}, {Ravi}, {Reichart}, {Smith}, {Terreran}, \& {Vink{\'o}}}]{2023arXiv230606097H}
{Hosseinzadeh}, G., {Farah}, J., {Shrestha}, M., {et~al.} 2023, arXiv e-prints, arXiv:2306.06097, \dodoi{10.48550/arXiv.2306.06097}

\bibitem[{Hsu {et~al.}(2024)Hsu, Smith, Goldberg, Bostroem, Hosseinzadeh, Sand, Pearson, Hiramatsu, Andrews, Beasor, Dong, Farah, Galbany, Gomez, Gonzalez, Gutiérrez, Howell, Könyves-Tóth, McCully, Newsome, Shrestha, Terreran, Villar, \& Wang}]{hsu_one_2024}
Hsu, B., Smith, N., Goldberg, J.~A., {et~al.} 2024, One {Year} of {SN} 2023ixf: {Breaking} {Through} the {Degenerate} {Parameter} {Space} in {Light}-{Curve} {Models} with {Pulsating} {Progenitors},  arXiv, \dodoi{10.48550/arXiv.2408.07874}

\bibitem[{Huang \& Chevalier(2018)}]{huang__electron_2018}
Huang, C., \& Chevalier, R.~A. 2018, Monthly Notices of the Royal Astronomical Society, 475, 1261, \dodoi{10.1093/mnras/stx3163}

\bibitem[{Jacobson-Galan {et~al.}(2023)Jacobson-Galan, Dessart, Margutti, Chornock, Foley, Kilpatrick, Jones, Taggart, Angus, Bhattacharjee, Braff, Brethauer, Burgasser, Cao, Carlile, Chambers, Coulter, Dominguez-Ruiz, Dickinson, de~Boer, Gagliano, Gall, Gao, Gates, Gomez, Guolo, Halford, Hjorth, Huber, Johnson, Karpoor, Laskar, LeBaron, Li, Lin, Loch, Lynam, Magnier, Maloney, Matthews, McDonald, Miao, Milisavljevic, Pan, Pradyumna, Ransome, Rees, Rest, Rojas-Bravo, Sandford, Ascencio, Sanjaripour, Savino, Sears, Sharei, Smartt, Softich, Theissen, Tinyanont, Tohfa, Villar, Wang, Wainscoat, Westerling, Wiston, Wozniak, Yadavalli, \& Zenati}]{jacobson-galan_sn_2023}
Jacobson-Galan, W.~V., Dessart, L., Margutti, R., {et~al.} 2023, {SN} 2023ixf in {Messier} 101: {Photo}-ionization of {Dense}, {Close}-in {Circumstellar} {Material} in a {Nearby} {Type} {II} {Supernova},  arXiv, \dodoi{10.48550/arXiv.2306.04721}

\bibitem[{{Jacobson-Galan} {et~al.}(2023){Jacobson-Galan}, {Dessart}, {Margutti}, {Chornock}, {Foley}, {Kilpatrick}, {Jones}, {Taggart}, {Angus}, {Bhattacharjee}, {Braff}, {Brethauer}, {Burgasser}, {Cao}, {Carlile}, {Chambers}, {Coulter}, {Dominguez-Ruiz}, {Dickinson}, {de Boer}, {Gagliano}, {Gall}, {Gao}, {Gates}, {Gomez}, {Guolo}, {Halford}, {Hjorth}, {Huber}, {Johnson}, {Karpoor}, {Laskar}, {LeBaron}, {Li}, {Lin}, {Loch}, {Lynam}, {Magnier}, {Maloney}, {Matthews}, {McDonald}, {Miao}, {Milisavljevic}, {Pan}, {Pradyumna}, {Ransome}, {Rees}, {Rest}, {Rojas-Bravo}, {Sandford}, {Sandoval Ascencio}, {Sanjaripour}, {Savino}, {Sears}, {Sharei}, {Smartt}, {Softich}, {Theissen}, {Tinyanont}, {Tohfa}, {Villar}, {Wang}, {Wainscoat}, {Westerling}, {Wiston}, {Wozniak}, {Yadavalli}, \& {Zenati}}]{2023arXiv230604721J}
{Jacobson-Galan}, W.~V., {Dessart}, L., {Margutti}, R., {et~al.} 2023, arXiv e-prints, arXiv:2306.04721, \dodoi{10.48550/arXiv.2306.04721}

\bibitem[{Jacobson-Galán {et~al.}(2024)Jacobson-Galán, Dessart, Davis, Kilpatrick, Margutti, Foley, Chornock, Terreran, Hiramatsu, Newsome, Gonzalez, Pellegrino, Howell, Filippenko, Anderson, Angus, Auchettl, Bostroem, Brink, Cartier, Coulter, de~Boer, Drout, Earl, Ertini, Farah, Farias, Gall, Gao, Gerlach, Guo, Haynie, Hosseinzadeh, Ibik, Jha, Jones, Langeroodi, LeBaron, Magnier, Piro, Raimundo, Rest, Rest, Rich, Rojas-Bravo, Sears, Taggart, Villar, Wainscoat, Wang, Wasserman, Yan, Yang, Zhang, \& Zheng}]{jacobson-galan_final_2024}
Jacobson-Galán, W.~V., Dessart, L., Davis, K.~W., {et~al.} 2024, Final {Moments} {II}: {Observational} {Properties} and {Physical} {Modeling} of {CSM}-{Interacting} {Type} {II} {Supernovae},  arXiv.
\newblock \url{http://arxiv.org/abs/2403.02382}

\bibitem[{Kasen {et~al.}(2006)Kasen, Thomas, \& Nugent}]{kasen_time-dependent_2006}
Kasen, D., Thomas, R.~C., \& Nugent, P. 2006, The Astrophysical Journal, 651, 366, \dodoi{10.1086/506190}

\bibitem[{Kasen {et~al.}(2003)Kasen, Nugent, Wang, Howell, Wheeler, Hoflich, Baade, Baron, \& Hauschildt}]{kasen_analysis_2003}
Kasen, D., Nugent, P., Wang, L., {et~al.} 2003, The Astrophysical Journal, 593, 788, \dodoi{10.1086/376601}

\bibitem[{Khazov {et~al.}(2016)Khazov, Yaron, Gal-Yam, Manulis, Rubin, Kulkarni, Arcavi, Kasliwal, Ofek, Cao, Perley, Sollerman, Horesh, Sullivan, Filippenko, Nugent, Howell, Cenko, Silverman, Ebeling, Taddia, Johansson, Laher, Surace, Rebbapragada, Wozniak, \& Matheson}]{khazov_flash_2016}
Khazov, D., Yaron, O., Gal-Yam, A., {et~al.} 2016, The Astrophysical Journal, 818, 3, \dodoi{10.3847/0004-637X/818/1/3}

\bibitem[{Kumar {et~al.}(2025)Kumar, Dastidar, Maund, Singleton, \& Sun}]{kumar_signatures_2025}
Kumar, A., Dastidar, R., Maund, J.~R., Singleton, A.~J., \& Sun, N.-C. 2025, Signatures of the {Shock} {Interaction} as an {Additional} {Power} {Source} in the {Nebular} {Spectra} of {SN} 2023ixf,  arXiv, \dodoi{10.48550/arXiv.2412.03509}

\bibitem[{Kumar {et~al.}(2014)Kumar, Pandey, Eswaraiah, \& Gorosabel}]{kumar_broad-band_2014}
Kumar, B., Pandey, S.~B., Eswaraiah, C., \& Gorosabel, J. 2014, Monthly Notices of the Royal Astronomical Society, 442, 2, \dodoi{10.1093/mnras/stu811}

\bibitem[{Kumar {et~al.}(2019)Kumar, Eswaraiah, Singh, Sahu, Anupama, Kawabata, Yamanaka, Otsubo, Pandey, Nakaoka, Kawabata, Aryan, \& Akitaya}]{kumar_observational_2019}
Kumar, B., Eswaraiah, C., Singh, A., {et~al.} 2019, Monthly Notices of the Royal Astronomical Society, 488, 3089, \dodoi{10.1093/mnras/stz1914}

\bibitem[{Landsman(1993)}]{landsman_idl_1993}
Landsman, W.~B. 1993, 52, 246.
\newblock \url{https://ui.adsabs.harvard.edu/abs/1993ASPC...52..246L}

\bibitem[{{Leonard} {et~al.}(2001){Leonard}, {Filippenko}, {Ardila}, \& {Brotherton}}]{leonard_is_2001}
{Leonard}, D.~C., {Filippenko}, A.~V., {Ardila}, D.~R., \& {Brotherton}, M.~S. 2001, \apj, 553, 861, \dodoi{10.1086/320959}

\bibitem[{Leonard {et~al.}(2000)Leonard, Filippenko, Barth, \& Matheson}]{leonard_evidence_2000}
Leonard, D.~C., Filippenko, A.~V., Barth, A.~J., \& Matheson, T. 2000, ApJ, 536, 239, \dodoi{10.1086/308910}

\bibitem[{Leonard {et~al.}(2006)Leonard, Filippenko, Ganeshalingam, Serduke, Li, Swift, Gal-Yam, Foley, Fox, Park, Hoffman, \& Wong}]{leonard_non-spherical_2006}
Leonard, D.~C., Filippenko, A.~V., Ganeshalingam, M., {et~al.} 2006, Nature, 440, 505, \dodoi{10.1038/nature04558}

\bibitem[{Li {et~al.}(2024)Li, Hu, Li, Yang, Wang, Yan, Hu, Zhang, Mao, Riise, Gao, Sun, Liu, Xiong, Wang, Mo, Iskandar, Xi, Xiang, Wang, Sun, Zhang, Chen, Lin, Guo, Liu, Cai, Zhou, Zhao, Chen, Zheng, Li, Zhang, Xu, Lyu, Castro-Tirado, Chufarin, Potapov, Ionov, Korotkiy, Nazarov, Sokolovsky, Hamann, \& Herman}]{li_shock_2024}
Li, G., Hu, M., Li, W., {et~al.} 2024, Nature, 627, 754, \dodoi{10.1038/s41586-023-06843-6}

\bibitem[{Livne(1993)}]{livne_implicit_1993}
Livne, E. 1993, The Astrophysical Journal, 412, 634, \dodoi{10.1086/172950}

\bibitem[{Maeder \& Meynet(2010)}]{maeder_evolution_2010}
Maeder, A., \& Meynet, G. 2010, New Astronomy Reviews, 54, 32, \dodoi{10.1016/j.newar.2010.09.017}

\bibitem[{Mauerhan {et~al.}(2014)Mauerhan, Williams, Smith, Smith, Filippenko, Hoffman, Milne, Leonard, Clubb, Fox, \& Kelly}]{mauerhan_multi-epoch_2014}
Mauerhan, J., Williams, G.~G., Smith, N., {et~al.} 2014, MNRAS, 442, 1166, \dodoi{10.1093/mnras/stu730}

\bibitem[{Mauerhan {et~al.}(2017)Mauerhan, Dyk, Johansson, Hu, Fox, Wang, Graham, Filippenko, \& Shivvers}]{mauerhan_asphericity_2017}
Mauerhan, J.~C., Dyk, S. D.~V., Johansson, J., {et~al.} 2017, ApJ, 834, 118, \dodoi{10.3847/1538-4357/834/2/118}

\bibitem[{Mauerhan {et~al.}(2024)Mauerhan, Smith, Williams, Smith, Filippenko, Bilinski, Zheng, Brink, Hoffman, Leonard, Milne, Jeffers, Modak, Stegman, \& Zhang}]{mauerhan_record-breaking_2024}
Mauerhan, J.~C., Smith, N., Williams, G.~G., {et~al.} 2024, Monthly Notices of the Royal Astronomical Society, 527, 6090, \dodoi{10.1093/mnras/stad3579}

\bibitem[{Maund {et~al.}(2007)Maund, Wheeler, Patat, Wang, Baade, \& Hoflich}]{maund_spectropolarimetry_2007}
Maund, J.~R., Wheeler, J.~C., Patat, F., {et~al.} 2007, ApJ, 671, 1944, \dodoi{10.1086/523261}

\bibitem[{Mauron \& Josselin(2011)}]{mauron_mass-loss_2011}
Mauron, N., \& Josselin, E. 2011, Astronomy \& Astrophysics, 526, A156, \dodoi{10.1051/0004-6361/201013993}

\bibitem[{Miller \& Stone(1994)}]{miller_stone_1994}
Miller, J., \& Stone, R. 1994, The Kast Double Spectrograph

\bibitem[{Miller {et~al.}(1988)Miller, Robinson, \& Goodrich}]{miller_ccd_1988}
Miller, J.~S., Robinson, L.~B., \& Goodrich, R.~W. 1988, in Instrumentation for {Ground}-{Based} {Optical} {Astronomy}, ed. L.~B. Robinson, Santa {Cruz} {Summer} {Workshops} in {Astronomy} and {Astrophysics} (New York, NY: Springer), 157--171, \dodoi{10.1007/978-1-4612-3880-5_14}

\bibitem[{Morozova {et~al.}(2017)Morozova, Piro, \& Valenti}]{morozova_unifying_2017}
Morozova, V., Piro, A.~L., \& Valenti, S. 2017, The Astrophysical Journal, 838, 28, \dodoi{10.3847/1538-4357/aa6251}

\bibitem[{Morozova {et~al.}(2018)Morozova, Piro, \& Valenti}]{morozova_measuring_2018}
---. 2018, The Astrophysical Journal, 858, 15, \dodoi{10.3847/1538-4357/aab9a6}

\bibitem[{Nagao {et~al.}(2019)Nagao, Cikota, Patat, Taubenberger, Bulla, Faran, Sand, Valenti, Andrews, \& Reichart}]{nagao_aspherical_2019}
Nagao, T., Cikota, A., Patat, F., {et~al.} 2019, MNRAS: Letters, 489, L69, \dodoi{10.1093/mnrasl/slz119}

\bibitem[{Nagao {et~al.}(2021)Nagao, Patat, Taubenberger, Baade, Faran, Cikota, Sand, Bulla, Kuncarayakti, Maund, Tartaglia, Valenti, \& Reichart}]{nagao_evidence_2021}
Nagao, T., Patat, F., Taubenberger, S., {et~al.} 2021, MNRAS, 505, 3664, \dodoi{10.1093/mnras/stab1582}

\bibitem[{Nagao {et~al.}(2024{\natexlab{a}})Nagao, Maeda, Mattila, Kuncarayakti, Kawabata, Taguchi, Nakaoka, Cikota, Bulla, Vasylyev, Gutiérrez, Yamanaka, Isogai, Uno, Ogawa, Inutsuka, Tsurumi, Imazawa, \& Kawabata}]{nagao_evidence_2024}
Nagao, T., Maeda, K., Mattila, S., {et~al.} 2024{\natexlab{a}}, Astronomy \& Astrophysics, 687, L17, \dodoi{10.1051/0004-6361/202450191}

\bibitem[{Nagao {et~al.}(2024{\natexlab{b}})Nagao, Patat, Cikota, Baade, Mattila, Kotak, Kuncarayakti, Bulla, \& Ayala}]{nagao_spectropolarimetry_2024}
Nagao, T., Patat, F., Cikota, A., {et~al.} 2024{\natexlab{b}}, Astronomy \& Astrophysics, 681, A11, \dodoi{10.1051/0004-6361/202346715}

\bibitem[{Nayana {et~al.}(2024)Nayana, Margutti, Wiston, Chornock, Campana, Laskar, Murase, Krips, Migliori, Tsuna, Alexander, Chandra, Bietenholz, Berger, Chevalier, Colle, Dessart, Diesing, Grefenstette, Jacobson-Galan, Maeda, Marcote, Matthews, Milisavljevic, Ray, Reguitti, \& Polzin}]{nayana_dinosaur_2024}
Nayana, A.~J., Margutti, R., Wiston, E., {et~al.} 2024, Dinosaur in a {Haystack} : {X}-ray {View} of the {Entrails} of {SN} 2023ixf and the {Radio} {Afterglow} of {Its} {Interaction} with the {Medium} {Spawned} by the {Progenitor} {Star} ({Paper} 1),  arXiv, \dodoi{10.48550/arXiv.2411.02647}

\bibitem[{Oke \& Gunn(1983)}]{oke_secondary_1983}
Oke, J.~B., \& Gunn, J.~E. 1983, The Astrophysical Journal, 266, 713, \dodoi{10.1086/160817}

\bibitem[{Patat(2017)}]{patat_introduction_2017}
Patat, F. 2017, in Handbook of {Supernovae}, ed. A.~W. Alsabti \& P.~Murdin (Cham: Springer International Publishing), 1017--1049

\bibitem[{Patat {et~al.}(2011)Patat, Taubenberger, Benetti, Pastorello, \& Harutyunyan}]{patat_asymmetries_2011}
Patat, F., Taubenberger, S., Benetti, S., Pastorello, A., \& Harutyunyan, A. 2011, A \& A, 527, L6, \dodoi{10.1051/0004-6361/201016217}

\bibitem[{Patra {et~al.}(2021)Patra, Yang~(杨轶), Brink, Höflich, Wang, Filippenko, Kasen, Baade, Foley, Maund, Zheng, Hung, Cikota, Wheeler, \& Bulla}]{patra_spectropolarimetry_2021}
Patra, K.~C., Yang~(杨轶), Y., Brink, T.~G., {et~al.} 2021, MNRAS, 509, 4058, \dodoi{10.1093/mnras/stab3136}

\bibitem[{Reilly {et~al.}(2017)Reilly, Maund, Baade, Wheeler, Höflich, Spyromilio, Patat, \& Wang}]{reilly_spectropolarimetry_2017}
Reilly, E., Maund, J.~R., Baade, D., {et~al.} 2017, Monthly Notices of the Royal Astronomical Society, 470, 1491, \dodoi{10.1093/mnras/stx1228}

\bibitem[{Roberge \& Lazarian(1999)}]{roberge_davisgreenstein_1999}
Roberge, W.~G., \& Lazarian, A. 1999, Monthly Notices of the Royal Astronomical Society, 305, 615, \dodoi{10.1046/j.1365-8711.1999.02464.x}

\bibitem[{Serkowski {et~al.}(1975)Serkowski, Mathewson, \& Ford}]{serkowski_wavelength_1975}
Serkowski, K., Mathewson, D.~S., \& Ford, V.~L. 1975, The Astrophysical Journal, 196, 261, \dodoi{10.1086/153410}

\bibitem[{Shivvers {et~al.}(2015)Shivvers, Groh, Mauerhan, Fox, Leonard, \& Filippenko}]{shivvers_early_2015}
Shivvers, I., Groh, J.~H., Mauerhan, J.~C., {et~al.} 2015, ApJ, 806, 213, \dodoi{10.1088/0004-637X/806/2/213}

\bibitem[{Shrestha {et~al.}(2024)Shrestha, DeSoto, Sand, Williams, Hoffman, Smith, Smith, Milne, McCall, Maund, Steele, Wiersema, Andrews, Bilinski, Anche, Bostroem, Hosseinzadeh, Pearson, Leonard, Hsu, Dong, Hoang, Janzen, Jencson, Jha, Lundquist, Mehta, Retamal, Valenti, Farah, Howell, McCully, Newsome, Gonzalez, Pellegrino, \& Terreran}]{shrestha_spectropolarimetry_2024}
Shrestha, M., DeSoto, S., Sand, D.~J., {et~al.} 2024, Spectropolarimetry of {SN} 2023ixf reveals both circumstellar material and helium core to be aspherical,  arXiv, \dodoi{10.48550/arXiv.2410.08199}

\bibitem[{Singh {et~al.}(2024)Singh, Teja, Moriya, Maeda, Kawabata, Tanaka, Imazawa, Nakaoka, Gangopadhyay, Yamanaka, Swain, Sahu, Anupama, Kumar, Anche, Sano, Raj, Agnihotri, Bhalerao, Bisht, Bisht, Belwal, Chakrabarti, Fujii, Nagayama, Matsumoto, Hamada, Kawabata, Kumar, Kumar, Malkan, Smith, Sakagami, Taguchi, Tominaga, \& Watanabe}]{singh_unravelling_2024}
Singh, A., Teja, R.~S., Moriya, T.~J., {et~al.} 2024, Unravelling the asphericities in the explosion and multi-faceted circumstellar matter of {SN} 2023ixf,  arXiv.
\newblock \url{http://arxiv.org/abs/2405.20989}

\bibitem[{Smith(2014)}]{smith_mass_2014}
Smith, N. 2014, Annual Review of Astronomy and Astrophysics, 52, 487, \dodoi{10.1146/annurev-astro-081913-040025}

\bibitem[{Smith {et~al.}(2008)Smith, Chornock, Li, Ganeshalingam, Silverman, Foley, Filippenko, \& Barth}]{smith_sn_2008}
Smith, N., Chornock, R., Li, W., {et~al.} 2008, The Astrophysical Journal, 686, 467, \dodoi{10.1086/591021}

\bibitem[{Smith {et~al.}(2011)Smith, Li, Filippenko, \& Chornock}]{smith_observed_2011}
Smith, N., Li, W., Filippenko, A.~V., \& Chornock, R. 2011, Monthly Notices of the Royal Astronomical Society, 412, 1522, \dodoi{10.1111/j.1365-2966.2011.17229.x}

\bibitem[{Smith {et~al.}(2023)Smith, Pearson, Sand, Ilyin, Bostroem, Hosseinzadeh, \& Shrestha}]{smith_high-resolution_2023}
Smith, N., Pearson, J., Sand, D.~J., {et~al.} 2023, The Astrophysical Journal, 956, 46, \dodoi{10.3847/1538-4357/acf366}

\bibitem[{{Smith} {et~al.}(2023){Smith}, {Pearson}, {Sand}, {Ilyin}, {Bostroem}, {Hosseinzadeh}, \& {Shrestha}}]{2023arXiv230607964S}
{Smith}, N., {Pearson}, J., {Sand}, D.~J., {et~al.} 2023, arXiv e-prints, arXiv:2306.07964, \dodoi{10.48550/arXiv.2306.07964}

\bibitem[{Soker(2023)}]{soker_pre-explosion_2023}
Soker, N. 2023, RA\&A, 23, 081002, \dodoi{10.1088/1674-4527/ace51f}

\bibitem[{Stevance {et~al.}(2016)Stevance, Maund, Baade, Höflich, Patat, Spyromilio, Wheeler, Clocchiatti, Wang, Yang, \& Zelaya}]{stevance_spectropolarimetry_2016}
Stevance, H.~F., Maund, J.~R., Baade, D., {et~al.} 2016, Monthly Notices of the Royal Astronomical Society, 461, 2019, \dodoi{10.1093/mnras/stw1479}

\bibitem[{Tanaka {et~al.}(2017)Tanaka, Maeda, Mazzali, Kawabata, \& Nomoto}]{tanaka_three-dimensional_2017}
Tanaka, M., Maeda, K., Mazzali, P.~A., Kawabata, K.~S., \& Nomoto, K. 2017, ApJ, 837, 105, \dodoi{10.3847/1538-4357/aa6035}

\bibitem[{Valenti {et~al.}(2016)Valenti, Howell, Stritzinger, Graham, Hosseinzadeh, Arcavi, Bildsten, Jerkstrand, McCully, Pastorello, Piro, Sand, Smartt, Terreran, Baltay, Benetti, Brown, Filippenko, Fraser, Rabinowitz, Sullivan, \& Yuan}]{valenti_diversity_2016}
Valenti, S., Howell, D.~A., Stritzinger, M.~D., {et~al.} 2016, MNRAS, 459, 3939, \dodoi{10.1093/mnras/stw870}

\bibitem[{Vasylyev {et~al.}(2022)Vasylyev, Filippenko, Vogl, Brink, Brown, Jaeger, Matheson, Gal-Yam, Mazzali, Modjaz, Patra, Rowe, Smith, Dyk, Williamson, Yang, Zheng, deGraw, Fox, Gates, Jennings, \& Rich}]{vasylyev_early-time_2022}
Vasylyev, S.~S., Filippenko, A.~V., Vogl, C., {et~al.} 2022, ApJ, 934, 134, \dodoi{10.3847/1538-4357/ac7220}

\bibitem[{Vasylyev {et~al.}(2023)Vasylyev, Yang, Filippenko, Patra, Brink, Wang, Chornock, Margutti, Gates, Burgasser, Karpoor, LeBaron, Softich, Theissen, Wiston, \& Zheng}]{vasylyev_early_2023}
Vasylyev, S.~S., Yang, Y., Filippenko, A.~V., {et~al.} 2023, The Astrophysical Journal Letters, 955, L37, \dodoi{10.3847/2041-8213/acf1a3}

\bibitem[{Vasylyev {et~al.}(2024)Vasylyev, Yang, Patra, Filippenko, Baade, Brink, Hoeflich, Maund, Patat, Wang, Wheeler, \& Zheng}]{vasylyev_spectropolarimetry_2024}
Vasylyev, S.~S., Yang, Y., Patra, K.~C., {et~al.} 2024, Monthly Notices of the Royal Astronomical Society, 527, 3106, \dodoi{10.1093/mnras/stad3352}

\bibitem[{Wang {et~al.}(2001)Wang, Howell, Höflich, \& Wheeler}]{wang_bipolar_2001}
Wang, L., Howell, D.~A., Höflich, P., \& Wheeler, J.~C. 2001, The Astrophysical Journal, 550, 1030, \dodoi{10.1086/319822}

\bibitem[{Wang \& Wheeler(2008)}]{wang_spectropolarimetry_2008}
Wang, L., \& Wheeler, J.~C. 2008, Annual Review of A \& A, 46, 433, \dodoi{10.1146/annurev.astro.46.060407.145139}

\bibitem[{{Wang} {et~al.}(2002){Wang}, {Wheeler}, {H{\"o}flich}, {Khokhlov}, {Baade}, {Branch}, {Challis}, {Filippenko}, {Fransson}, {Garnavich}, {Kirshner}, {Lundqvist}, {McCray}, {Panagia}, {Pun}, {Phillips}, {Sonneborn}, \& {Suntzeff}}]{Wang1987A2002ApJ...579..671W}
{Wang}, L., {Wheeler}, J.~C., {H{\"o}flich}, P., {et~al.} 2002, \apj, 579, 671, \dodoi{10.1086/342824}

\bibitem[{Wang {et~al.}(2002)Wang, Wheeler, Höflich, Khokhlov, Baade, Branch, Challis, Filippenko, Fransson, Garnavich, Kirshner, Lundqvist, McCray, Panagia, Pun, Phillips, Sonneborn, \& Suntzeff}]{wang_axisymmetric_2002}
Wang, L., Wheeler, J.~C., Höflich, P., {et~al.} 2002, The Astrophysical Journal, 579, 671, \dodoi{10.1086/342824}

\bibitem[{Wongwathanarat {et~al.}(2015)Wongwathanarat, Müller, \& Janka}]{wongwathanarat_three-dimensional_2015}
Wongwathanarat, A., Müller, E., \& Janka, H.-T. 2015, A \& A, 577, A48, \dodoi{10.1051/0004-6361/201425025}

\bibitem[{Yaron {et~al.}(2017)Yaron, Perley, Gal-Yam, Groh, Horesh, Ofek, Kulkarni, Sollerman, Fransson, Rubin, Szabo, Sapir, Taddia, Cenko, Valenti, Arcavi, Howell, Kasliwal, Vreeswijk, Khazov, Fox, Cao, Gnat, Kelly, Nugent, Filippenko, Laher, Wozniak, Lee, Rebbapragada, Maguire, Sullivan, \& Soumagnac}]{yaron_confined_2017}
Yaron, O., Perley, D.~A., Gal-Yam, A., {et~al.} 2017, Nature Physics, 13, 510, \dodoi{10.1038/nphys4025}

\bibitem[{Yoon \& Cantiello(2010)}]{yoon_evolution_2010}
Yoon, S.-C., \& Cantiello, M. 2010, The Astrophysical Journal Letters, 717, L62, \dodoi{10.1088/2041-8205/717/1/L62}

\bibitem[{Zhang {et~al.}(2012)Zhang, Wang, Wu, Chen, Chen, Liu, Huang, Liang, Zhao, Lin, Wang, Dennefeld, Zhang, Zhai, Wu, Fan, Zou, Zhou, \& Ma}]{zhang_type_2012}
Zhang, T., Wang, X., Wu, C., {et~al.} 2012, The Astronomical Journal, 144, 131, \dodoi{10.1088/0004-6256/144/5/131}

\bibitem[{Zheng {et~al.}(2025)Zheng, Dessart, Filippenko, Yang, Brink, Jaeger, Vasylyev, Dyk, Patra, Jacobson-Galan, Stewart, III, Arikatla, Beddow, Betz, Born, Bostow, Burgasser, Caceres, Carrasco, Chuang, DeGraw, Gates, Gendreau-Distler, Jacobus, Jennings, Karpoor, Lynam, Mina, Mora, Pichay, Ravi, Rees, Rich, Risin, Sandford, Savino, Softich, Theissen, Vidal, Wu, \& Zeng}]{zheng_sn_2025}
Zheng, W., Dessart, L., Filippenko, A.~V., {et~al.} 2025, {SN} 2023ixf in the {Pinwheel} {Galaxy} {M101}: {From} {Shock} {Breakout} to the {Nebular} {Phase},  arXiv, \dodoi{10.48550/arXiv.2503.13974}

\bibitem[{Zimmerman {et~al.}(2024)Zimmerman, Irani, Chen, Gal-Yam, Schulze, Perley, Sollerman, Filippenko, Shenar, Yaron, Shahaf, Bruch, Ofek, De~Cia, Brink, Yang, Vasylyev, Ben~Ami, Aubert, Badash, Bloom, Brown, De, Dimitriadis, Fransson, Fremling, Hinds, Horesh, Johansson, Kasliwal, Kulkarni, Kushnir, Martin, Matuzewski, McGurk, Miller, Morag, Neil, Nugent, Post, Prusinski, Qin, Raichoor, Riddle, Rowe, Rusholme, Sfaradi, Sjoberg, Soumagnac, Stein, Strotjohann, Terwel, Wasserman, Wise, Wold, Yan, \& Zhang}]{zimmerman_complex_2024}
Zimmerman, E.~A., Irani, I., Chen, P., {et~al.} 2024, Nature, 627, 759, \dodoi{10.1038/s41586-024-07116-6}

\end{thebibliography}

\listofchanges
\end{document}